\documentclass[fleqn,usenatbib]{mnras}

\usepackage{newtxtext,newtxmath}
\usepackage{txfonts}
\usepackage{xspace}
\usepackage[T1]{fontenc}
\usepackage{ae,aecompl}
\usepackage{url}

\usepackage{appendix}
\usepackage{comment}
\usepackage{subcaption}
\usepackage{pifont}
\usepackage{xcolor}
\usepackage{lastpage}

\usepackage{graphicx}	% Including figure files
\usepackage{amsmath}	% Advanced maths commands
\usepackage{amssymb}	% Extra maths symbols
\usepackage{float}

\usepackage[utf8]{inputenc}

\title[UGS spectra]{Spectroscopy of a Sample of Unidentified Gamma-ray Fermi Sources}

\author[A. Ulgiati et al.]{Alberto Ulgiati$^{1,2}$\thanks{E-mail:
alberto.ulgiati@inaf.it}, Simona Paiano$^{1}$\thanks{E-mail:
simona.paiano@inaf.it}, Aldo Treves$^{3,4}$, Renato Falomo$^{5}$, Boris Sbarufatti$^{4}$, Fabio Pintore$^{1}$,
\newauthor
Thomas D. Russell$^{1}$, Giancarlo Cusumano$^{1}$ \\
$^{1}$INAF - IASF Palermo, via Ugo La Malfa, 153, I-90146, Palermo, Italy \\ 
$^{2}$Universit\`a degli Studi di Palermo, Dipartimento di Fisica e Chimica, via Archirafi 36, I-90123 Palermo, Italy\\
$^{3}$Universit\`a dell'Insubria, via Valeggio, 22100, Como, Italy\\
$^{4}$INAF - Osservatorio Astronomico di Brera, via Bianchi 46, I-23807, Merate (Lecco), Italy\\
$^{5}$INAF - Osservatorio Astronomico di Padova, vicolo dell'Osservatorio 5, I-35122, Padova, Italy\\
}
\date{Accepted: 2024 February 19; Received 2024 February 5; in original form 2023 October 24.}

\begin{document}
\label{firstpage}
\pagerange{\pageref{firstpage}--\pageref{lastpage}}
\maketitle

\begin{abstract}
The fourth-DR3 version (4FGL-DR3) of the Fermi/LAT catalogue of $\gamma$-ray sources contains $\sim1000$ objects at a galactic latitude $|b| > 10^{\circ}$ which are not identified with an optical counterpart (UGS). %1125
We performed a systematic study of these sources, focusing on 190 objects that have a unique X-ray counterpart in the available \textit{Swift}/XRT observations. 
Optical counterparts are then selected, and for 33 sources optical spectra were found in the literature.
%For other $\sim$60 UGS counterparts, spectroscopy at large telescopes has been already performed by us and for the rest ongoing programs are in progress. 
%In this first paper we consider the sample of 33 literature objects. 
We found that 21 can be classified as BL Lac objects.
Among these we were able to provide the redshift for 8 of them while for 2 others we established a lower limit to the redshift by detecting intervening absorption.
The other 12 objects display optical spectra with prominent emission lines (0.036$<$z$<$1.65). 
These spectra  are characterized by both broad and narrow emission lines with the exception of 3 sources. One of them displays only broad emission lines, while the other two exclusively exhibit narrow lines.
On the basis of the radio/optical flux ratio, all BL Lac objects in this study are radio loud. Four sources out of the 12 with prominent emission lines can be classified as radio loud, while at least 5 of the 12 sources with prominent lines are radio quiet. This is somewhat unexpected comparing with the radio-loudness distribution of the 4FGL-associated blazars.
%On the basis of the radio/optical flux ratio, all BL Lac objects appear radio loud, except one; 3 sources out of the 12 with prominent emission lines can be classified as radio loud, and the rest are radio quiet.
%We compare our results with the statistics of the 4FGL-associated blazars. % as well as with UGS selected from the 2FGL/3FGL catalogs of our previous works.
\end{abstract}


\section{Introduction}  
\label{sec:introduction}

In the last 15 years the \textit{Fermi} satellite contributed dramatically to the progress of high-energy astrophysics.  
%In June 2022 \textbf{\citet[][]{4fgldr3} released 4FGL-DR3, the third version of the Fermi-LAT Fourth Source Catalogue \citep[4FGL,][]{4fgl},} based on 12 years of observations and containing 6659 $\gamma$-ray detections.
%%%%%\textbf{In June 2022, the third version of the Fermi-LAT Fourth Source Catalogue \citep[4FGL-DR3,][]{4fgl,4fgldr3},} based on 12 years of observations and containing 6659 $\gamma$-ray detections, was released.
The third version of the Fermi-LAT Fourth Source Catalogue \citep[4FGL-DR3,][]{4fgl,4fgldr3}, based on 12 years of observations and containing 6659 $\gamma$-ray detections, was released in June 2022. 

About 4500 sources are associated or identified with targets at other wavelengths thanks to a positional overlap in the sky, measurements of correlated variability at other wavelengths, and/or multi-wavelength spectral properties \citep[][]{4fgl,4fgldr3}.
The majority of the \textit{Fermi} extragalactic population (at $|b|>10^{\circ}$) belongs to the blazar class, radio jetted Active Galactic Nuclei (AGN) with relativistic jets oriented close to the line of sight of the observer.  
Therefore, the $\gamma$-rays band presents an efficient and suitable energy band to detect this class of sources.

Blazar radiation, from radio to $\gamma$-ray band, is dominated by non-thermal emission and the typical Spectral Energy Distribution (SED) exhibits a doubled-bump shape. 
The first low energy peak, ranging from radio to X-rays energy band, is due to a synchrotron emission from relativistic electrons, while for the second one at high energy (in the MeV~-~TeV range) the origin remains still under debate  and several scenarios are proposed in the framework of hadronic and/or leptonic models \citep[e.g.][]{Cerruti2011, costamante2018, Rodrigues2019, Gao2019, Cerruti2020}. %\red{REFERENZE CAMBIATE}.

Based on the properties of their optical spectra, blazars are divided in two main classes: i) Flat Spectrum Radio Quasars (FSRQs) characterised by prominent emission lines and ii) BL Lacertae objects (BLL) for which the spectral features are very weak or even absent \citep[e.g.][]{falomo2014}. 
This implies that for a large fraction of BLL the redshift is unknown or highly uncertain,  making their characterization in terms of physical properties and modelling quite difficult.

In any case, the firm classification of the Fermi sources requires spectroscopy of the optical counterpart. 
%\textbf{Every time a new Fermi catalog is released, it prompts various optical campaigns.}
After the release of the first Fermi catalogs and subsequent versions, a number of optical campaigns were activated.
One of the first substantial contributions came from \citet[][]{Shaw_2012, Shaw_2013} that collected optical spectra of $\sim$500 blazars reported in the \textit{Fermi} catalogs.
The results of the search on blazar candidates proposed on the basis of the colours of the infrared (IR) counterparts in the Wide-field Infrared Survey Explorer (WISE) survey and subsequent optical spectroscopy were presented by \citet[]{D'Abrusco_2013}, \citet[][]{Massaro_2015, Massaro_2016} and references therein, who found $\sim$600 blazar candidates and established the blazar nature for $\sim$200 of them. %through the analysis of their optical spectra}.

Our group has carried out several campaigns of optical spectroscopy (some of them still on-going) and, in particular, we focused on the analysis of optical spectra of various samples of GeV-TeV blazars and neutrino candidate blazars, with the aim to determine their classification, the  redshift and/or   an upper limit of the redshift \citep[see details in ][]{Paiano_2017_TeV}. 
In particular \citet[][]{Paiano_2017_TeV, Paiano_hz, Paiano_TeV} analysed 87 Fermi BLL (or candidates) detected in the VHE band. 
In \citet[][]{Paiano_2021,Paiano_2023} and \citet[][]{Padovani_2022} optical spectroscopy of $\sim$50 blazars candidates to be the counterparts of IceCube neutrino events, is reported. From the analysis of these spectra, we derived  the properties of the emission lines, crucial for the source characterization, and from the spectral decomposition of the continuum  we estimated the properties of the host galaxy and the mass of the central super massive black hole.  
No significant spectral differences were found for this group of objects with respect to the rest of Fermi BLL.\\
A recent summary of spectroscopic observations on blazar \textit{Fermi} sources can be found in \citet{Paliya_2021}.

%In particular, in \citet[][]{Paiano_hz} were analysed 10 BLLs whose previous works suggested they were high redshift objects. The result obtained from this work was the confirmation of the redshifts of two BLLs and the estimate of 8 new redshifts for the remaining sample. In \citet[][]{Paiano_TeV} a sample of 55 $\gamma$-ray sources detected with high significance above 50 GeV, previously classified by Fermi team as BLL, blazar candidate or blazar of uncertain type, was studied. All sources were classified as BLL, for 25 of them it was possible to determine a redshift, while for another 5 only a lower limit. Finally, in \citep[][]{Paiano_2021,Paiano_2023} the optical spectroscopy of 36 BLLs and two quasars, candidates to be the counterparts of IceCube neutrino events, is reported. For them, an estimate of the redshift or a lower limit is produced, in cases where it was not possible to identify features from the object.\\}
%Our group focused on the classification and the redshift estimate of various samples of GeV-TeV blazars \citep[][\textbf{ where 10 BLLs candidates be at high redshift and 55 $\gamma$-ray sources detected with high significance above 50 GeV, classified by Fermi team as BLL, blazar candidate or blazar of uncertain type, were analyzed respectively}]{Paiano_hz,Paiano_TeV} and neutrino candidate blazars \citep[][\textbf{in which the optical spectroscopy of 36 BLLs and two quasars, candidates to be the counterparts of IceCube neutrino events, is reported}]{Paiano_2021,Paiano_2023}. 

About 25\% of the extragalactic sources reported in the \textit{Fermi} catalogue still remains unassociated, either due to the lack of X-ray and/or radio counterparts or because of multiple possible associations at other wavelengths can be found inside the positional error box (on average circles of about 6 arcmins of radius).
These unassociated $\gamma$-ray sources (UGSs) represent a key component of the very high-energy sky.  
Their identification with lower energy counterparts and classification is crucial for population studies and the interpretation of the cosmic evolution of the gamma-ray sources \citep[e.g.][]{Ajello_2014,Ghisellini_2017}.
UGSs may hide new blazars and/or new AGN classes emitting at GeV energies (as Narrow Line Seyfert-1 and Seyfert-like objects, for which only a few are known as $\gamma$-ray emitters). 
Moreover, since most UGSs have weaker $\gamma$-ray fluxes (on average $\sim 2 \times 10^{-12}$ erg cm$^{-2}$ s$^{-1}$) than the identified sources (with an average flux of $\sim 1.6 \times 10^{-11}$ erg cm$^{-2}$ s$^{-1}$) in the 100 MeV to 100 GeV range, they can represent a more distant extragalactic population, and/or lower luminosity sources. 
%UFO- mean=2.3E-12  min=4.1E-13 max=4.4E-11 median=1.8E-12
%AGN- mean=7.8E-12  min=3.7E-13 max=8.4E-10 median=2.9E-12
%AGNobcu- mean=1.0E-11  min=3.9E-13 max=8.4E-10 median=3.8E-12

%\textbf{Since 2015 we carried out a systematic study focused on the search for UGS lower energy counterparts, respect to the $\gamma$-ray \textit{Fermi} energy band, starting from the analysis of the available \textit{Swift}/XRT images that cover the $\gamma$-ray error-box \citep[e.g.][]{stephen2010,takahashi2012,acero2013,takeuchi2013,landi2015,paiano2017_sed,paiano2017_ufo1,paiano2019_ufo2,kerby2021,kaur2023}.}

Since 2015 we carried out a systematic study, using \textit{Swift}/XRT images, with the aim of finding lower energy counterparts, with respect to the $\gamma$-ray energy band, of UGS based on available X-ray data that cover the $\gamma$-ray error-box \citep[e.g.][]{stephen2010,takahashi2012,acero2013,takeuchi2013,landi2015,paiano2017_sed,paiano2017_ufo1,paiano2019_ufo2,kerby2021,kaur2023}. 

To assess the nature of the source, we use optical spectroscopy available in the literature or obtained by our dedicated observation campaign using telescopes in the 8-10m class (such as Gran Telescopio Canarias and Large Binocular Telescope). 
This allows us to estimate or constrain the redshifts confirming the extragalactic nature of the proposed UGS counterparts. 
%Moreover this yields crucial \textbf{information to constrain the region and physical mechanism of the source emission, to determine the main physical parameters of the emitting region (such as the size, the temperature, the chemical composition...) and to shed light on the extreme physical processes, e.g. the neutrino production.}
Moreover this yields crucial information to constrain the source emission, to determine the main physical parameters of the emitting region (as the nucleus luminosity and the nucleus-to-host ratio), of the host galaxy (morphology, size, and luminosity) and to shed light on the extreme physical processes, e.g. the neutrino production.

In previous papers \citep[][]{paiano2017_ufo1, paiano2019_ufo2}, we studied a sample of 48 UGSs selected from the 2FGL and 3FGL catalogs and with at least one X-ray source detected inside the UGS error box. 
All sources exhibited an AGN optical spectrum (44 are BLLs, 1 QSO, 1 NLSy1 and 2 objects with a Seyfert 2 type spectrum).

In this work, we report the results of the association study for a first sample of 33 UGSs of the 4FGL-DR3 catalogue that have only one X-ray counterpart in the Fermi error ellipse and for which an optical spectrum is already available in literature or in public surveys, providing details about the main spectral properties and the classification.

This paper is structured as follows: in Section \ref{sec:sample}, we describe our procedure of search for UGS counterparts and the sample of this work, in Section \ref{sec:results} we present the main results from the analysis of the optical spectra and multi-wavelength data, in the Section \ref{sec:sum_con} we summarize and discuss the main properties of the sample in a multi-wavelength point of view, and finally in Section \ref{sec:note_ind_sources} we give notes on individual objects.

We adopted concordance cosmology \citep[e.g. ][]{Seehars2016} assuming  Hubble constant $H_0$ = $70$ km s$^{-1}$ Mpc$^{-1}$, matter density $\Omega_{\rm m,0}$ = $0.3$, and dark energy density $\Omega_{\Lambda,0}$ = $0.7$.

\section{Search for UGS counterparts and definition of sample} 
\label{sec:sample} 

\subsection{Search for UGS counterparts}
\subsubsection{X-ray band}
Over the last 10 years, the \textit{Swift} satellite has been involved in a campaign dedicated to the observation of UGSs with the XRT telescope \citep[][]{Stroh_2013, Falcone_2014}, and all data are available in the public archive \footnote{https://www.swift.ac.uk/swift\_portal/}. 
We searched for UGS X-ray counterparts by selecting all the Swift/XRT data that cover the UGS positions in the field of view. A total of 697 (over a total of the 1125) high latitude ($|b|>10^{\circ}$) UGSs are covered by at least one \textit{Swift}/XRT observation. For each of them we obtained an X-ray skymap covering a sky region of $\sim$15 arcmins (see an example in Fig. \ref{fig:Xskymap_2207} and for the remaining sources see Appendix A1). 
We reduced all the available Swift/XRT observations with the on-line tool provided by the \textit{Swift} consortium\footnote{https://www.swift.ac.uk/user\_objects/} \citep{Goad2007, Evans2009, Evans2020} that, for each given UGS, created a 0.3--10 keV stacked image and performed a source detection. 
We considered as X-ray sources detection having signal-to-noise (SNR) $ \geq 3.0$ inside the 3$\sigma$ \textit{Fermi} positional error ellipses\footnote{Note the two axes of the 4FGL-DR3 error ellipses at 95\% confidence level have been increased by 50\% in order to yield the $\sim$99\% confidence level (see examples in Fig. \ref{fig:Xskymap}).}. The online tool provides as output the source detection list, with the position\footnote{ If data acquired with the UV and Optical Telescope (UVOT) on board of Swift are available, the on-line \textit{Swift} analysis tool gives the enhanced position corrected for astrometry.} and the signal-to-noise ratio. %If data acquired with the UV and Optical Telescope (UVOT) on board of Swift are available, it gives the enhanced position corrected for astrometry. 
%Our search for UGS counterparts at lower energy is based on the data reduction and analysis of these Swift observations, exploiting the on-line tool provided by the \textit{Swift} consortium\footnote{https://www.swift.ac.uk/user\_objects/} \citep{Evans2009, Evans2020, Goad2007}, of all XRT {\bf observations} that cover the UGS sky regions and stacking them all \citep[details in our previous works ][]{paiano2017_ufo1, paiano2019_ufo2}.
%For each stacked image, we performed an analysis in order to find possible X-ray sources with a detection signal-to-noise (SNR) $ \geq 3.0$ inside the 3$\sigma$ \textit{Fermi} positional error ellipses {\bf mention the tools you used to perform the source detection and the energy range where you calculated the significance}. 

We find that 265 UGSs have at least one X-ray detection inside the $\gamma$-ray error box, and in particular 190 of them have only one possible X-ray counterpart inside the $\gamma$-ray error box, while the remaining 75 have multiple number of X-ray sources.
These numbers obviously depends on the chosen significance of the X-ray detection. 

\begin{figure}%[htbp]
%\vspace{2.2cm}
\hspace{-0.5cm}
\centering
   \includegraphics[width=7.35truecm]{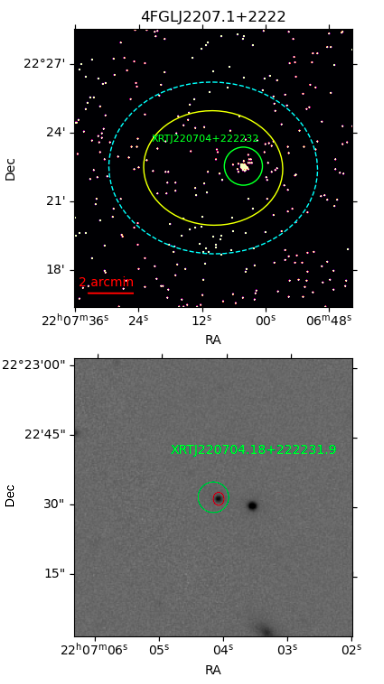}
\caption{\textit{Upper panel}: X-ray skymap of 4FGLJ22017.1+2222. The yellow and cyan ellipses are respectively the $2\sigma$ and $3\sigma$ \textit{Fermi} $\gamma$-ray error regions. X-ray detection, found through \textit{Swift}/XRT analysis, is reported in green.\\
\textit{Bottom panel}: Optical r-band PanSTARRs image of 4FGLJ2207.1+2222 counterpart. The green circle represent the error box of the X-ray counterpart and the red ellipses the error box of radio counterparts found within VLASS catalog.} 
\label{fig:Xskymap_2207}
\end{figure}%[htbp]

\subsubsection{Radio band}
Starting from the X-ray positions and error boxes (on average of the order of 4 arcsecs), we looked for radio and optical counterparts. % (see an example in Fig. \ref{fig:Xskymap_2207}).}
We used two radio catalogs, provided by the Very Large Array Sky Survey \citep[VLASS, ][]{Gordon_2021} and the Rapid ASKAP Continuum Survey \citep[RACS, ][]{Hale_2021}.
Furthermore, we performed a dedicated search\footnote{The radio detection was performed by applying two circular regions: one around the source (found through the search for the centroid, or in the absence of an obvious source on the position of the X-ray source) of size 10 arcsec, and the other in a region without of radio sources, of size 50 arcsec, to estimate the background. The ratio between the maximum flux in the source region and the rms of the background region gives the SNR of the detection.} for any uncatalogued radio sources using public radio images provided through the LOw-Frequency ARray (LOFAR) Two-metre Sky Survey \citep[LoTSS, ][]{LoTSS_2022} in the northern hemisphere and RACS in the southern hemisphere.
%With these images we searched coincident radio emission, detect any extended structure, or place accurate upper-limits on the source brightness. 
For those sources that have a marginal detection (SNR $\leq$~3~$\sigma$), that show an extended structure (non-circularly symmetric radio source) or no radio images are available in literature, we asked observations with the Australian Telescope Compact Array (ATCA). Full details on the ATCA program, will be presented in a future publication (Ulgiati et al., in prep.). 
For the case of 4FGLJ2030.0$-$0310, the details are reported in the source notes (see Section \ref{sec:note_ind_sources}).

\subsubsection{Optical band}
\label{sec:O_counterpart_sel}
For the optical band, we searched counterparts using catalogues by the Sloan Digital Sky Survey \citep[SDSS, ][]{SDSS_2020}, the Panoramic Survey Telescope and Rapid Response System \citep[PanSTARRS, ][]{Chambers_2016}) database, the Dark Energy Survey \citep[DES, ][]{Abbott_2021} and the United States Naval Observatory \citep[USNO, ][]{Monet_2003} survey (see an example in Fig. \ref{fig:Xskymap_2207}-\textit{bottom panel} and for the remaining sources see Appendix B). 

For 15 objects (with redshift $<$0.5) and/or with a substantial contribution from the host galaxies), we analysed the \textit{r}-band images taken from the PANSTARRs, SDSS and DES database. 
We have used the AIDA software \citep[][]{uslenghi2011} in order to separate the nuclear and host galaxy components and to determine the main photometric and morphological characteristics of the host galaxies (absolute magnitude, effective radius, and  Sersic index $n$ \footnote{The surface brightness profile of a galaxy is described by the Sersic law: \textit{ $\frac{I(r)}{I(e)}$ = exp( - b$_n$ ( ($\frac{r}{R_e}$)$^{\frac{1}{n}}$ - 1)) }.
where I$_e$ is the light intensity at the effective radius R$_e$, the major-axis effective radius encompassing half of the total flux of the source, $n$ is the Sersic index and $b_n$ a constant depending on $n$.}).

\subsection{Definition of the sample under consideration}

The necessary step to probe the nature of the sources and to provide a classification is to study their optical spectra. 
Considering the UGSs with only one X-ray detection, we found that 33 objects have a X-ray counterpart which has already an optical spectrum in literature. For other $\sim$60 UGS counterparts, optical spectroscopic data from large telescopes (e.g. Large Binocular Telescope (LBT) and Gran Telescopio Canarias (GTC)) has been already granted to our group (Paiano et al., in prep.).
%Considering the UGSs with only one X-ray detection,  we obtained optical spectroscopy at large diameter telescopes  for $\sim$60 UGS counterparts (Paiano et al., in prep.), while for 33 objects an optical spectrum is already available in literature. 

In this paper we report the main results for this sub-sample of 33 sources, obtained by the study of the X-ray images and the analysis of the literature spectrum, when available in ASCII or FITS format.
The main information of the \textit{Swift}/XRT observations and results of the X-ray analysis are summarized in Tab. \ref{tab:33_UGS_X}. 
In Tab. \ref{tab:33_UGS_Multi} we list the multiwawelength counterparts of each UGS.
In Appendix A we show the \textit{Fermi} error boxes superimposed to the X-ray images of this sample. 
Appendix B contains the optical images (Fig. \ref{fig:Oskymap}) taken from the PANSTARRs database (with the exception of two cases in which the images were taken from the Dark Energy Survey (DES) and NOIRLab Astro Data archives) with the overlay of the proposed X-ray and radio counterparts.
Details of the multiwavelength study, including the X-ray and the optical skymaps for the entire UGS sample with at least one X-ray detection will be presented in Ulgiati et al., (in prep.).

\begin{table*}
\begin{center}
\caption{X-ray info of the 33 UGSs with only one X-ray counterpart within the 3$\sigma$ Fermi error box} 
%(XXX objects)
%\resizebox{16cm }{!}{
\resizebox{19cm }{!}{
\begin{tabular}{lccccccc}
\hline
4FGL Name & $\gamma$-ray    & Swift exp.  & XRT counterpart         & RA      & DEC  & X-ray positional & X-ray  \\   
          & significance ($\sigma$) & time (ks)  &                & (J2000) & (J2000) & error radius (") & significance ($\sigma$)  \\  
\hline
4FGLJ0023.6-4209 & 5.1 & 3.4 & XRTJ002303.5-420509.6 & 5.76496 & -42.08601  & 2.9 & 5.0  \\
4FGLJ0112.0+3442 & 4.8 & 4.7 & XRTJ011124.8+344154.1 & 17.85359 & 34.69837 & 4.7 & 4.0   \\
4FGLJ0117.9+1430 & 6.1 & 6.6 & XRTJ011804.7+143159.5 &  19.51996 & 14.53322 & 3.0 &  4.7 \\
4FGLJ0202.7+3133 & 4.3 & 10.5 & XRTJ020242.1+313211.3 & 30.67554 & 31.53649 & 2.5 &  10.3 \\
4FGLJ0251.1-1830 & 11.0 & 7.7 & XRTJ025111.7-183111.1 & 42.79877 & -18.51976 & 2.7 &  8.9 \\
4FGLJ0259.0+0552 & 23.2 & 4.9 & XRTJ025857.5+055244.4 & 44.73982 & 5.87900 & 3.4 &  5.2 \\
4FGLJ0641.4+3349 & 4.4 & 3.6 & XRTJ064111.2+334502.0 & 100.29683 & 33.75056 & 3.7 & 13.8  \\
4FGLJ0838.5+4013 & 4.1 & 3.9 & XRTJ083902.9+401546.9 & 129.76240 & 40.26303 & 3.0 &  4.4 \\
4FGLJ0938.8+5155 & 6.1 & 10.6 & XRTJ093834.5+515454.7 & 144.64375 & 51.91522 & 6.1 & 3.8  \\
4FGLJ1016.1-4247 & 12.8 & 6.1 & XRTJ101620.7-424723.2 & 154.08659 & -42.78978 & 2.7 & 8.6  \\
4FGLJ1039.2+3258 & 9.8 & 5.1 & XRTJ103852.1+325651.9 & 159.71738 & 32.94776 & 3.0 &  4.6 \\
4FGLJ1049.8+2741 & 6.5 & 5.5 & XRTJ104938.7+274212.0 & 162.41124 & 27.70335 & 2.9 & 5.8  \\
4FGLJ1125.1+4811 & 4.5 & 50.5 & XRTJ112526.0+480922.8 & 171.35839 & 48.15634 & 5.4 & 4.3 \\
4FGLJ1131.6+4657 & 6.3 & 4.6 & XRTJ113142.3+470009.2 & 172.92651 & 47.00256 & 2.6 & 6.0  \\
4FGLJ1146.0-0638 & 13.5 & 3.3 & XRTJ114600.8-063853.9 & 176.50361 & -6.64831 & 3.1 & 6.7 \\
4FGLJ1256.8+5329 & 5.6 & 5.6 & XRTJ125630.5+533202.2 & 194.12725 & 53.53397 & 6.4 & 4.2  \\
4FGLJ1308.7+0347 & 9.7 & 3.6 & XRTJ130832.2+034405.3 & 197.13445 & 3.73483 & 4.0 &  5.5 \\
4FGLJ1346.5+5330 & 10.3 & 3.9 & XRTJ134545.1+533252.4 & 206.43811 & 53.54791 & 2.2 & 13.5  \\
4FGLJ1410.7+7405 & 22.9 & 11.7 & XRTJ141045.6+740509.8 & 212.69026 & 74.08608 & 7.2 &  4.5 \\
4FGLJ1430.6+1543 & 4.8 & 1.6 & XRTJ143057.9+154556.0 & 217.74133 & 15.76529 & 3.1 &  5.4 \\
4FGLJ1535.9+3743 & 18.1 & 10.7 & XRTJ153550.5+374056.8 & 233.96065 & 37.68245 & 4.6 &  4.1 \\
4FGLJ1539.1+1008 & 5.6 & 6.2 & XRTJ153848.5+101841.7 & 234.70214 & 10.31159 & 3.2 &  5.0 \\
4FGLJ1544.9+3218 & 6.5 & 13.9 & XRTJ154433.1+322148.5 & 236.13813 & 32.36349 & 2.6 & 10.1  \\
4FGLJ1554.2+2008 & 10.6 & 8.2 & XRTJ155424.1+201125.3 & 238.60069 & 20.19041 & 2.0 & 34.5  \\
4FGLJ1555.3+2903 & 5.1 & 3.5 & XRTJ155513.0+290328.0 & 238.80422 & 29.05779 & 3.5 &  10.9 \\
4FGLJ1631.8+4144 & 8.0 & 3.7 & XRTJ163146.8+414631.8 & 247.94510 & 41.77550 & 2.6 &  8.9 \\
4FGLJ1648.7+4834 & 5.7 & 4.2 & XRTJ164900.5+483409.1 & 252.25233 & 48.56921 & 2.8 & 5.7  \\
4FGLJ2030.0-0310 & 4.5 & 4.5 & XRTJ203014.3-030722.8 & 307.55974 & -3.12276 & 2.8 & 11.0  \\
4FGLJ2207.1+2222 & 7.5 & 4.8 & XRTJ220704.1+222231.8 & 331.76740 & 22.37552 & 3.3 &  5.0 \\
4FGLJ2240.3-5241 & 11.8 & 5.8 & XRTJ224017.55-524112.3 & 340.07314 & -52.68676 & 3.6 &  3.4 \\
4FGLJ2317.7+2839 & 10.7 & 14.1 & XRTJ231740.1+283955.4 & 349.41730 & 28.66540 & 5.8 & 4.3 \\
4FGLJ2323.1+2040 & 6.2 & 3.3 & XRTJ232320.30+203523.6 & 350.83459 & 20.58990 & 4.2 & 3.5\\
4FGLJ2353.2+3135 & 9.0 & 28.0 & XRTJ235319.39+313616.9 & 358.3308 & 31.6047 & 3.5 & 8.2  \\
%\hline
%& & & Objects with prominent emission lines & & & &\\
%\hline
\hline
\end{tabular}
}
\label{tab:33_UGS_X}
\end{center}
\raggedright
\footnotesize{\textbf{Note.} Column 1: 4FGL Name of the target; Column 2: $\gamma$-ray detection significance as reported in the 4FGL catalog;  Column 3: \textit{Swift}/XRT exposure time; Column 4: X-ray counterpart detected from our Swift/XRT analysis (Names report the acronym of the XRT detector plus the J2000 sexagesimal coordinates); Column 5-6: Coordinates of the proposed X-ray counterpart in degrees; Column 7: X-ray positional error radius in arcsecs; Column 8: Detection significance of the X-ray counterpart.\\
}
%\tablenotetext{}{
%\raggedright
% } 
\end{table*}

\begin{table*}
\begin{center}
\caption{Multiwavelength counterparts proposed for the sample of 33 UGSs } %with only one X-ray counterpart within the 3$\sigma$ Fermi error box 
%(XXX objects)
%\resizebox{16cm }{!}{
\begin{tabular}{llllll}
\hline 
4FGL Name & XRT         & Radio         & Optical       & RA      & DEC \\   
          & counterpart & counterpart   & counterpart   &  J2000  &  J2000    \\  
\hline
%\hline
% & & & Objects with BLL-like spectrum & & & &\\
% & & & objects & & & &\\
%\hline
4FGLJ0023.6-4209 & XRTJ002303.5-420509.6 & RACSJ002303.61-420509.57 & DESJ002303.74-420508.4 & 5.765596 & -42.08569 \\
4FGLJ0112.0+3442 & XRTJ011124.8+344154.1 & VLASS1QLCIRJ011124.83+344154.5 & SDSSJ011124.86+344154.6 & 17.853583 & 34.698500 \\
4FGLJ0117.9+1430 & XRTJ011804.7+143159.5 & - & SDSSJ011804.83+143158.6 & 19.520125 & 14.532944 \\
4FGLJ0202.7+3133 & XRTJ020242.1+313211.3 & VLASS1QLCIRJ020242.03+313211.0 & SDSSJ020242.06+313210.9 &  19.520125 & 14.532944 \\
4FGLJ0251.1-1830 & XRTJ025111.7-183111.1 & VLASS1QLCIRJ025111.53-183112.3 & PANJ025111.53-183112.7 & 42.798000 & -18.520167	 \\
4FGLJ0259.0+0552 & XRTJ025857.5+055244.4 & VLASS1QLCIRJ025857.55+055244.0 & SDSSJ025857.55+055243.9 & 44.739792 & 5.878861 \\
4FGLJ0641.4+3349 & XRTJ064111.24+334502.0 & VLASS1QLCIRJ064111.20+334459.6 & PANJ064111.22+334459.7 & 100.296750 & 33.749917 \\
4FGLJ0838.5+4013 & XRTJ083902.9+401546.9 & VLASS1QLCIRJ083903.07+401545.6 & SDSSJ083903.08+401545.6 & 129.762833 & 40.262667 \\
4FGLJ0938.8+5155 & XRTJ093834.5+515454.7 & LoTSS093834.68+515451.8 & SDSSJ093834.72+515452.3 & 144.644667 & 51.914528 \\
4FGLJ1016.1-4247 & XRTJ101620.7-424723.2 & ATCAJ101620.76-424723.1 & USNOJ101620.67-424722.6 & 154.086125 & -42.789611 \\
4FGLJ1039.2+3258 & XRTJ103852.1+325651.9 & VLASS1QLCIRJ103852.17+325651.9 & SDSSJ103852.17+325651.6 & 159.717375 & 32.947667 \\
4FGLJ1049.8+2741 & XRTJ104938.7+274212.0 & VLASS1QLCIRJ104938.81+274213.1 & SDSSJ104938.79+274213.0 & 162.411625 & 27.703611 \\
4FGLJ1125.1+4811 & XRTJ112526.0+480922.8 & - & SDSSJ112526.27+480922.0 & 171.359458 & 48.156111 \\
4FGLJ1131.6+4657 & XRTJ113142.3+470009.2 & VLASS1QLCIRJ113142.36+470009.4 & SDSSJ113142.27+470008.6 & 172.926125 & 47.002389 \\
4FGLJ1146.0-0638 & XRTJ114600.8-063853.9 & VLASS1QLCIR114600.87-063854.5 & USNOB1-0833-0250645 & 176.504000 & -6.648556 \\
4FGLJ1256.8+5329 & XRTJ125630.5+533202.2 & - & SDSSJ125630.43+533204.3 & 194.126792 & 53.534528	 \\
4FGLJ1308.7+0347 & XRTJ130832.2+034405.3 & - & SDSSJ130832.10+034403.9 & 197.133750 & 3.734417 \\
4FGLJ1346.5+5330 & XRTJ134545.1+533252.4 & VLASS1QLCIRJ134545.34+533252.1 & SDSSJ134545.36+533252.3 & 206.439000 & 53.547861 \\
4FGLJ1410.7+7405 & XRTJ141045.6+740509.8 & JVLAJ141046.00+740511.2$^*$  & PANJ141045.95+740510.8 &  212.691458 & 74.086333 \\
4FGLJ1430.6+1543 & XRTJ143057.9+154555.0 & - & SDSSJ143058.03+154555.6 &  217.741792 & 15.765444\\
4FGLJ1535.9+3743 & XRTJ153550.56+374056.8 & VLASS1QLCIR J153550.56+374055.5 & SDSSJ153550.54+374055.6 & 233.960583 & 37.682111 \\
4FGLJ1539.1+1008 & XRTJ153848.5+101841.7 & - & SDSSJ153848.47+101843.2 & 234.701958 & 10.312000 \\
4FGLJ1544.9+3218 & XRTJ154433.1+322148.5 & VLASS1QLCIRJ154433.20+322149.1 & SDSSJ154433.19+322149.1 & 236.138292 & 32.363639 \\
4FGLJ1554.2+2008 & XRTJ155424.1+201125.3 & VLASS1QLCIRJ155424.15+201125.5 & SDSSJ155424.12+201125.4 &  238.600500 & 20.190389 \\
4FGLJ1555.3+2903 & XRTJ155513.0+290328.0 & VLASS1QLCIRJ155512.89+290330.0 & SDSSJ155512.91+290329.9 & 238.803792 & 29.058306 \\
4FGLJ1631.8+4144 & XRTJ163146.8+414631.8 & VLASS1QLCIRJ163146.74+414632.7 & SDSSJ163146.72+414632.8 & 247.944667 & 41.775778 \\
4FGLJ1648.7+4834 & XRTJ164900.5+483409.1 & VLASS1QLCIRJ164900.35+483411.7 & SDSSJ164900.34+483411.8 & 252.251417 & 48.569944 \\
4FGLJ2030.0-0310 & XRTJ203014.3-030722.8 & ATCAJ203014.27-030721.8 & PANJ203014.27-030722.56 & 307.559458 & -3.122933 \\
4FGLJ2207.1+2222 & XRTJ220704.1+222231.8 & VLASS1QLCIRJ220704.09+222231.5 & SDSSJ220704.10+222231.4 & 331.767083 & 22.375389 \\
4FGLJ2240.3-5241 & XRTJ224017.55-524112.3 & RACS224017.79-524111.1 & DESJ224017.71-524113.7 &  340.073792 & -52.687139	 \\
4FGLJ2317.7+2839 & XRTJ231740.1+283955.4 & VLASS1QLCIRJ231740.21+283955.8 & SDSSJ231740.00+283955.7 & 349.416667 & 28.665472 \\
4FGLJ2323.1+2040 & XRTJ232320.30+203523.6 & VLASS1QLCIRJ232319.95+203523.7 & SDSSJ232320.34+203523.4 & 350.834772 & 20.589860 \\
4FGLJ2353.2+3135 & XRTJ235319.39+313616.9 & VLASS1QLCIRJ235319.50+313616.8 & SDSSJ235319.54+313616.7 & 358.331417 & 31.604639 \\
\hline
\end{tabular}
\label{tab:33_UGS_Multi}
\end{center}
\raggedright
\footnotesize{\textbf{Note.} Column 1: 4FGL Name of the target; Column 2: X-ray counterpart; Column 3: Radio counterpart (Names report the acronym of the radio facility plus the J2000 sexagesimal coordinates); Column 4: Optical counterpart; Column 5-6: Coordinates of the optical counterpart in degrees. \\
(*) Radio source proposed by \citep{marchesini2023}\\
%\textbf{Names of the uncatalogued radio sources were created by placing the acronym of the facility that observed them before the position in sexagesimal degrees in the J2000 coordinate system.}
 } 
\end{table*}

\section{Results}  
\label{sec:results} 

The available and downloaded optical spectra found in literature of our 33 UGS sample (see Table \ref{tab:33_UGS_Opt} for the references) are shown in Fig.\ref{fig:spectrum}. 
The spectra were dereddened for the Galaxy contribution, applying the extinction law by \citet{cardelli1989} and assuming the E(B-V) values provided by the NASA/IPAC infrared science archive \footnote{https://irsa.ipac.caltech.edu/applications/DUST/}. 
Note that for two objects (4FGLJ0023.6-4209 and 4FGLJ2030.0-0310) the spectra, taken from the 6dF survey, are not flux calibrated and dereddened. Although they are useful for the redshift determination, no information of the magnitude of the source and line luminosities can be derived.

From the redshifts deduced from the analysis of the optical spectra and/or the presence of a power-law component (typical of the BLL), it was possible to asses the extragalacic nature of the sources.

For 20 out of 33 objects, clear spectral features are found that allow us to derive their redshift (see Table \ref{tab:33_UGS_Opt}). 
%{\bf We determine the position of the emission and absorption lines performing accurate centroids for each feature by computing the barycenter of the lines after subtraction of the underlined continuum. Then, comparing all measured lines for each target, we checked their consistency within the measured errors and we derived the redshift of the source from the median of all measurements.}
We determine the position of the emission and absorption lines performing accurate search of the centroids, by computing the barycenter of each line after subtraction of the underlined continuum.
The redshift is determined by comparing the found wavelenght centroids of each line with the expected rest-frame wavelenghts. After checking that all lines provide consistent results, we derived the final redshift from the average.
11 sources do not reveal intrinsic spectral features. 
For 7 sources, the optical spectrum exhibits absorption lines of the host galaxy stellar population (Ca II 3934,3968, G-band 4305, Mg I5175, and Na I 5893), 12 objects present only emission lines, and for one case (4FGLJ0112.0+3442) both emission and absorption lines are present.

%From the spectroscopy point of view, 
We can classify 21 sources as BLL: 15 with spectra characterized by a power-law emission and another 6 with a strong signature of the stellar population due to the host galaxy (labelled as galaxy-dominated BLL (BLG) in Table \ref{tab:33_UGS_Opt}). 
Seven BLLs have their redshift determined through the detection of the host galaxy absorption lines. 
%while one object (4FGLJ0112.0+3442) exhibits both absorption and emission lines. 
The redshifts are in the interval 0.04~$<z<$~0.64 and the average value is $<$z$>$~=~$0.2 \pm 0.1$, in agreement with the typical BLL redshift distribution \citep[][]{Padovani_2017, Garofalo_2018}.
%For two BLLs we can set a spectroscopic redshift lower limit on the based of the detection of intervening absorption lines.
For two BLLs, 4FGLJ0251.0-1830 \citep[][for details]{paiano2019_ufo2} and 4FGLJ2353.0+3135 (see Fig. \ref{fig:spectrum}), we can detect only intervening absorption systems due to MgII 2899 that allow us to provide a spectroscopic redshift lower limit. 

The spectra of the remaining 12 objects all have prominent emission lines (mainly CIII], MgII, [OII], H$_{\beta}$, [OIII], H$_{\alpha}$, [N II] and [SII]), most of them showing a type-1 spectrum with broad and narrow emission lines. 
One high redshift source  ($z\geq$1) displays only broad emission lines.
4FGLJ0117.9+1430 can be classified as a Narrow Line Seyfert 1 (see details in Sec. \ref{sec:note_ind_sources}), while the other 11 have a Seyfert/QSO-like spectrum.
%\red{For 15 objects, with redshift z$<$0.5, we analysed the r-band images taken from the PANSTARRs, SDSS or DES database. We have used the AIDA software \citep[][]{uslenghi2011} in order to separate the nuclear and host galaxy components and to determine the main photometric and morphological characteristics of the host galaxies (absolute magnitude, effective radius, and Sersic\footnote{ln I(R) = ln I$_0$ - kR$^{\frac{1}{n}}$ } index ($n$) ).}

From the \textit{r}-band image decomposition analysis of the 15 objects with redshift z$<$0.5, all galaxies are resolved, except for 4FGLJ0112.0+3442, and the results are reported in Table \ref{tab:decomp}.

The optical properties and the multiwavelength emission data of the 33 UGS are summarized in Tab. \ref{tab:33_UGS_Flux}. 
While $\gamma$-ray, optical and radio fluxes come from catalogs, the X-ray fluxes are extracted through spectral fits. Spectra are fitted using an absorbed power-law, where the NH parameter, defined as the equivalent hydrogen column (in units of 10$^{22}$ atoms cm$^{-2}$), was set to the equivalent Galactic value in the direction of the source \citep[]{HI4PI_coll_2016}. Fluxes are estimated in the energy range 0.3 - 10 keV.
For each source, we evaluate a \textit{radio-loudness} parameter \textit{R} defined as the ratio between the radio flux (in the range 2 - 4~GHz) and the optical band flux of the nucleus component (see Table \ref{tab:33_UGS_Flux}).%\ref{tab:R_value}).
We consider a source as radio loud if R$>$10 \citep[][]{kellermann1989}. All BLL are radio loud sources. 

About the other 12 sources,  4 sources (4FGLJ0023.6-4209, 4FGLJ0938.8+5155, 4FGLJ1346.5+5330 and 4FGLJ1535.9+3743) can be classified as radio-loud, two have $R<10$, and  six are not detected  in the radio images, allowing us to put an upper limit on $R$. 

Based on the absolute magnitude \citep[M$_{abs} < $-23 for QSO, e.g. ][]{Osterbrock_1980}{}{} and the optical imaging analysis (see details in Section \ref{sec:O_counterpart_sel}), four sources have the typical luminosity of the QSO: 4FGLJ1535.9+3743 is a radio-loud quasar and 4FGLJ1125.1+4811, 4FGLJ1256.8+5329 and 4FGLJ1308.7+0347 radio quiet quasars.
The other eight objects can be classified as Seyfert galaxies.

\begin{table*}
\begin{center}
\caption{Optical properties of the 33 UGSs with only one X-ray counterpart within the 3$\sigma$ Fermi error box}
\begin{tabular}{llcclcll}
\hline
4FGL Name & Optical counterpart    & g & r & Spectrum Reference & Line type & Redshift & Spectrum Class\\  
\hline
4FGLJ0023.6-4209 & DESJ002303.74-420508.4  & 15.6 & 15.0 & 6dF         & e   & 0.053  & Type-2\\% only narrow line, ty-2, Sy2, RQ, spiral, \\%  & 11.6 (10.9) \\ 
4FGLJ0112.0+3442 & SDSSJ011124.86+344154.6 & 19.4 & 19.0 & SDSS        & e,g & 0.3997 & BLL \\
4FGLJ0117.9+1430 & SDSSJ011804.83+143158.6 & 18.7 & 18.3 & SDSS        & e & 0.129    & Type-1\\% n,b(?), ty-2,NLSy1 \\
4FGLJ0202.7+3133 & SDSSJ020242.06+313210.9 & 18.7 & 18.4 & SDSS         & -  & ? $^*$        & BLL \\
4FGLJ0251.1-1830 & PANJ025111.53-183112.7  & 20.2 & 19.6 & \citet{paiano2019_ufo2} & i  & $>$0.615 & BLL \\
4FGLJ0259.0+0552 & SDSSJ025857.55+055243.9 & 18.6 & 18.3 & \citet{paiano2019_ufo2} & -  & ?        & BLL \\ %redshift LL $>0.7
4FGLJ0641.4+3349 & PANJ064111.22+334459.7  & 17.1 & 16.4 & \citet{monroe2016}  & e & 0.1657 & Type-1\\% n,b ty-1, RQ, quasar?\\ %https://archive.stsci.edu/prepds/uvqs/
4FGLJ0838.5+4013 & SDSSJ083903.08+401545.6 & 18.2 & 17.0 & SDSS         & g & 0.1945 & BLG \\ 
4FGLJ0938.8+5155 & SDSSJ093834.72+515452.3 & 20.3 & 20.1 & SDSS         & e & 0.4168 & Type-1\\% ty-1, RL, quasar-Sy1?\\    
4FGLJ1016.1-4247 & USNOJ101620.67-424722.6 & 19.3 & 18.2 & \citet{rajagopal2023} & - & ? & BLL\\
4FGLJ1039.2+3258 & SDSSJ103852.17+325651.6 & 19.7 & 18.9 & SDSS         & - & ? $^*$    & BLL \\
4FGLJ1049.8+2741 & SDSSJ104938.79+274213.0 & 18.2 & 17.3 & SDSS, \citet{demenezes2019}        & g & 0.144  & BLG \\
4FGLJ1125.1+4811 & SDSSJ112526.27+480922.0 & 20.3 & 20.2 & SDSS         & e & 1.649  & Type-1\\% ty?,quasar, RQ?\\
4FGLJ1131.6+4657 & SDSSJ113142.27+470008.6 & 17.5 & 16.5 & SDSS         & g & 0.1255 & BLG \\
4FGLJ1146.0-0638 & USNOB1-0833-0250645     & 19.5 & 19.7 & \citet{paiano2019_ufo2}  & g & 0.6407 & BLL \\
4FGLJ1256.8+5329 & SDSSJ125630.43+533204.3 & 20.6 & 20.3 & SDSS         & e & 0.996  & Type-1 \\%ty1, quasar, RQ\\
4FGLJ1308.7+0347 & SDSSJ130832.10+034403.9 & 17.2 & 17.3 & SDSS         & e & 0.6193 & Type-1 \\%ty1, RS, QSO,  \\ %radio quiet (silent) from Rusinek-Abarca et al. 2021
4FGLJ1346.5+5330 & SDSSJ134545.36+533252.3 & 17.0 & 16.6 & SDSS         & e & 0.1359 & Type-1 \\% ty1, RL, quasar, FR1  \\
4FGLJ1410.7+7405 & PANJ141045.95+740510.8  & 19.2 & 19.3 & \citet{marchesini2023} & - & ? & BLL\\ %(<0.2?) 
4FGLJ1430.6+1543 & SDSSJ143058.03+154555.6 & 17.4 & 16.9 & SDSS         & e & 0.1633 &  Type-1 \\%ty1,Sy?, RQ\\%upper limit from Coziol et al. 2017
4FGLJ1535.9+3743 & SDSSJ153550.54+374055.6 & 19.7 & 19.4 & SDSS         & e & 0.6255 &  Type-1 \\%ty?, RL, Quasar \\
4FGLJ1539.1+1008 & SDSSJ153848.47+101843.2 & 18.3 & 18.0 & SDSS         & e & 0.2345 &   Type-1 \\%ty1, Sy?, RQ, \\ %upper limit from Coziol et al. 2017
4FGLJ1544.9+3218 & SDSSJ154433.19+322149.1 & 18.7 & 18.4 & SDSS         & - & ? $^*$ & BLL\\
4FGLJ1554.2+2008 & SDSSJ155424.12+201125.4 & 18.1 & 17.2 & SDSS         & g & 0.2225 & BLG\\
4FGLJ1555.3+2903 & SDSSJ155512.91+290329.9 & 18.2 & 17.2 & SDSS         & g & 0.1767 & BLG\\
4FGLJ1631.8+4144 & SDSSJ163146.72+414632.8 & 20.5 & 20.4 & SDSS         & - & ? $^*$ & BLL\\ %\red{0.721?}
4FGLJ1648.7+4834 & SDSSJ164900.34+483411.8 & 19.4 & 19.1 & SDSS         & - & ? $^*$ & BLL\\
4FGLJ2030.0-0310 & PANJ203014.27-030722.56 & 16.8 & 16.2 & 6dF          & e & 0.036 &  Type-2 \\%ty2,??, RQ, Sph\\
4FGLJ2207.1+2222 & SDSSJ220704.10+222231.4 & 20.4 & 19.9 & SDSS         & - & ? $^*$ & BLL \\
4FGLJ2240.3-5241 & DESJ224017.71-524113.7  & 18.1 & 17.4 & \citet{desai2019} & - & ? & BLL\\
4FGLJ2317.7+2839 & SDSSJ231740.00+283955.7 & 19.6 & 19.1 & SDSS         & - & ? $^*$ & BLL \\
4FGLJ2323.1+2040 & SDSSJ232320.34+203523.4 & 14.4 & 13.4 & \citet{marcha1996} & g & 0.038 & BLG\\ 
4FGLJ2353.2+3135 & SDSSJ235319.54+313616.7 & 20.5 & 20.5 & SDSS  & i & $>$0.8809$^*$ & BLL\\ 
\hline
\end{tabular}
%}
\label{tab:33_UGS_Opt}
\end{center}
\raggedright
\footnotesize{\textbf{Note.} Column 1: 4FGL Name of the target; Column 2: Optical counterpart; Column 3-4: magnitude in g and r band from PANSTARRs; Column 5: Reference of the optical spectrum; Column 6: Type of the detected lines: e = emission lines, g = absoption lines from the host galaxy, i = intervening absorption lines; Column 7: Redshift; Column 7: Classification based on the optical spectrum.\\
(*) New redshift estimates reported for the first time by this work. It is worth noting that, for a given number of sources, the literature redshift was disproved and then they are not known (except in one case, where we determined a lower limit). Details are reported in Sec. \ref{sec:note_ind_sources}
\\
%\tablenotetext{}{ , n = narrow emission line, b = broad emission line
%\raggedright
 } 
\end{table*}

\begin{table*}
\begin{center}
\caption{Results of the analysis and decomposition of the PANSTARRs images} 
%(XXX objects)
%\resizebox{16cm }{!}{
%\resizebox{19cm }{!}{
\begin{tabular}{lcccccccc}
\hline
4FGL Name  & $z$  & r & r$_{n}$ & r$_{h}$ & M(r)$_{h}$ & $R_{e}$ & $n$ & N/H\\  
\hline   
4FGLJ0023.6-4209$^{*}$  & 0.053 & 14.9 & $>$18.5 & 14.9 & -21.9 & 6.8 & 1.4 & $<$0.05  \\%0.2 \\% BLL \\
4FGLJ0112.0+3442  & 0.3997 & 18.7 & 19.0 & 20.4 & $>$-21.3 & 7* & 4 & 4 \\%0.2 \\% BLL \\
4FGLJ0117.9+1430  & 0.129  & 18.2 & 19.4 & 18.5 & -20.4 & 2.0 & 4 & 0.4 \\%0.7 \\ %Type-1\\%ty-2,NLSy1 
4FGLJ0641.4+3349  & 0.1657 & 16.3 & 17.1 & 16.7 &-22.8 & 7* & 5 & 0.7\\% n,b ty-1, RQ, \\ %https://archive.stsci.edu/prepds/uvqs/
4FGLJ0838.5+4013  & 0.1945 & 16.9 & 19.1 & 16.7 & -23.1 & 10.5 & 4 & 0.1\\ 
4FGLJ0938.8+5155  & 0.4168 & 19.8 & 20.4 & 20.6 & -21.2 & 10.7 & 4 & 1\\% ty-1, RL, quasar-Sy1\\    
4FGLJ1049.8+2741  & 0.144  & 17.2 & 19.0 & 17.2 & -21.9 & 5.1 & 4 & 0.25\\
4FGLJ1131.6+4657  & 0.1255 & 16.4 & 18.9 & 16.3 & -22.5 & 7.6 & 4 & 0.1\\%BLG \\
4FGLJ1346.5+5330  & 0.1359 & 16.4 & 17.8 & 16.7 & -22.3 & 4 & 4 & 0.4\\% Type-1 \\% ty1, RL, quasar, FR1 
4FGLJ1430.6+1543  & 0.1633 & 16.8 & 18.0 & 17.1 & -22.3 & 5.9 & 2.1 & 0.4\\%  Type-1 \\%ty1,Sy, RQ\\%upper limit from Coziol et al. 2017
4FGLJ1539.1+1008  & 0.2345 & 17.9 & 18.3 & 19.2 & -21.1 & 3.6 & 3.0 & 2.5\\%   Type-1 \\%ty1, Sy, RQ, \\ %upper limit from Coziol et al. 2017
4FGLJ1554.2+2008  & 0.2225 & 16.9 & 18.3 & 17.2 &-23.0 & 7.9 & 4 & 0.4\\% BLG\\
4FGLJ1555.3+2903  & 0.1767 & 17.0 & 19.3 & 17.0 &-22.6 & 7.3 & 4 & 0.1 \\%BLG\\
4FGLJ2030.0-0310  & 0.036  & 16.1 & 18.8  & 16.1 &-19.8 & 1.1 & 1.9 & 0.1 \\%  Type-2 \\%ty2, RQ,Sph\\
4FGLJ2323.1+2040$^{*}$  & 0.038  & 13.6 & $>$17.0 & 13.6 & -22.5 & 6.5 & 4 & $<$0.05 \\%  Type-2 \\%ty2, RQ,Sph\\

%2323 is satured
\hline
\end{tabular}
%}
\label{tab:decomp}
\end{center}
\raggedright
\footnotesize{\textbf{Note.} Column 1: Fermi name, Column 2: Redshift, Column 3: r from aperture photometry of the PANSTARRs image, Column 4-5: apparent magnitude of the nucleus (r$_{n}$) and of the host galaxy (r$_{h}$) derived by the imaging analysis, Column 6: Absolute magnitude of the host galaxy, Column 7: Effective radius (kpc), Column 8: Sersic index $n$, Column 9: Galaxy flux to nucleus flux ratio.\\
(*) For 4FGLJ0023.6-4209 the DES image is used and for 4FGLJ2323.1+2040 we used the SDSS image.
%\tablenotetext{}{
%\raggedright
 } 
\end{table*}

\section{Summary and conclusions}  
\label{sec:sum_con}
We have examined a set of 33 UGS counterparts for which an optical spectrum was recovered in the literature. 
From the spectroscopic analysis we found that all counterparts are extragalactic objects: 21 sources have been classified as BLL, while the others 12 are AGN with prominent emission lines. 

In particular we found that among the 21 BLL, 7 are notable for having their redshift determined solely through the detection of absorption lines due to the host galaxy, while one object (4FGLJ0112.0+3442) exhibits both absorption and emission lines.
%The redshifts are in the interval 0.04~$<z<$~0.64 and the average value is $<$z$>$~$\sim$~0.2. %\pm 0.18$. 
For two BLLs, spectroscopic redshift lower limit can be set on the based of the detection of intervening absorption lines. 
The others have a featureless spectrum, described by a power-law shape, and the redshift is unknown.

The average $\gamma$-ray luminosity of the BLL in the sample ($<$L$_{\gamma}$$>$~=~10$^{44}$ erg s$^{-1}$), is found significantly lower than that ($<$L$_{\gamma}$ $>$ $\sim$  10$^{45}$ erg s$^{-1}$)  of $\sim$900 BLL of the 4FGL-DR3 catalog (see also Table \ref{tab:targets}). 
While, regarding radio band, all BLL are radio-loud sources. 
It is worth noting that 4FGLJ1410.7+7405 identified as a candidate Radio Weak BL Lac (RWBL) by \citet[]{marchesini2023} and \citet[]{massaro2017}, turns out to instead be a radio loud BLL using the radio flux value estimated in the same work by \citet[]{marchesini2023}.
From the imaging decomposition (see Table \ref{tab:decomp}), the absolute magnitude of their host galaxies in the optical r-band is in the range -23.1$<$M(r)$<$-21.2, with $<$M(r)$>$=-22.4. 
These values are consistent with that typical of BLL host galaxies $<$Mr$>$=-22.9 \citep[][]{urry2000, sbarufatti2005}. 
The indication is therefore that these UGSs identified as BLL belong to the same population of already identified \textit{Fermi} BLL, covering the faint tail of this luminosity \textit{Fermi}  BLL distribution (see Table \ref{tab:targets}). 
% {\bf with low N/H and a lower $\gamma$-ray luminosity.} 
%\red{N/H più basso}
%\red{corresponding to the higher part of the redshift distribution.}

%The other 12 UGS have spectra characterized by prominent emission lines. 
The other 12 UGS of this study are characterized by optical spectra with strong emission lines and constitute $\sim$40\% of the total of our studied sources. 
%They are $\sim$40\% of the total of our studied sample. 
This fraction is in fact in agreement with the larger statistics of the Fourth Catalog of Active Galactic Nuclei detected by \textit{Fermi} - Data Release 3 \citep[4LAC-DR3, ][]{Ajello_2020, Ajello_2022} ($\sim 1400$ BLL and $\sim$ 800 FSRQ) and it is much greater than $\sim$10\% found in our earlier studies of UGS \citep[][]{paiano2017_ufo1, paiano2019_ufo2} for which only 3 objects showed a spectrum not compatible with a BLL classification. 
Note that they were based on the 2FGL and 3FGL catalogues \citep[][]{2fgl,3fgl} which explored fluxes sensibly higher than those examined here (see Table \ref{tab:targets}).

More specifically, we found one radio-loud quasar (4FGLJ1535.9+3743), three radio-quiet quasars (4FGLJ1125.1+4811, 4FGLJ1256.8+5329 and 4FGLJ1308.7+0347) and eight Seyfert-like objects. 
4FGLJ0023.6-4209 and 4FGLJ2030.0-0310, two Type-2 AGN, are the closest sources of the sample with z$\leq$0.05 and have the smallest $\gamma$-ray luminosity (see Table \ref{tab:33_UGS_Flux}).
4FGLJ0117.9+1430 is classified as a NLSy1 (see the note of the source in Section \ref{sec:note_ind_sources}). This increases the total number of objects classified as Seyfert of the 4FGL-DR3 catalog (8 NLSy and 3 Seyfert).
%This increases the total of recognized objects classified as NLSy1 in the 4FGL-DR3 catalog to nine.}
%11 previous
%8 this work

It is of interest to compare the values of the \textit{radio-loudness (R)} found in our sample of 33 objects with the results of 4LAC, which are summarized in the distribution reported in Fig. \ref{fig:R_distr}. 
It is apparent that our 21 BLL have \textit{R} indexes which are in the low part of the distribution. 
The median value is \textit{R}=120 to be compared with that R=230 for the 4LAC BLL. 
%The mean and median values are R$_{mean}$=260, R$_{median}$=120 to be compared with R$_{mean}$=2200 and R$_{median}$=220 for the 4LAC BLL. 
We also compared the distribution of R values of the 12 objects with prominent emission lines with the 4LAC FSRQ one. We found a significant  difference and, as expected for the case of Seyfert galaxies, their \textit{radio-loudness} value is well below the 4LAC FSRQ one (median R=5600; see Figure \ref{fig:R_distr}).
%and in particular $\geq \, 5$ of them have R smaller than 10 (see Figure \ref{fig:R_distr}). 

The presence of a sizeable fraction of radio faint objects, in particular $\geq \, 5$ sources have R smaller than 10, among the UGS counterparts is somewhat unexpected but not implausible \citep[][]{massaro2017,jarvela2021}{}{}. 
The obvious comment is that our current search for UGS counterparts is completely independent of the radio brightness, which is gathered \textit{a posteriori}, it rather depends on the choose of the significance of the X-ray detection.
In spite of the rather small sample the issue of radio quiet counterparts of $\gamma$-ray sources is of potentially great interest since they may represent a poorly explored type of $\gamma$-ray objects. %\textbf{A more significant result could be highlighted in a follow-up paper (Paiano et al.,in prep.), in which a larger sample of UGS will be analyzed.}
%A better  understanding of these sources, however,  requires the study of a larger sample.

%\setcounter{figure}{2}
\begin{figure*}%[htbp]
\center
\includegraphics[width=0.33\textwidth, angle=0]{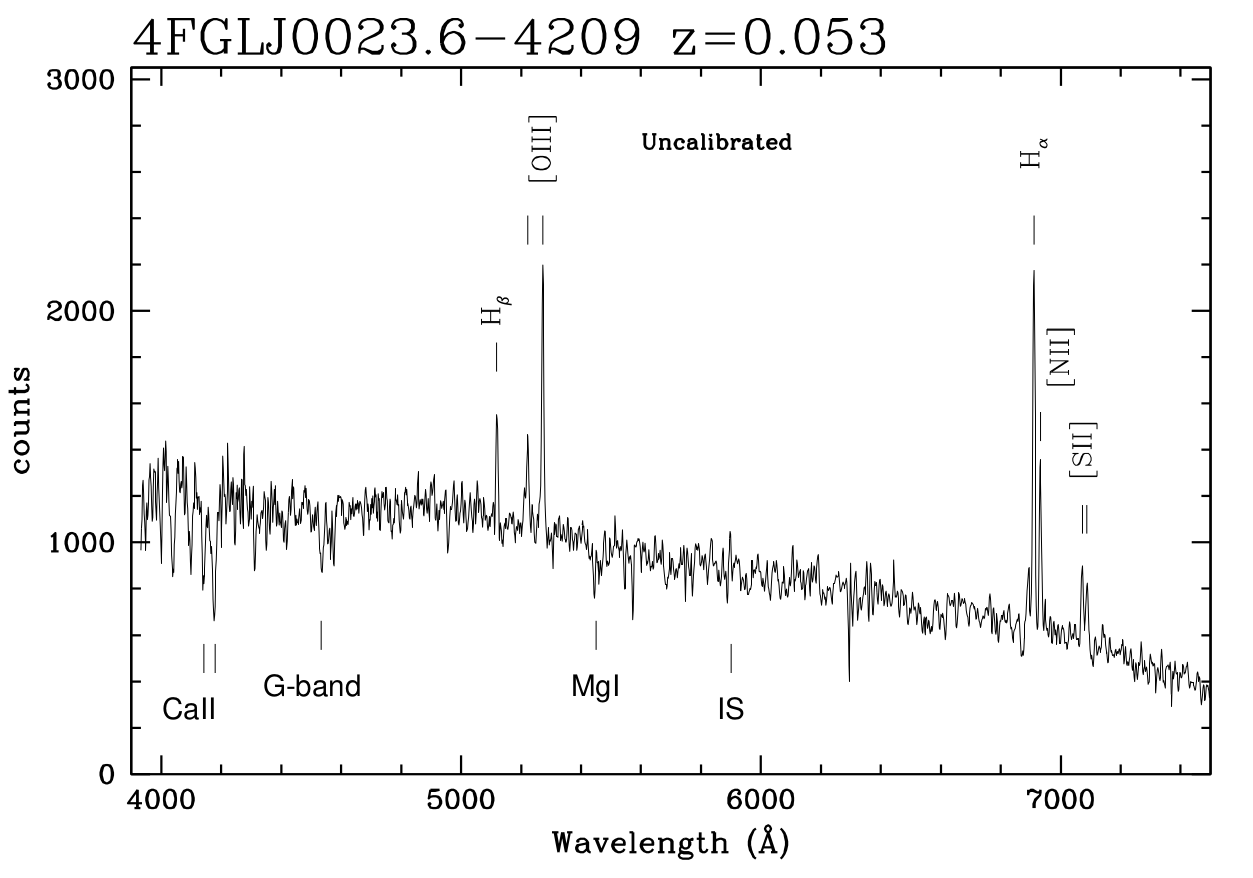}
\includegraphics[width=0.33\textwidth, angle=0]{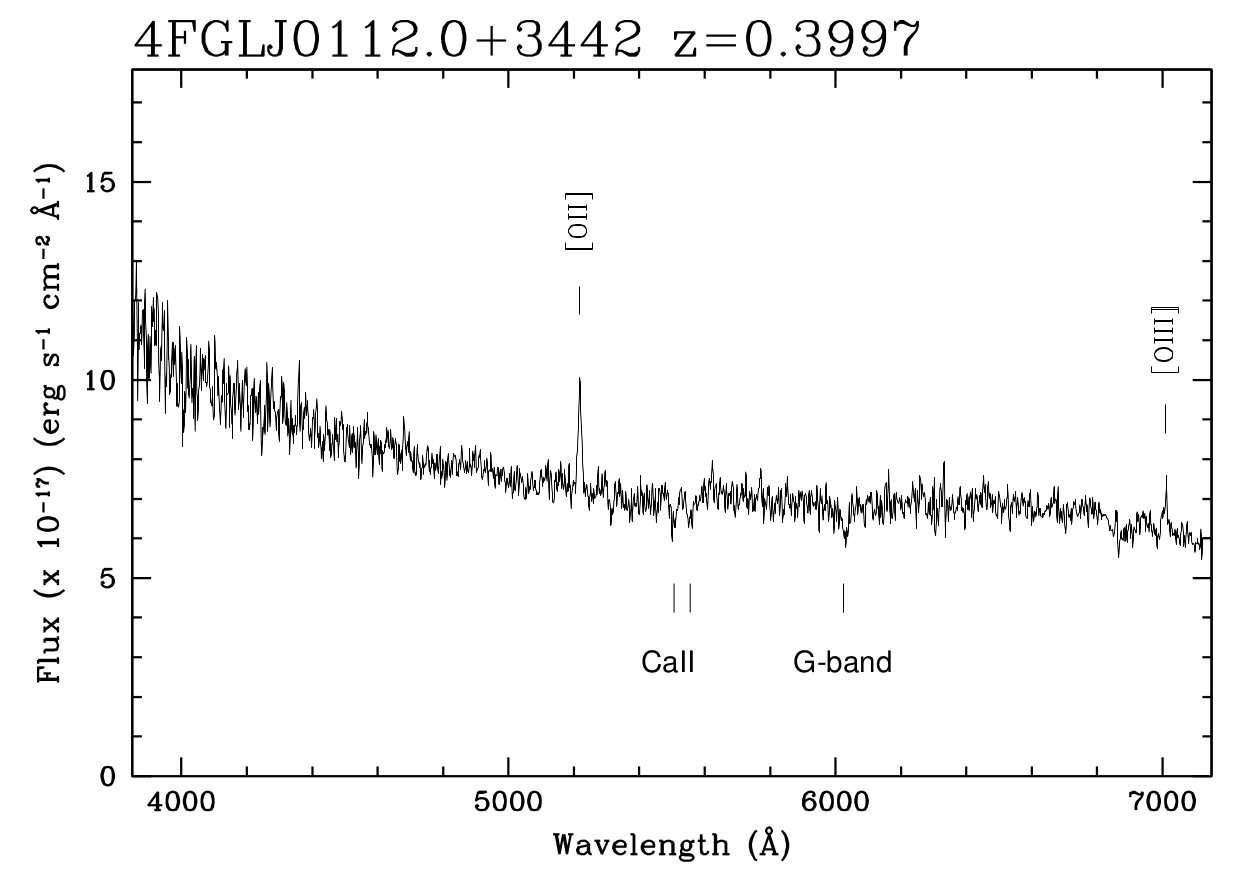}
\includegraphics[width=0.33\textwidth, angle=0]{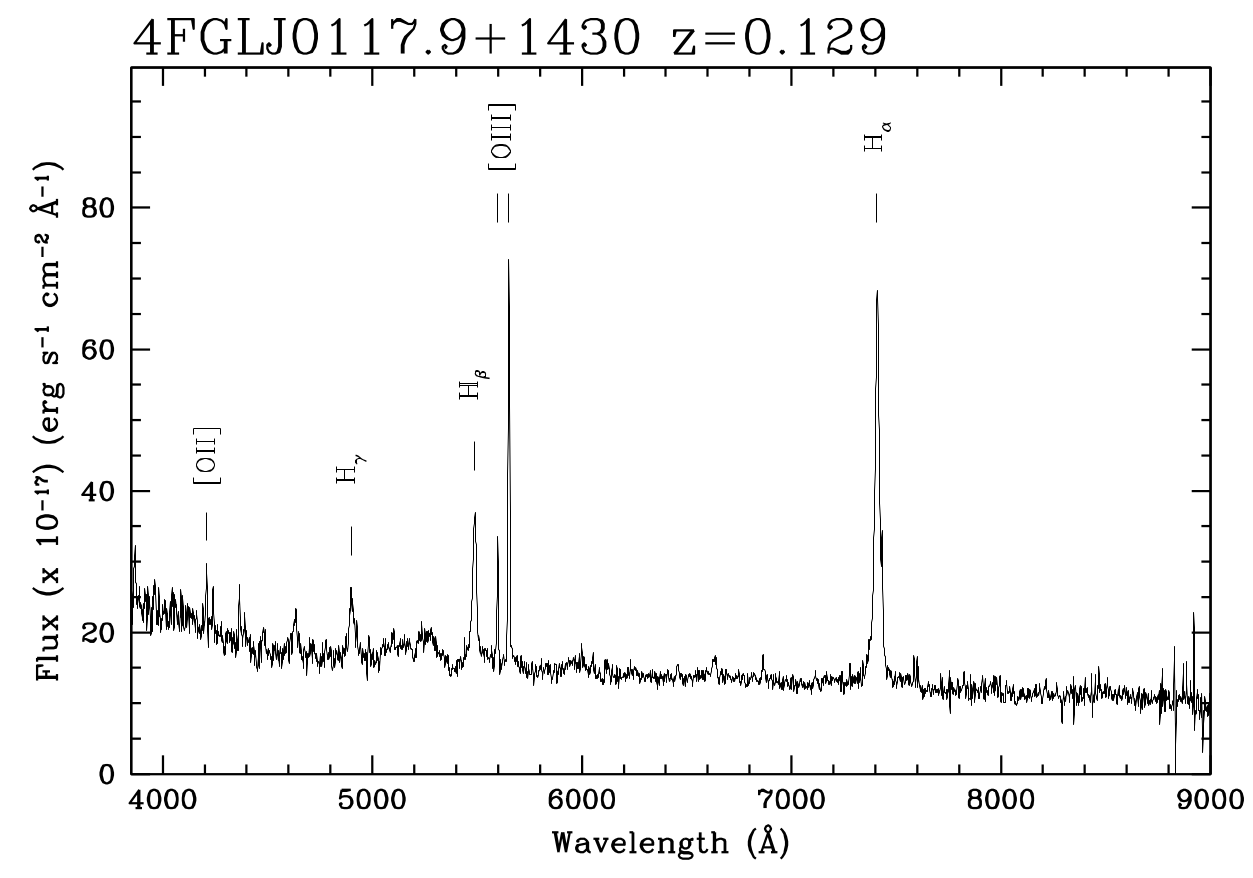}
\includegraphics[width=0.33\textwidth, angle=0]{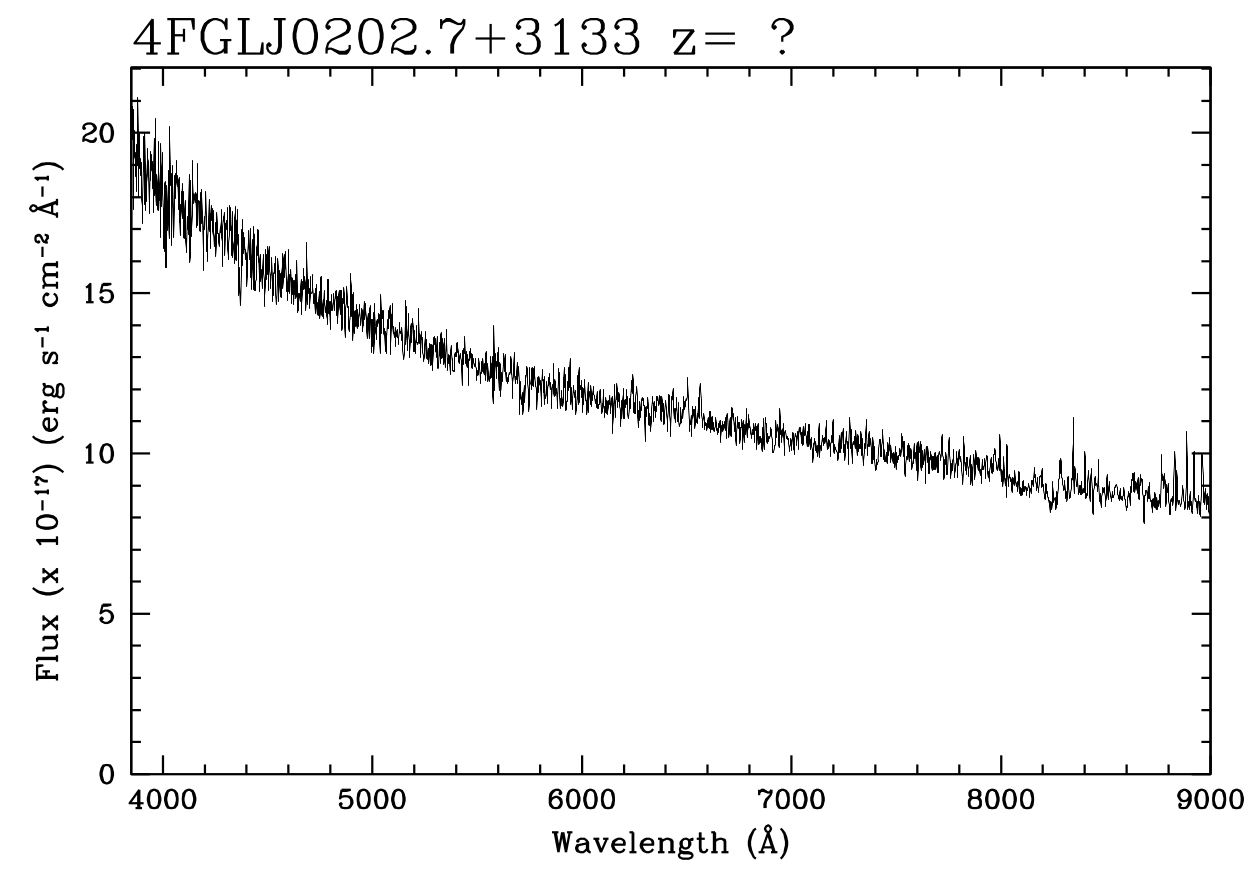}
\includegraphics[width=0.33\textwidth, angle=0]{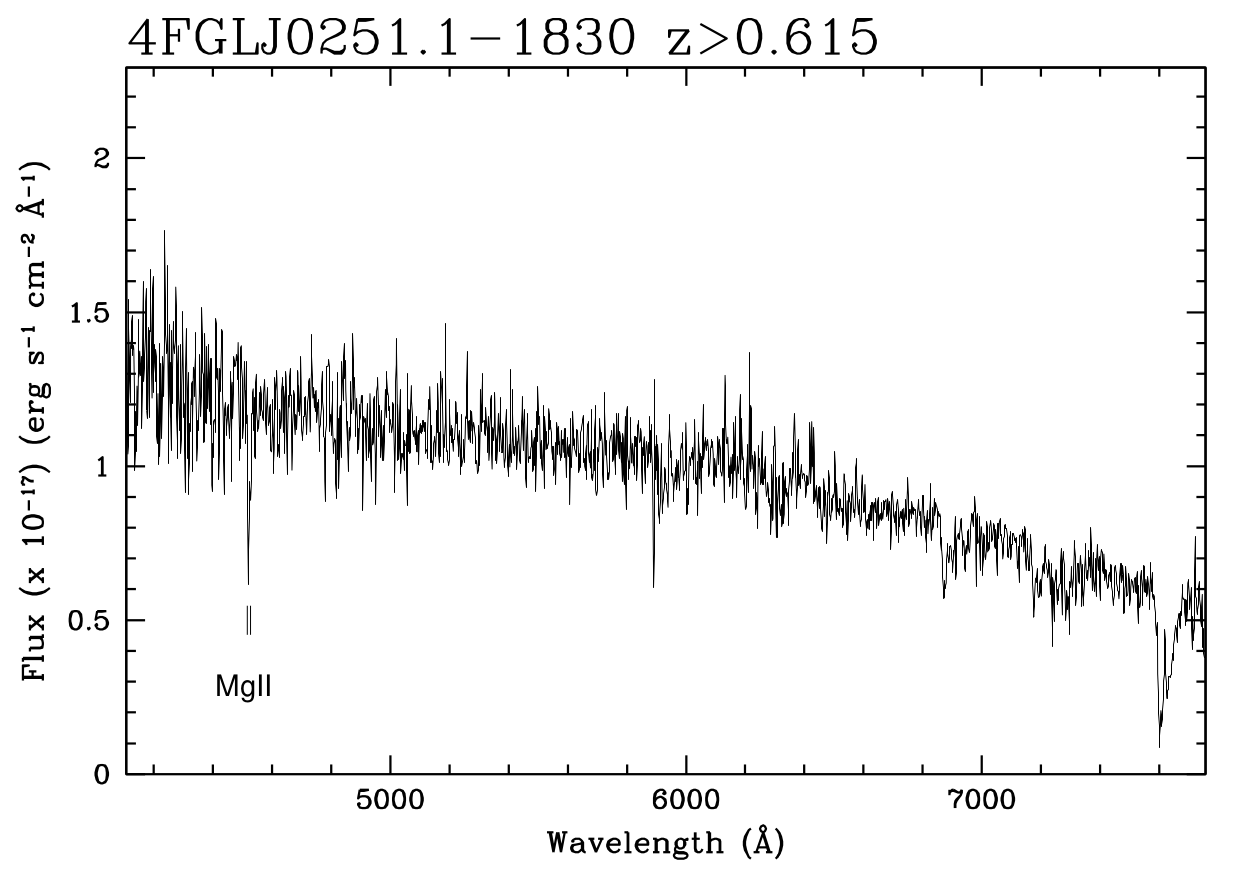}
\includegraphics[width=0.33\textwidth, angle=0]{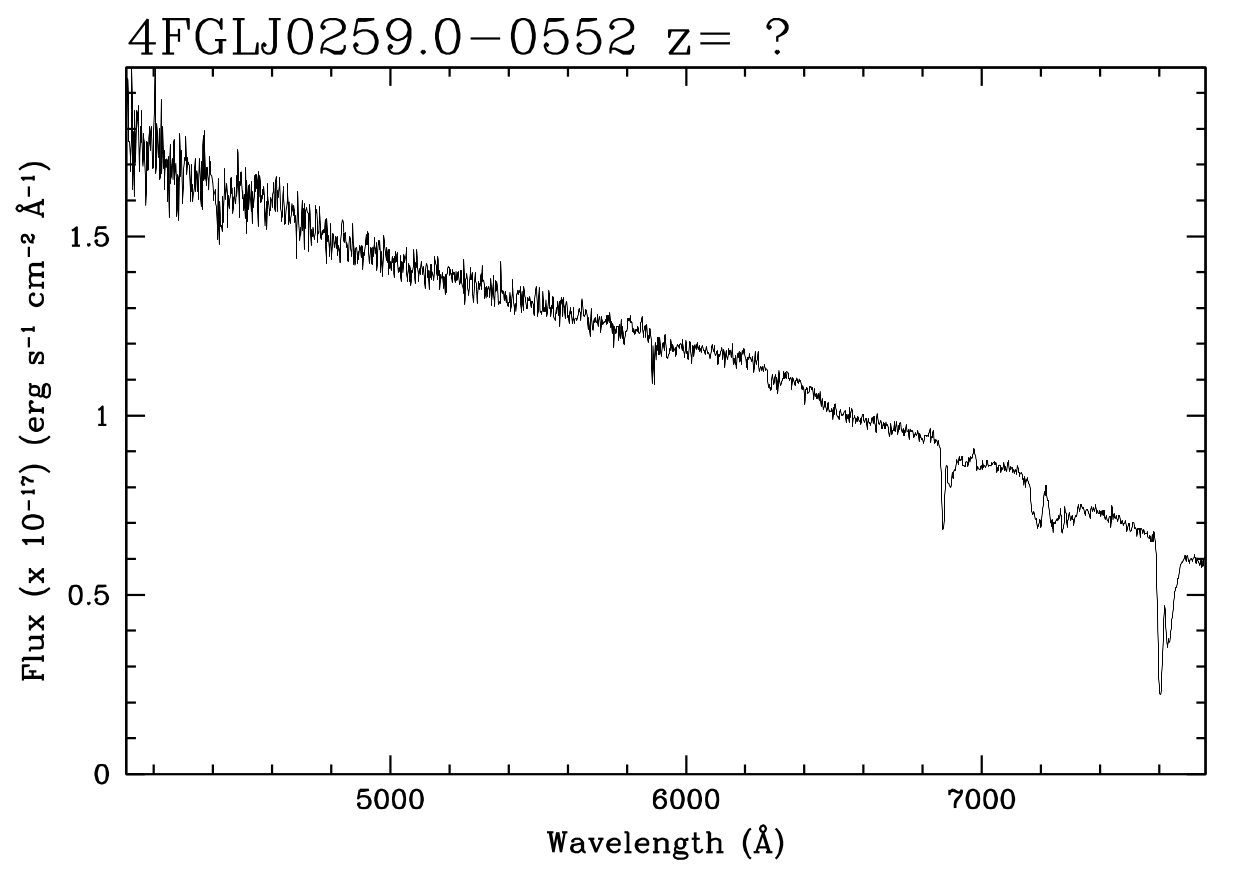}
\includegraphics[width=0.33\textwidth,height=0.172\textheight, angle=0]{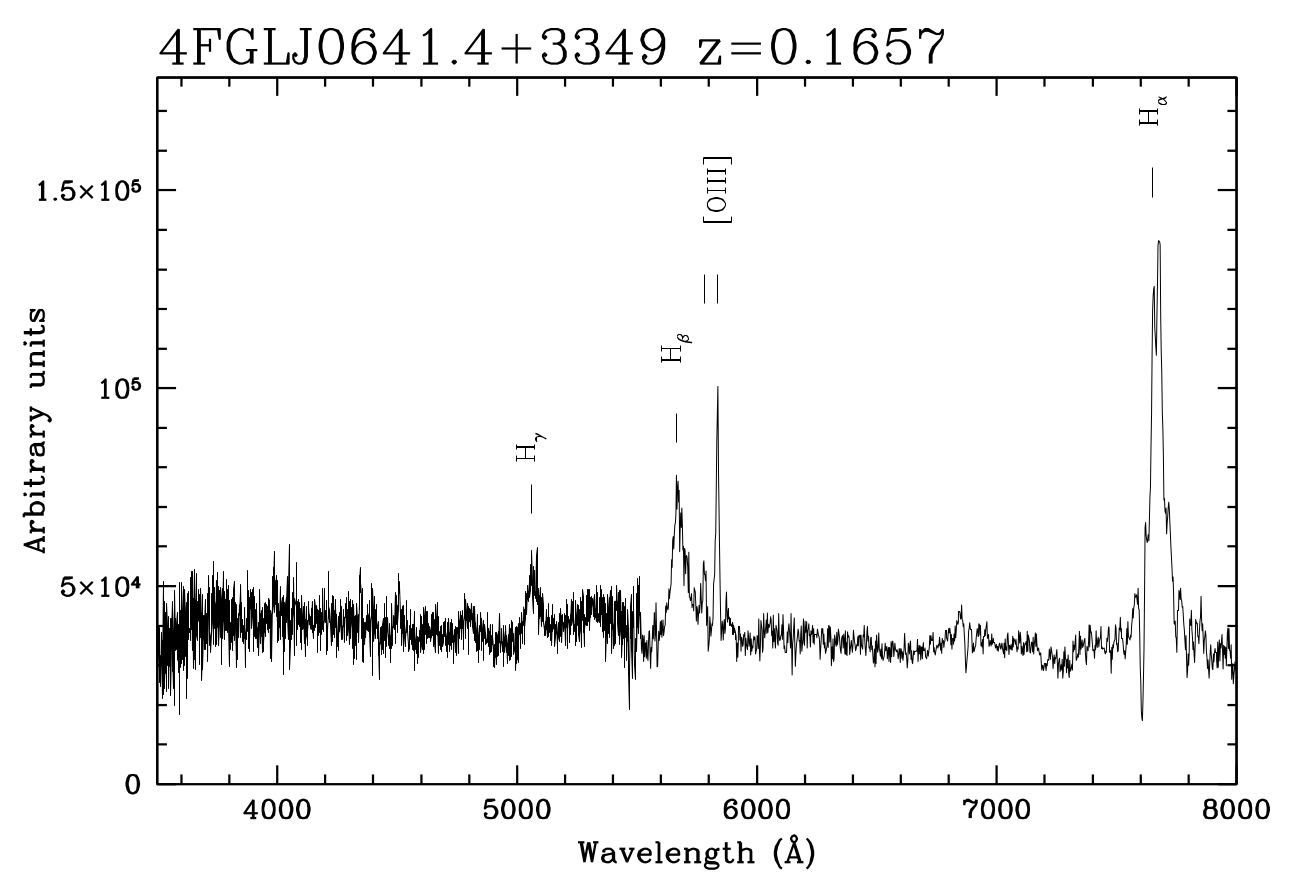}
\includegraphics[width=0.33\textwidth, angle=0]{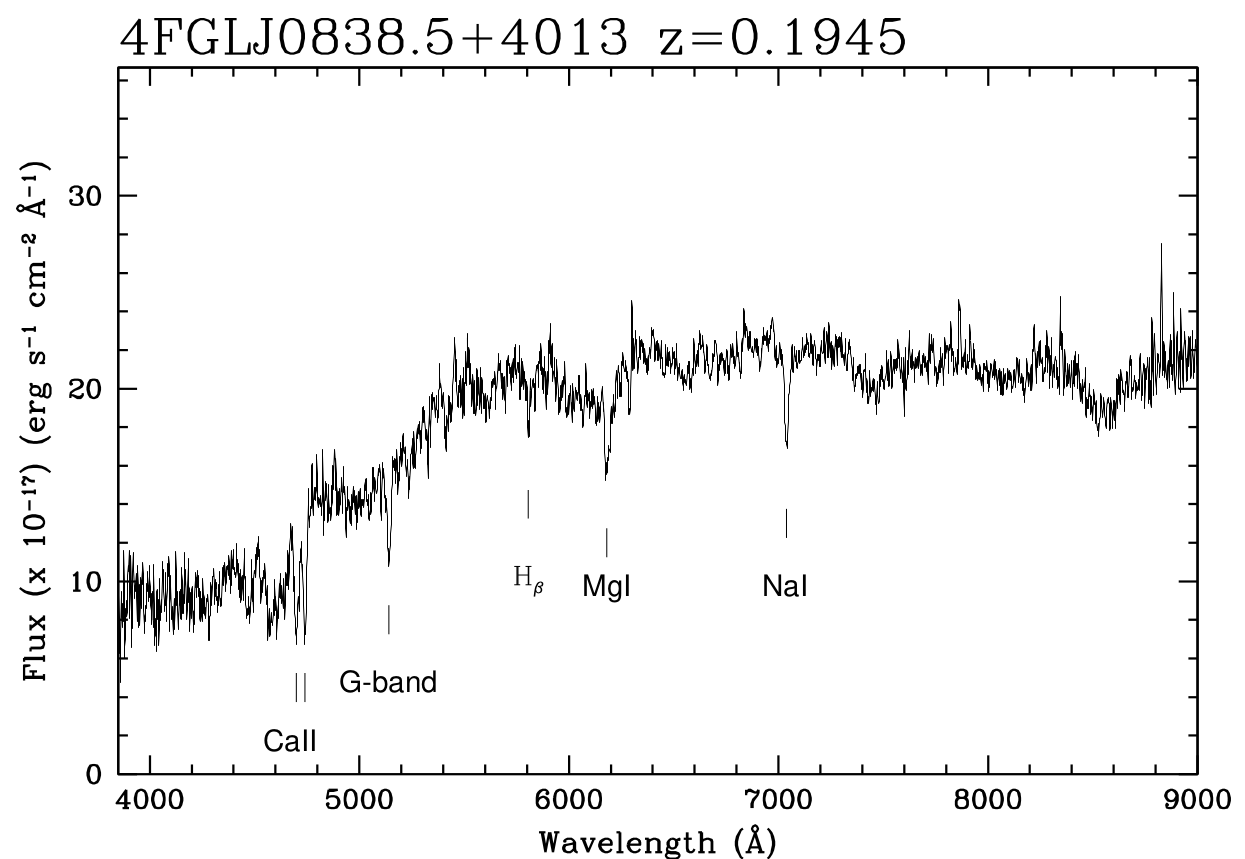}
\includegraphics[width=0.33\textwidth, angle=0]{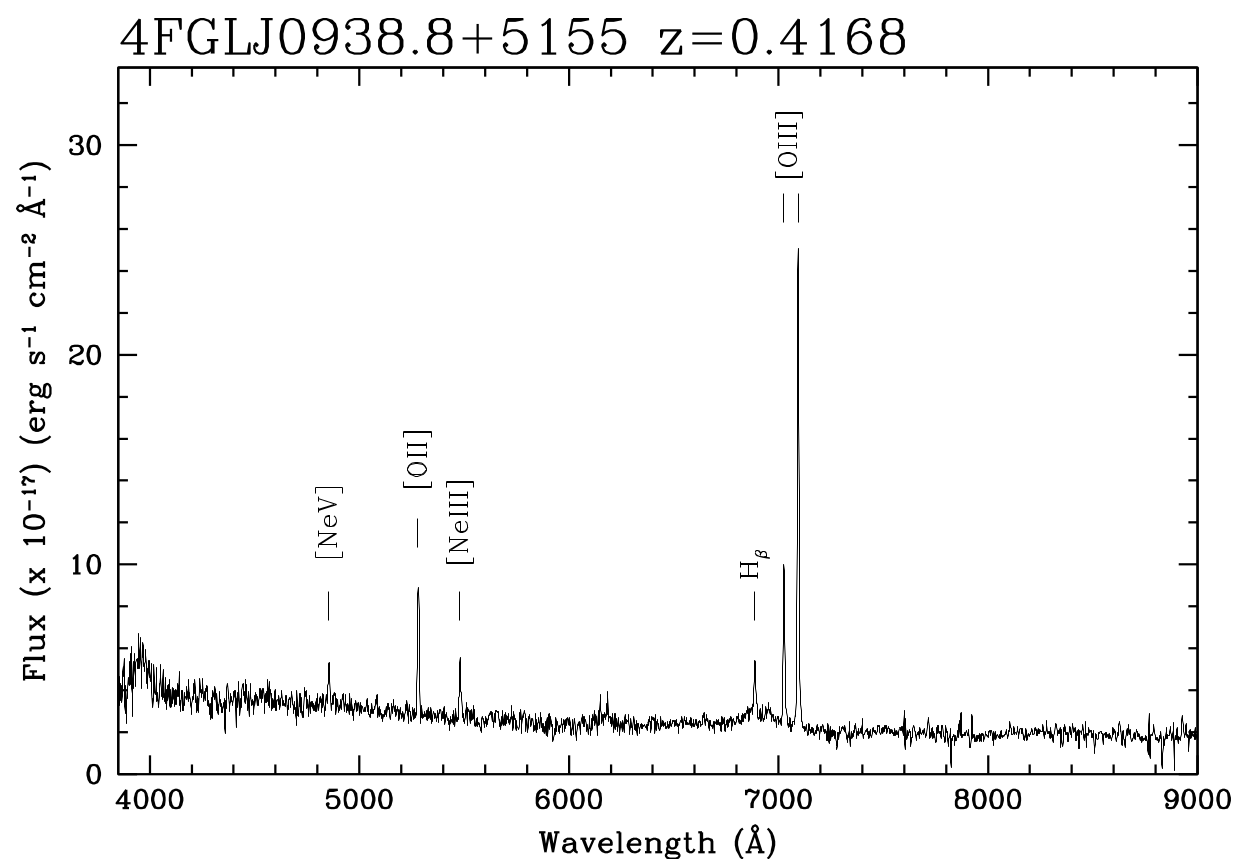}
\includegraphics[width=0.33\textwidth, angle=0]{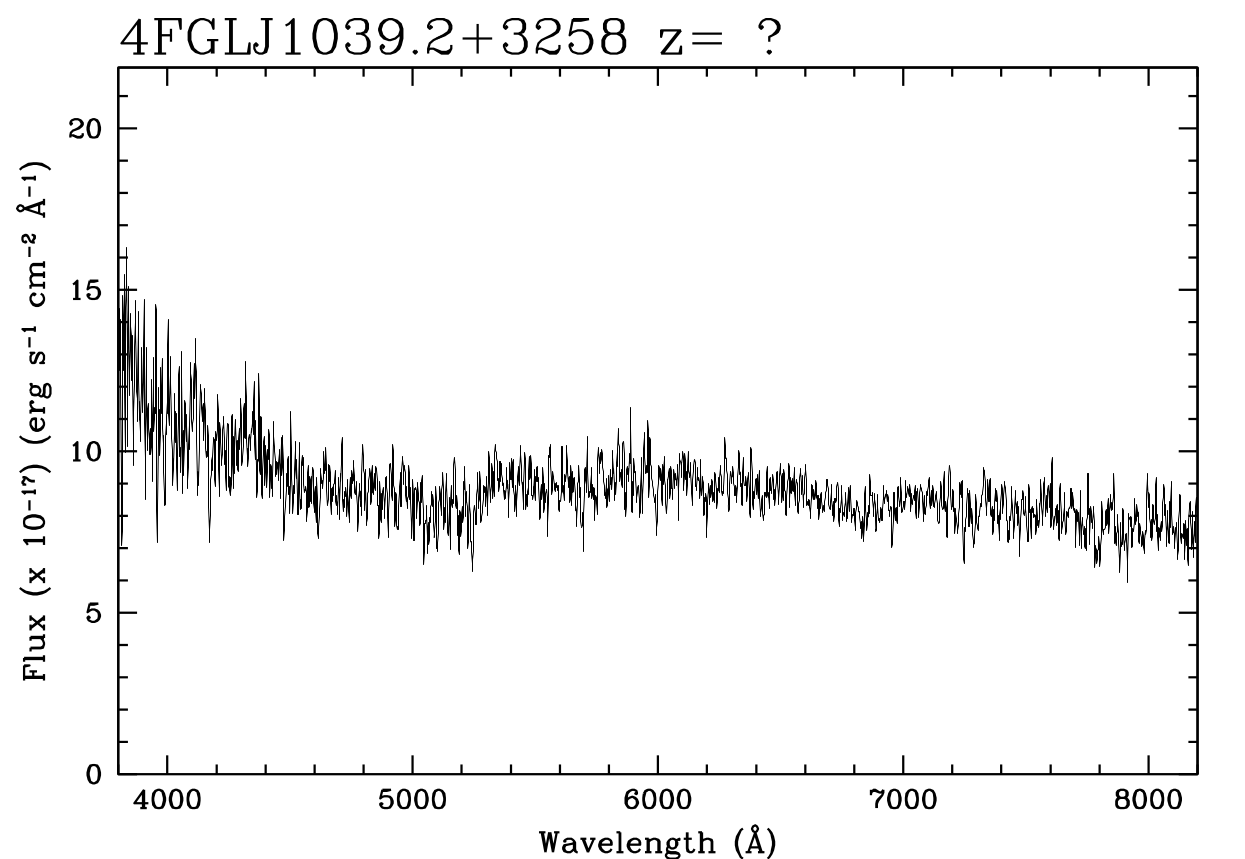}
\includegraphics[width=0.33\textwidth, angle=0]{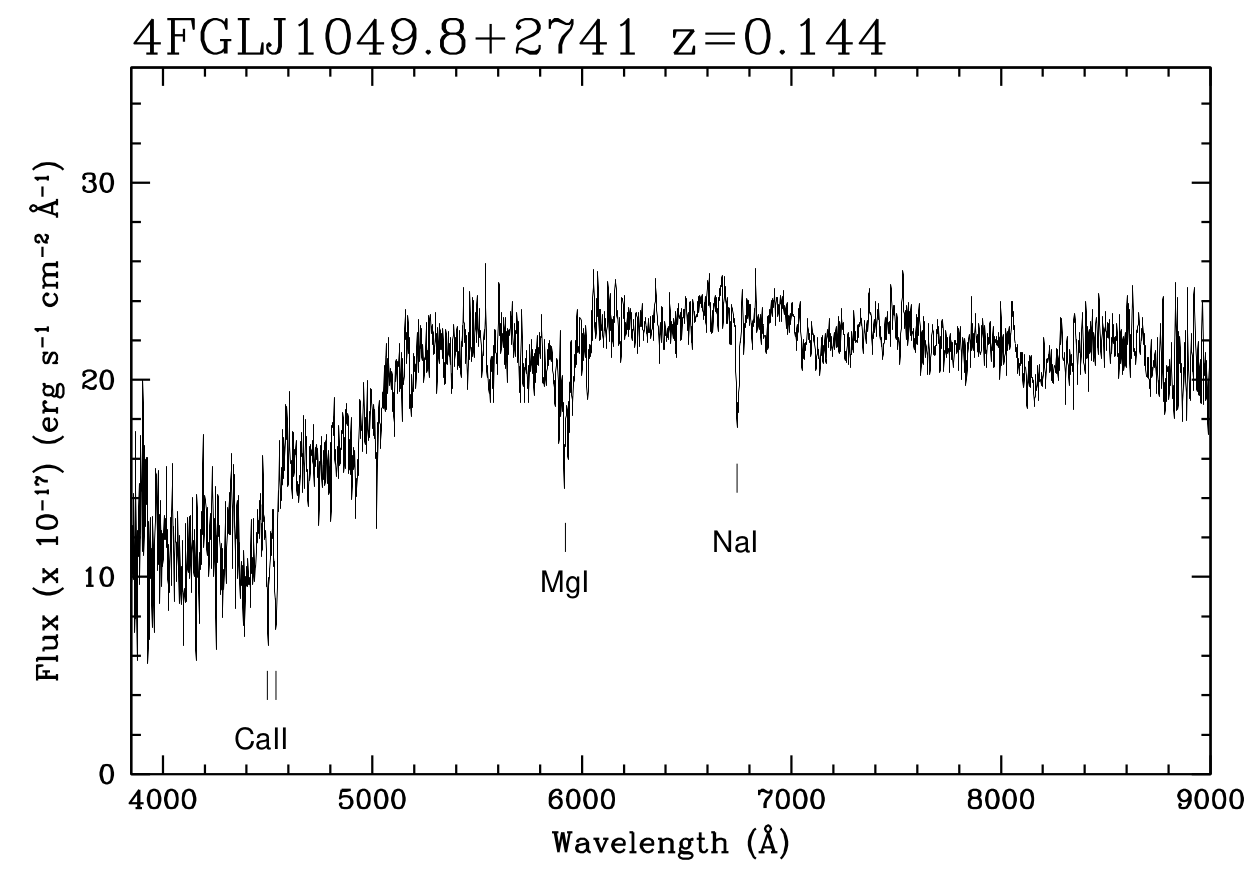}
\includegraphics[width=0.33\textwidth, angle=0]{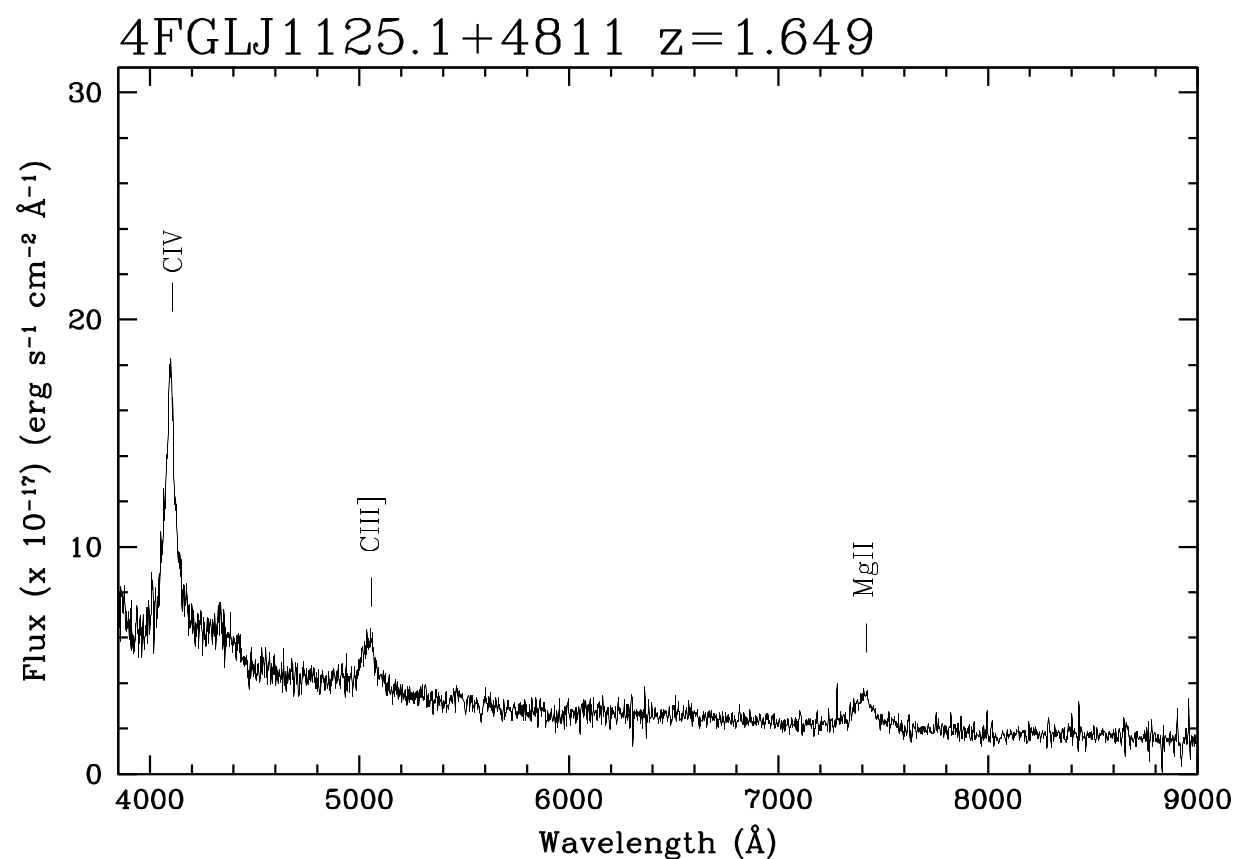}
\includegraphics[width=0.33\textwidth, angle=0]{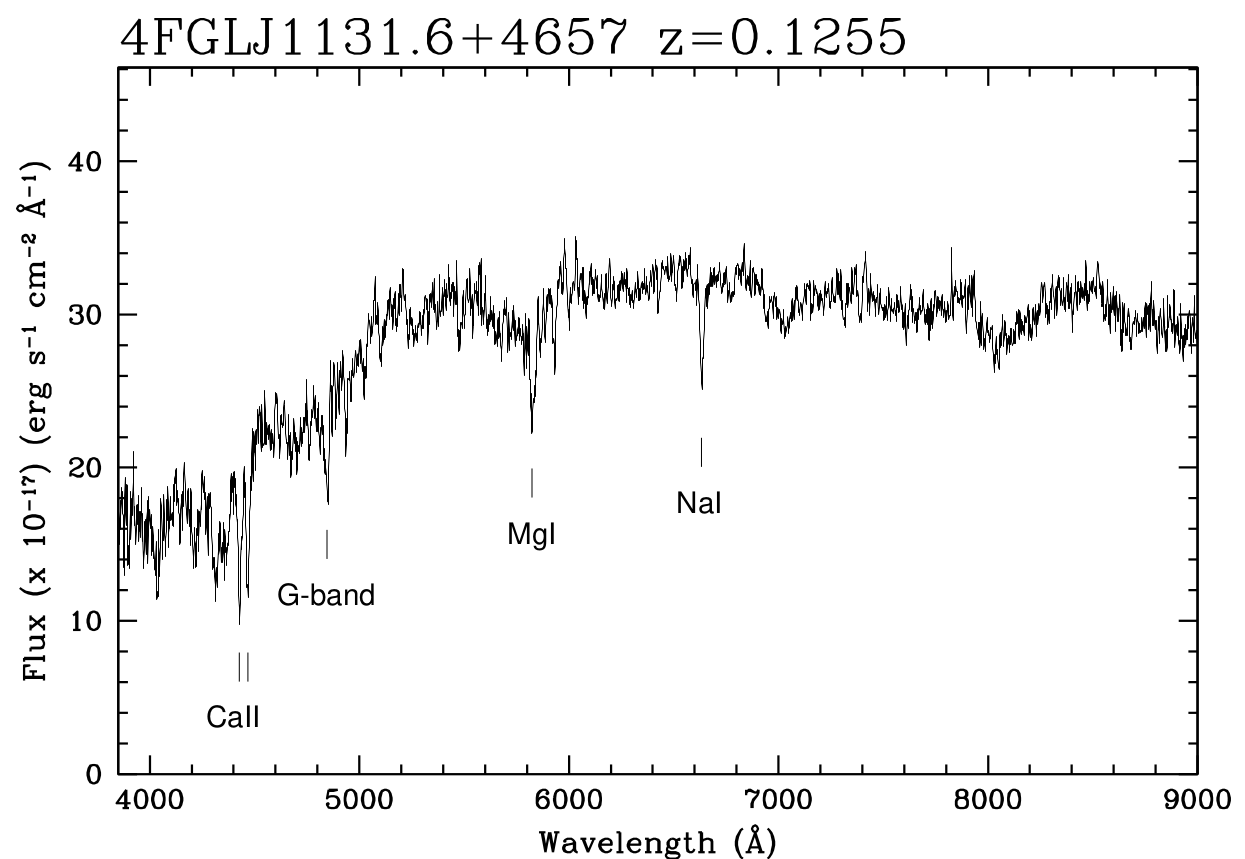}
\includegraphics[width=0.33\textwidth, angle=0]{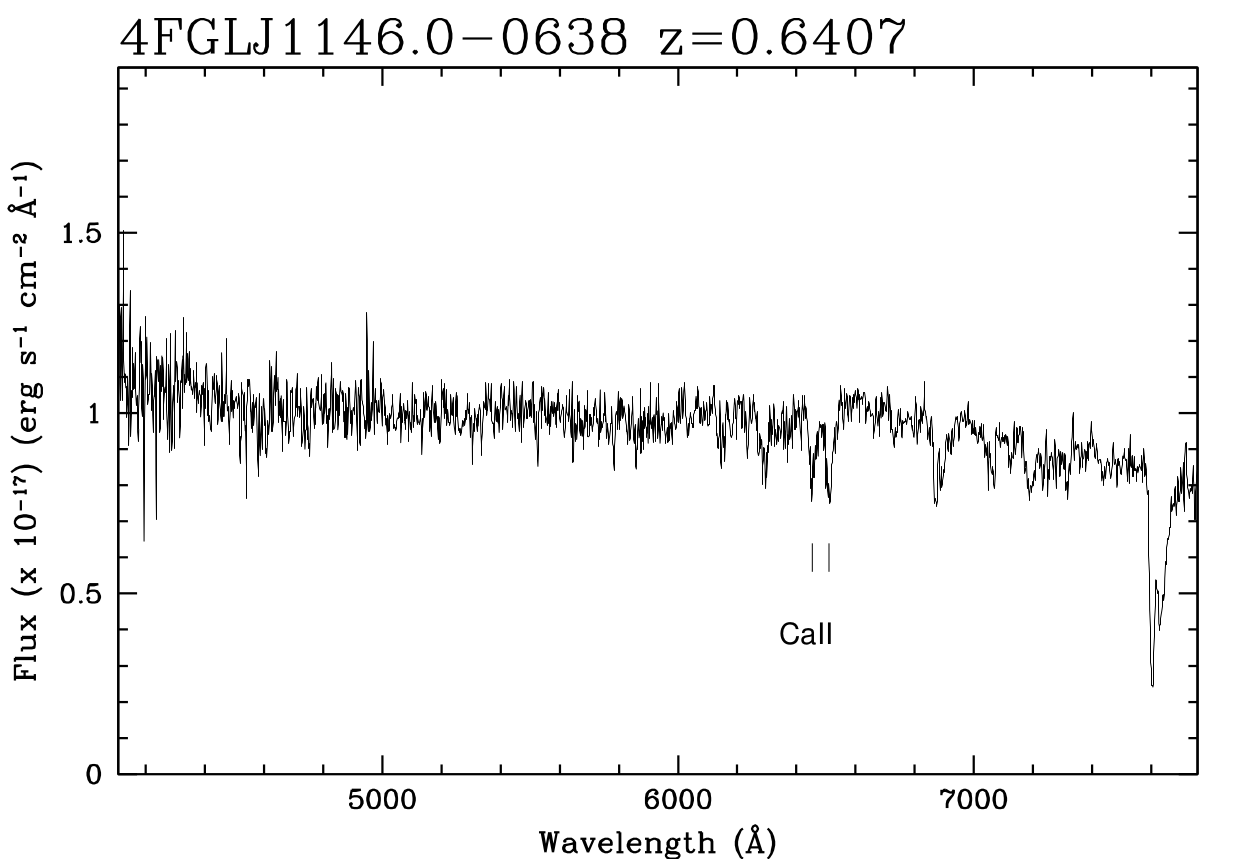}
\includegraphics[width=0.33\textwidth, angle=0]{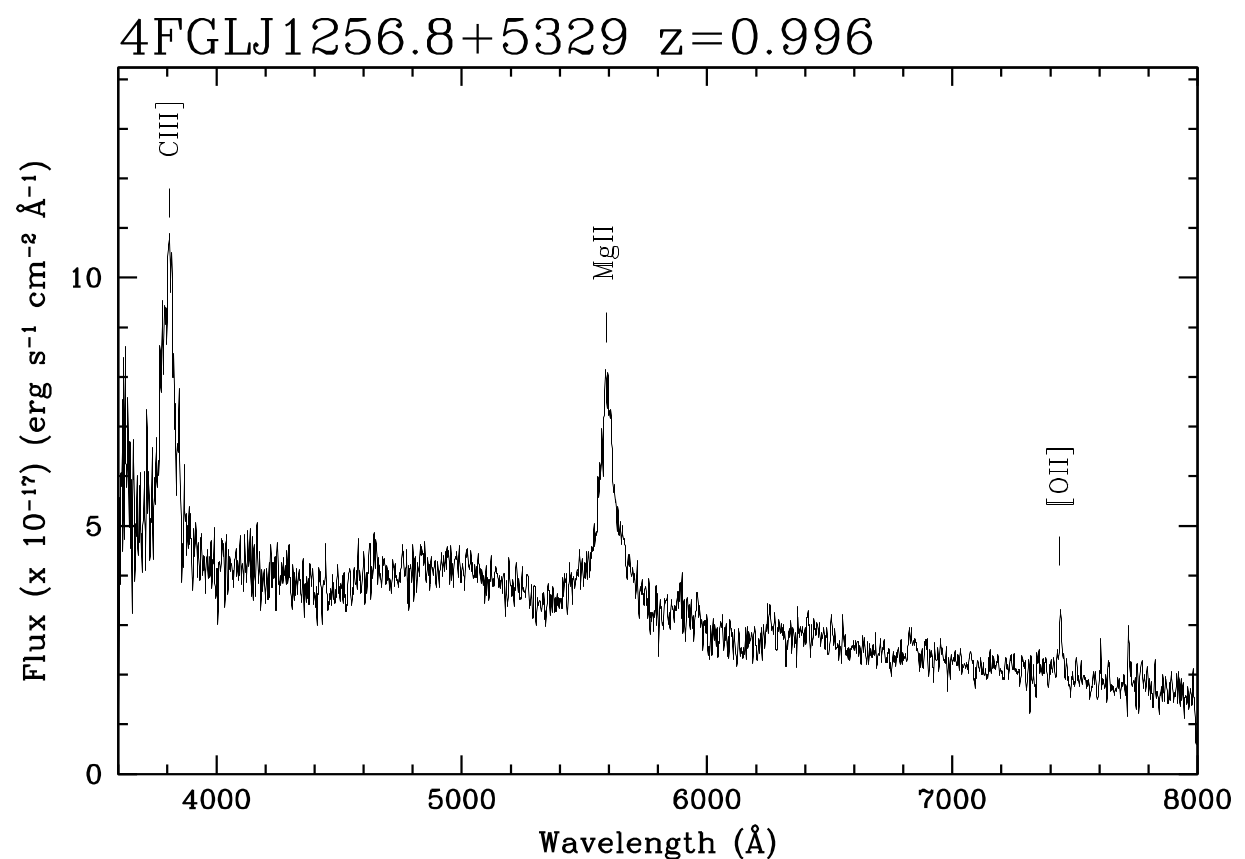}
\caption{Optical spectra of the counterparts of UGS with only one X-ray detection within the 3$\sigma$ Fermi error ellipses (see text and Table 2 and 3 for details).} 
%\red{MANCANO QUATTRO SPETTRI:  1410 DI MARCHESINI (ma di quest'ultimo non abbiamo il fits probabilmente) E 1016 DI RAJAGOPAL, 2240 DI DESAI E 2323 DI MARCHA}} 
\label{fig:spectrum}
\end{figure*}%[htbp]

\setcounter{figure}{1}
\begin{figure*}%[htbp]
\center
\includegraphics[width=0.33\textwidth, angle=0]{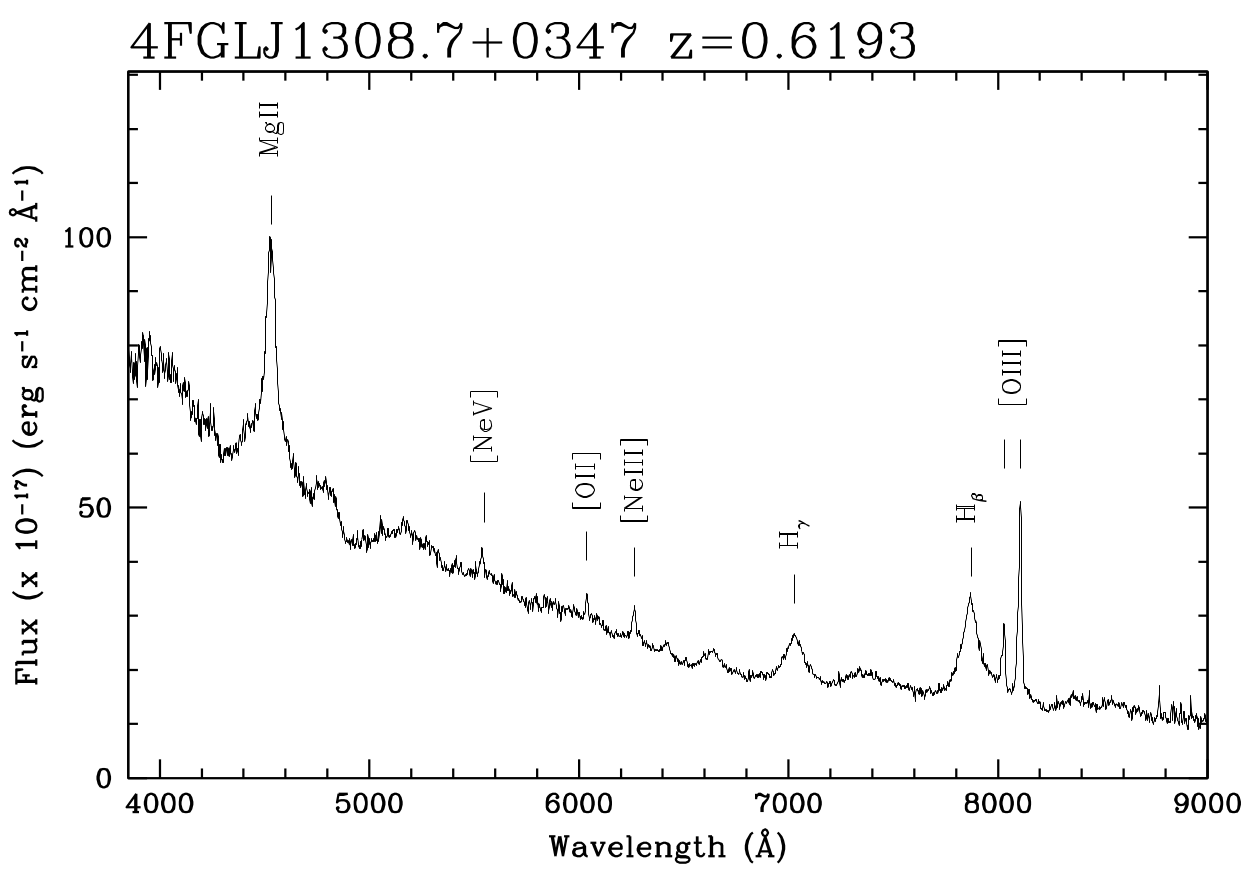}
\includegraphics[width=0.33\textwidth, angle=0]{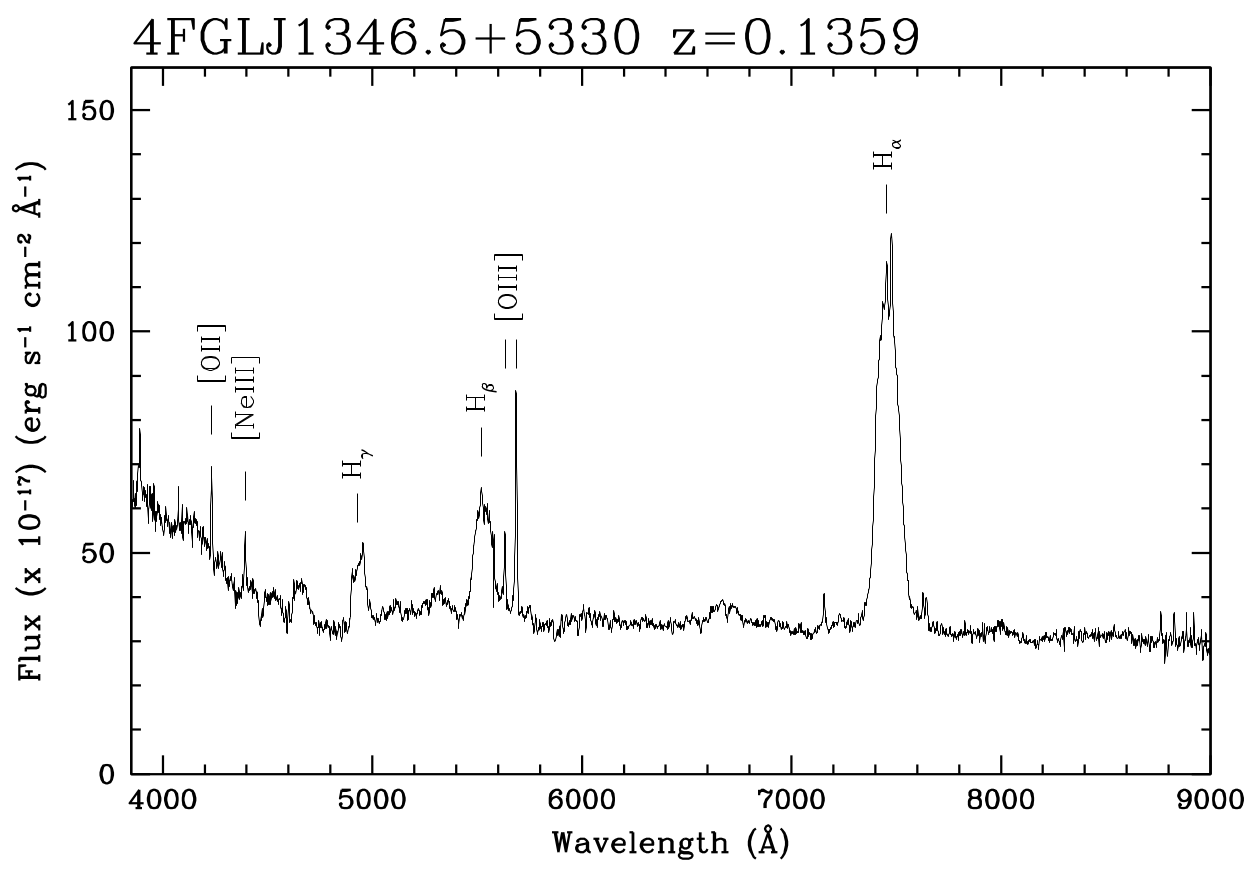}
\includegraphics[width=0.33\textwidth, angle=0]{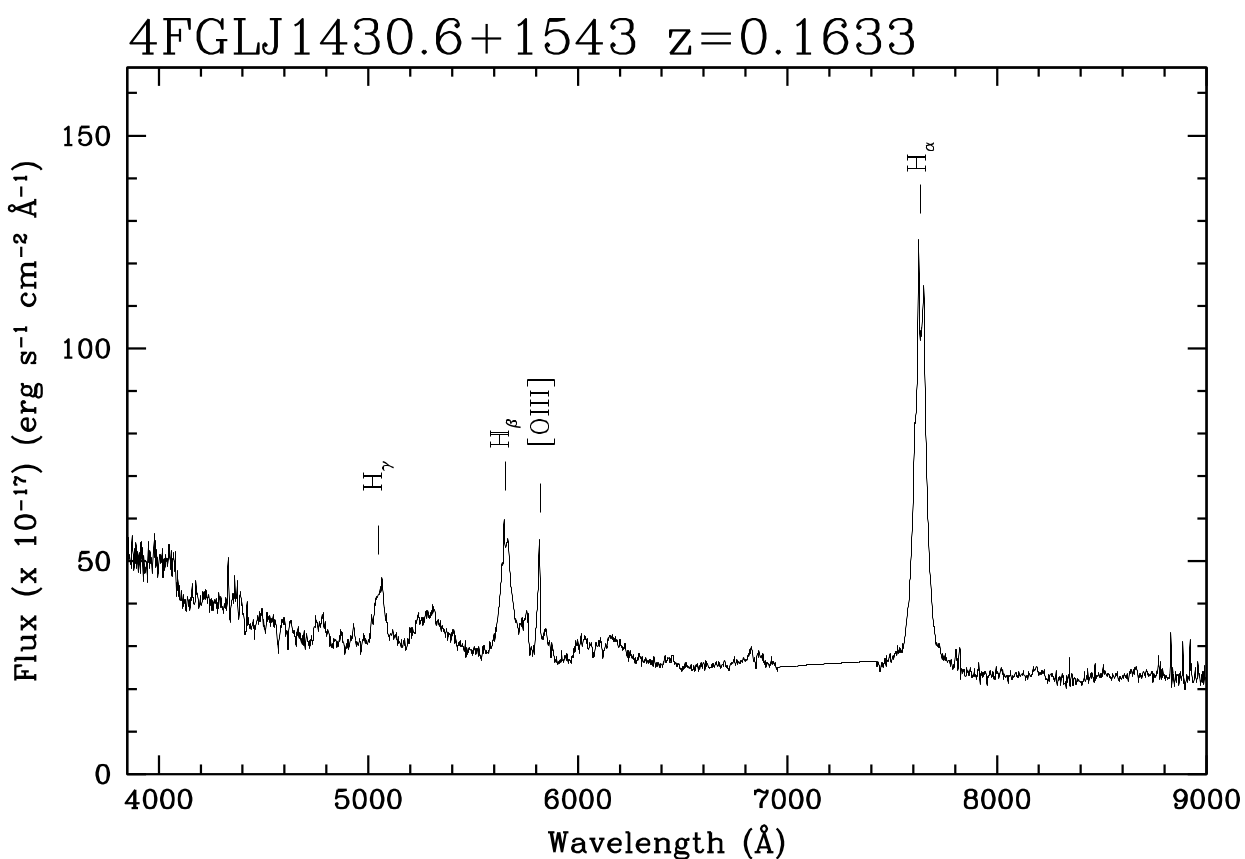}
\includegraphics[width=0.33\textwidth, angle=0]{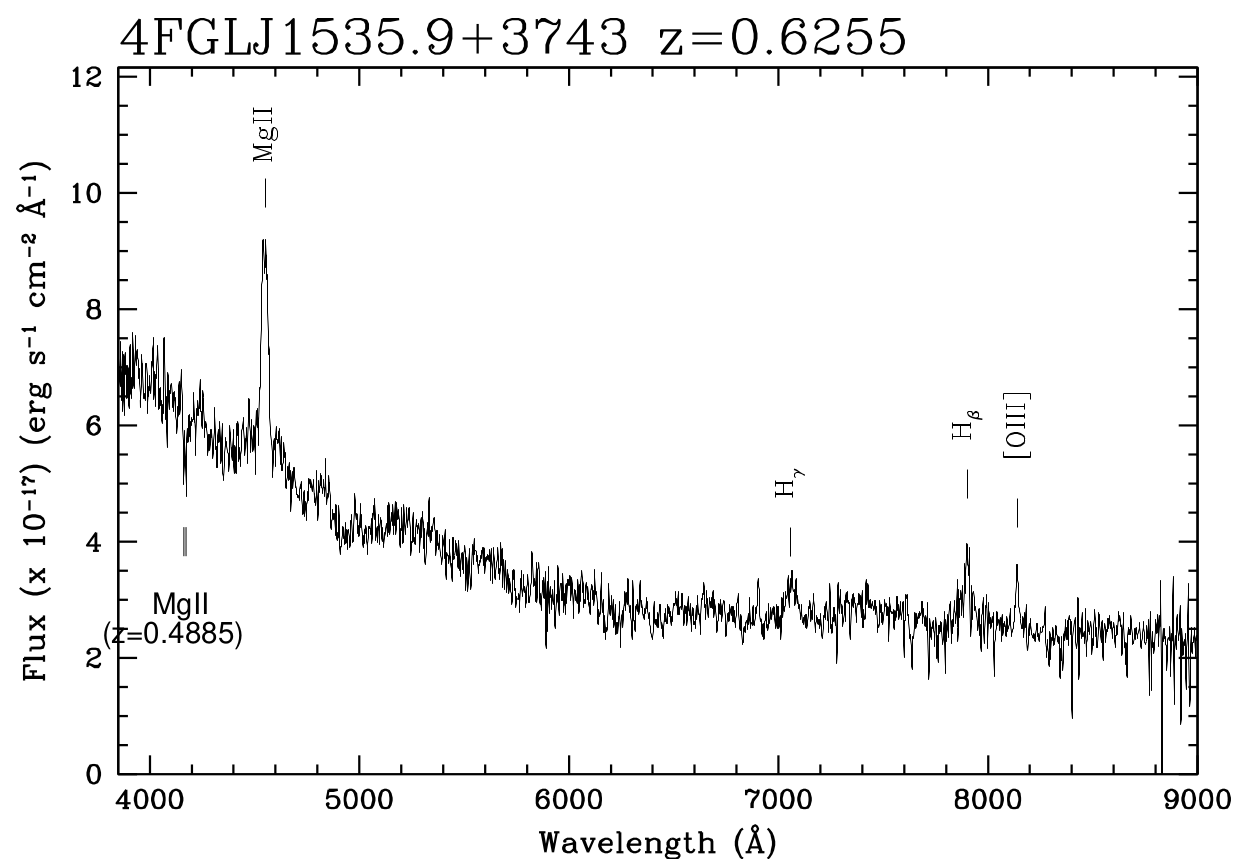}
\includegraphics[width=0.33\textwidth, angle=0]{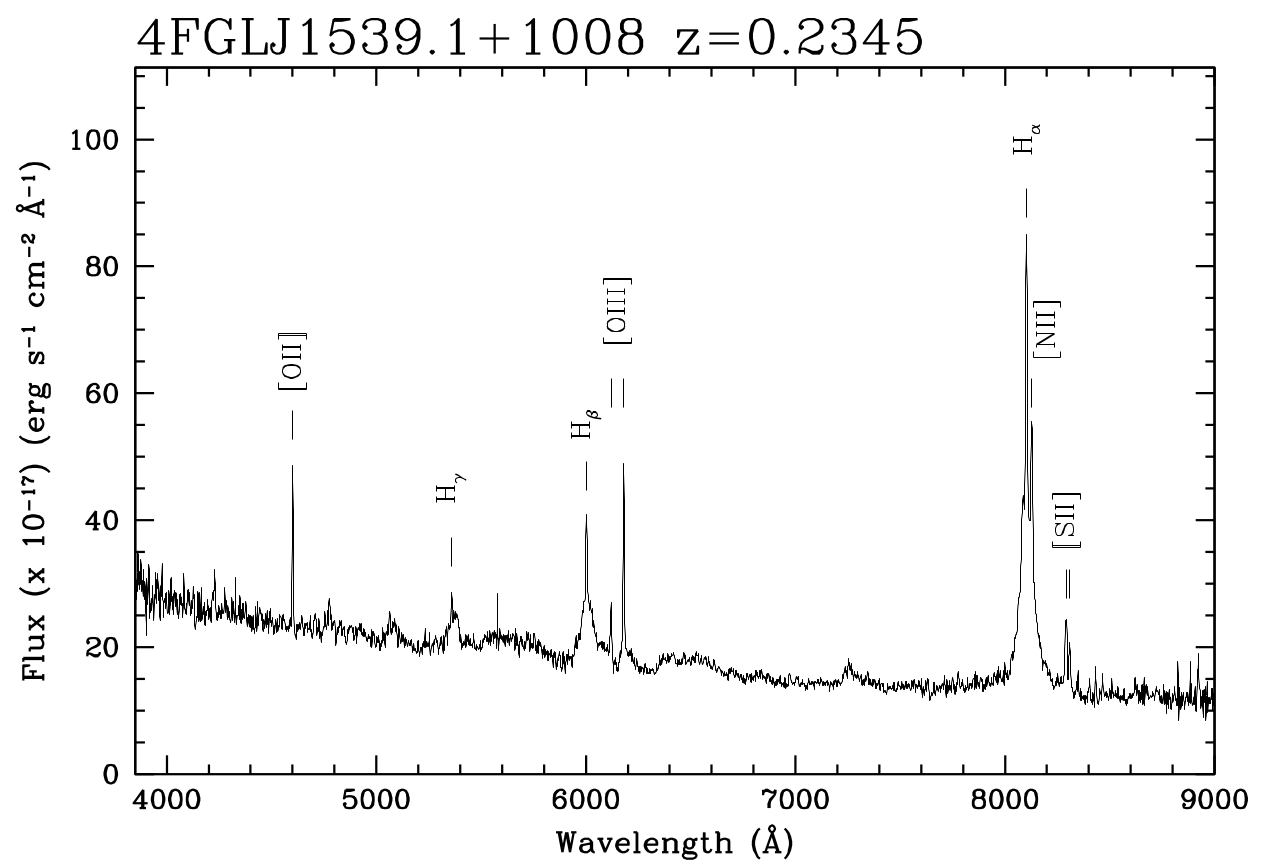}
\includegraphics[width=0.33\textwidth, angle=0]{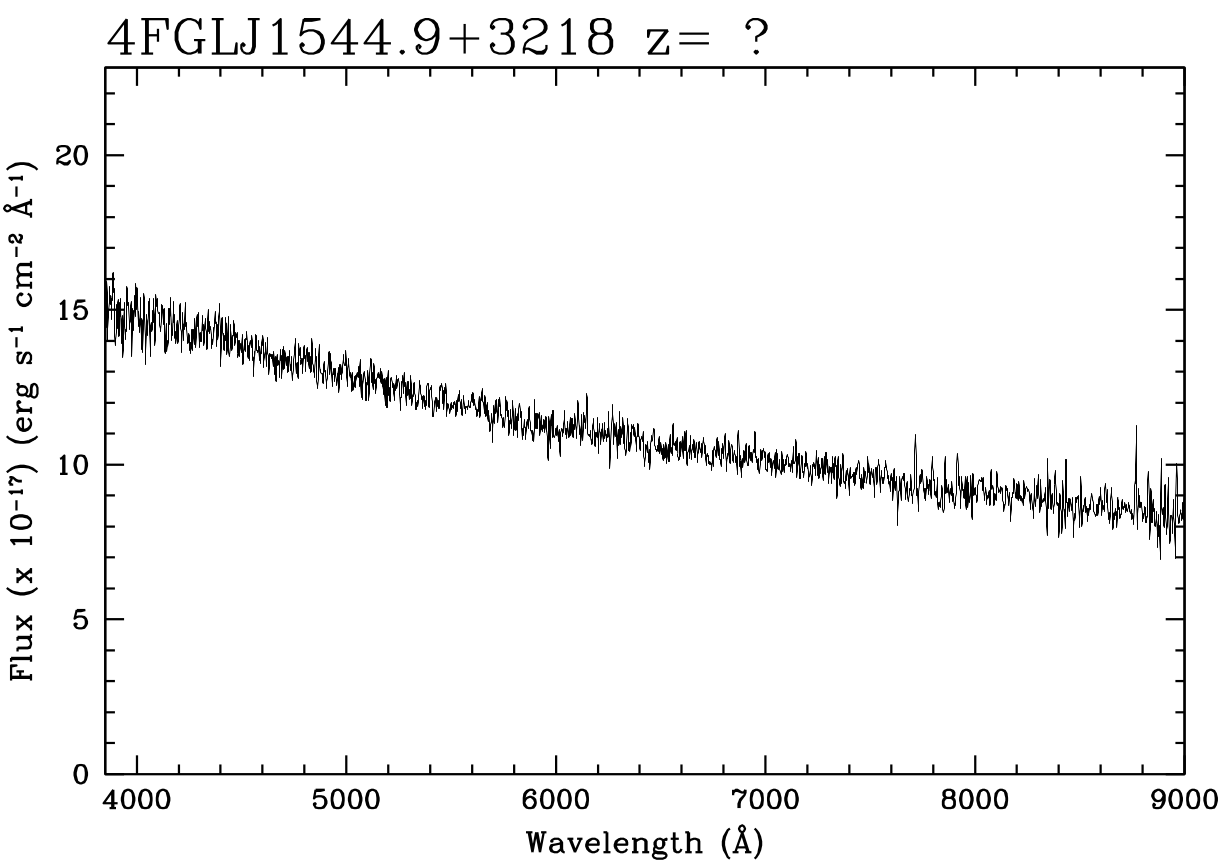}
\includegraphics[width=0.33\textwidth,height=0.17\textheight, angle=0]{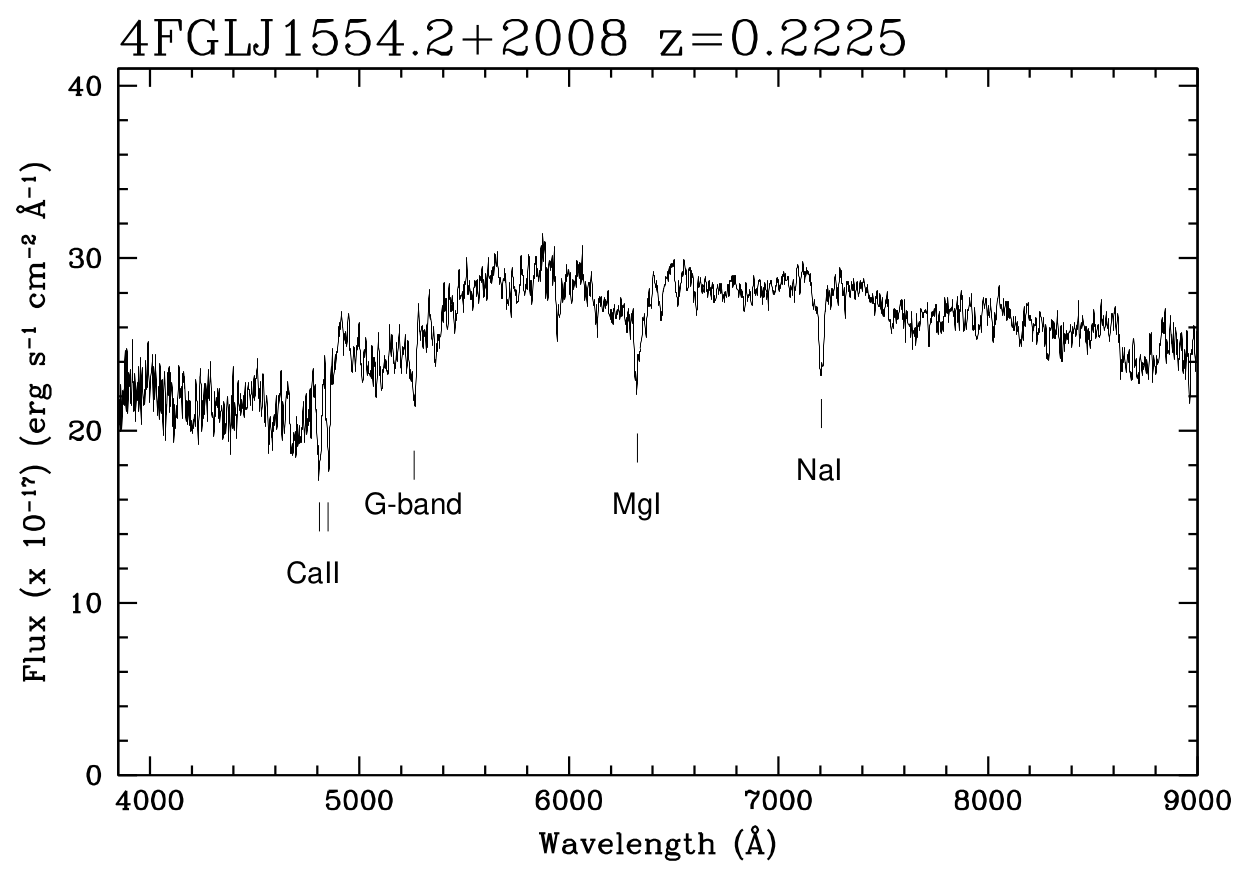}
\includegraphics[width=0.33\textwidth, angle=0]{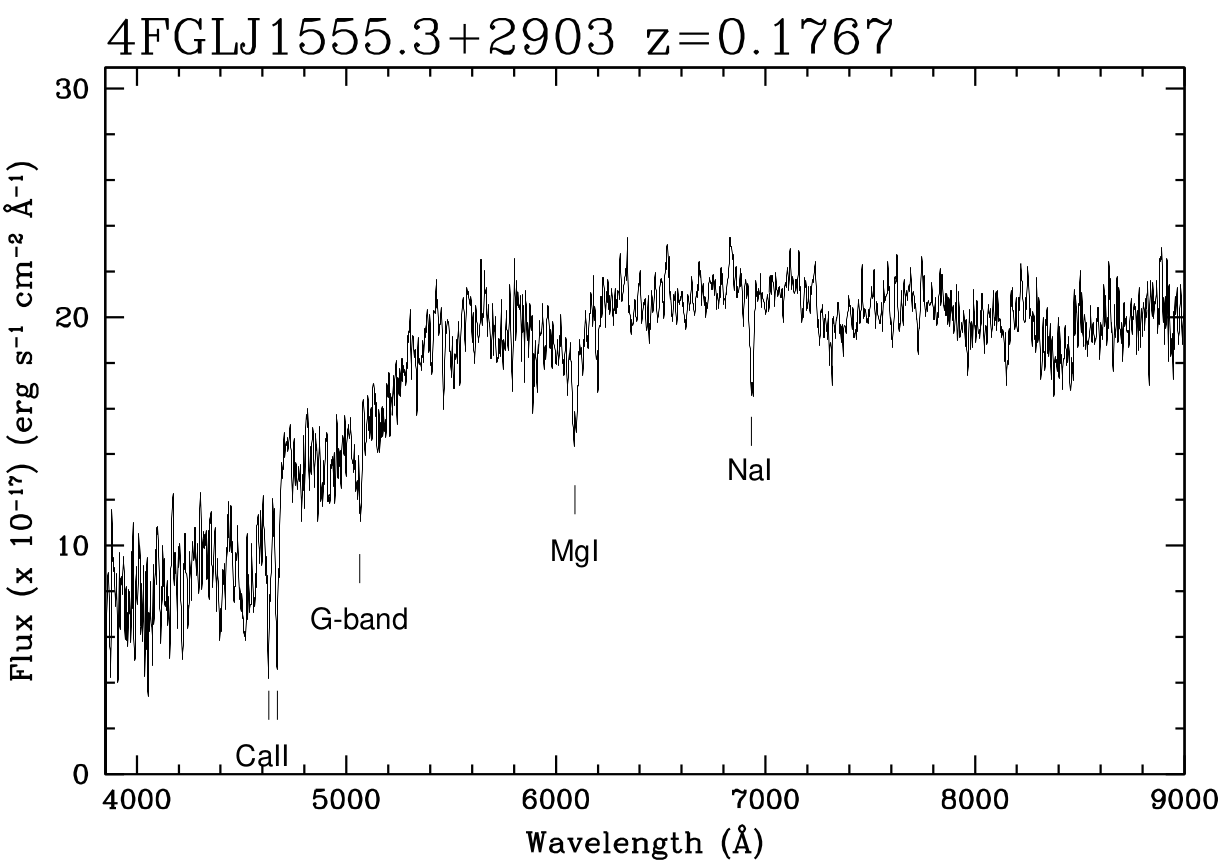}
\includegraphics[width=0.33\textwidth, angle=0]{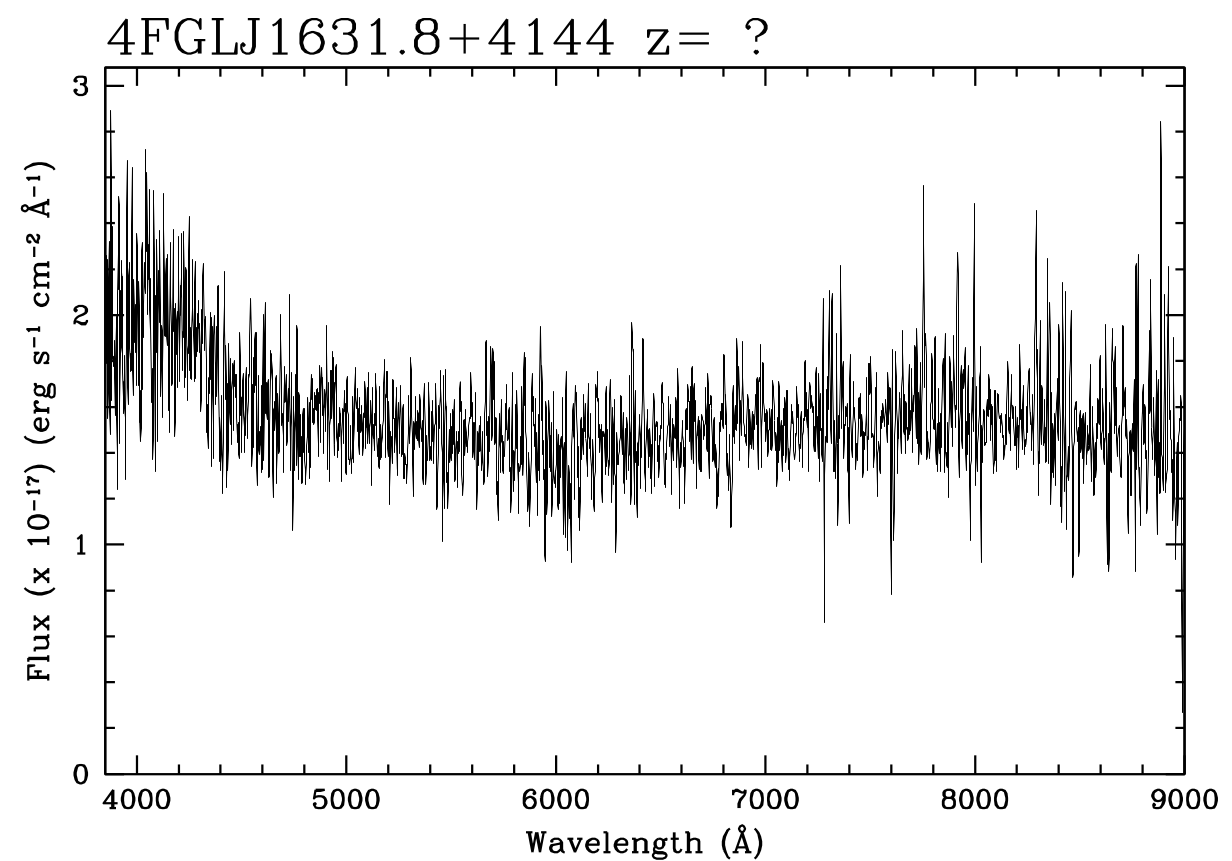}
\includegraphics[width=0.33\textwidth,height=0.17\textheight, angle=0]{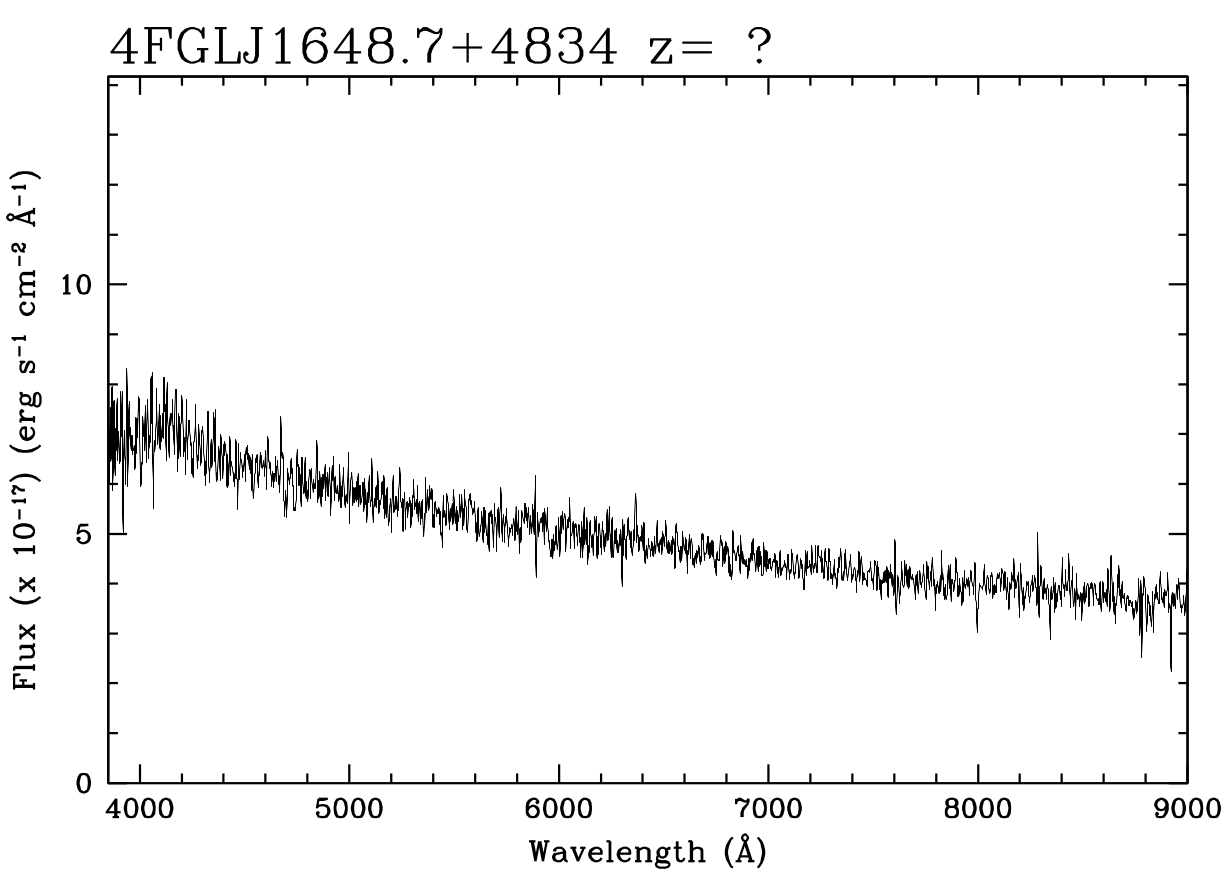}
\includegraphics[width=0.33\textwidth, angle=0]{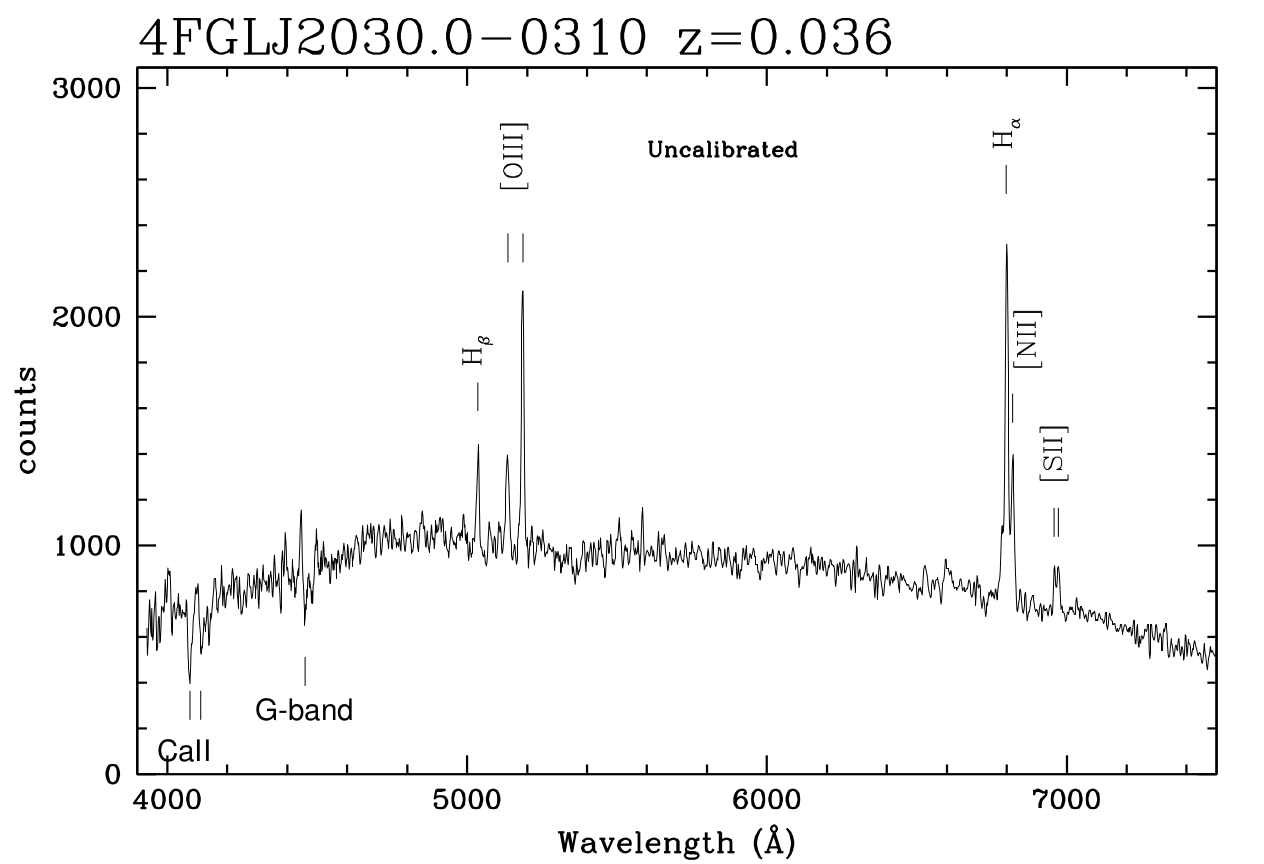}
\includegraphics[width=0.33\textwidth, angle=0]{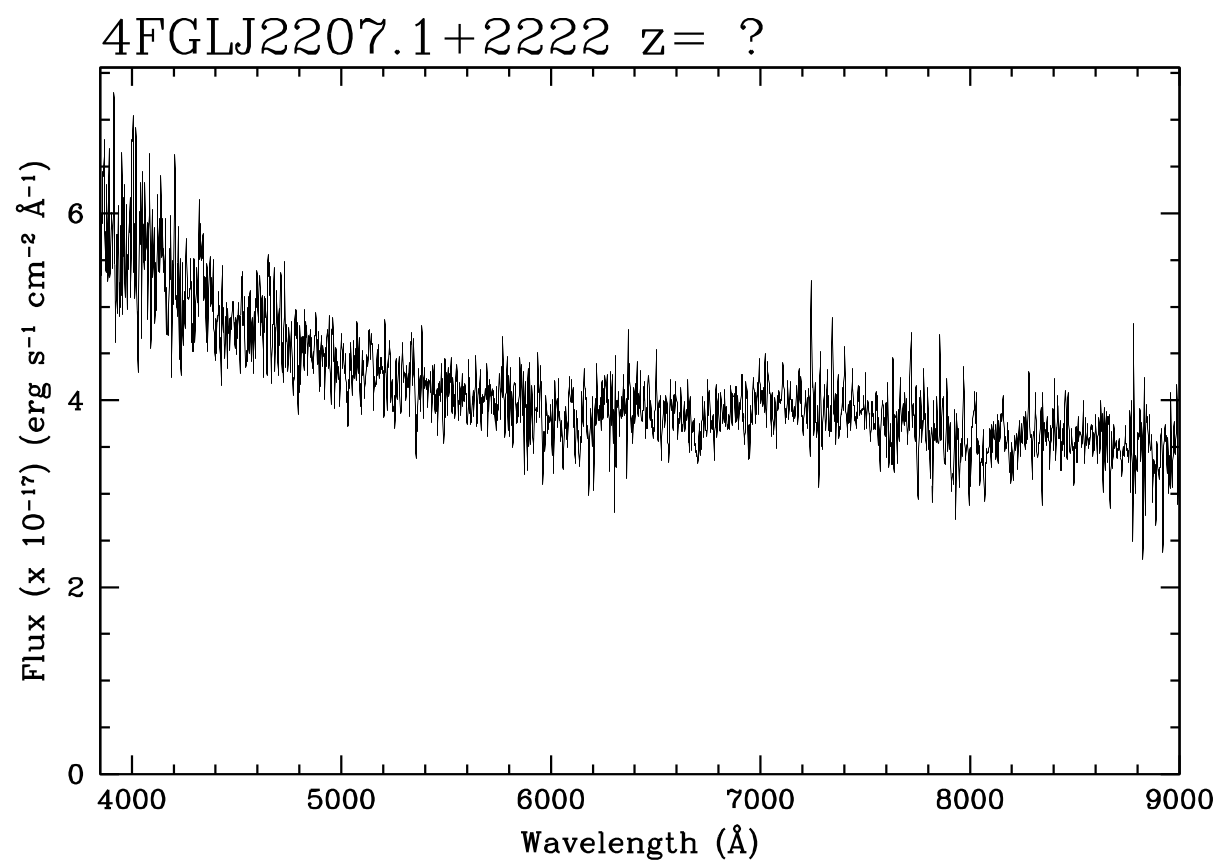}
\includegraphics[width=0.33\textwidth, angle=0]{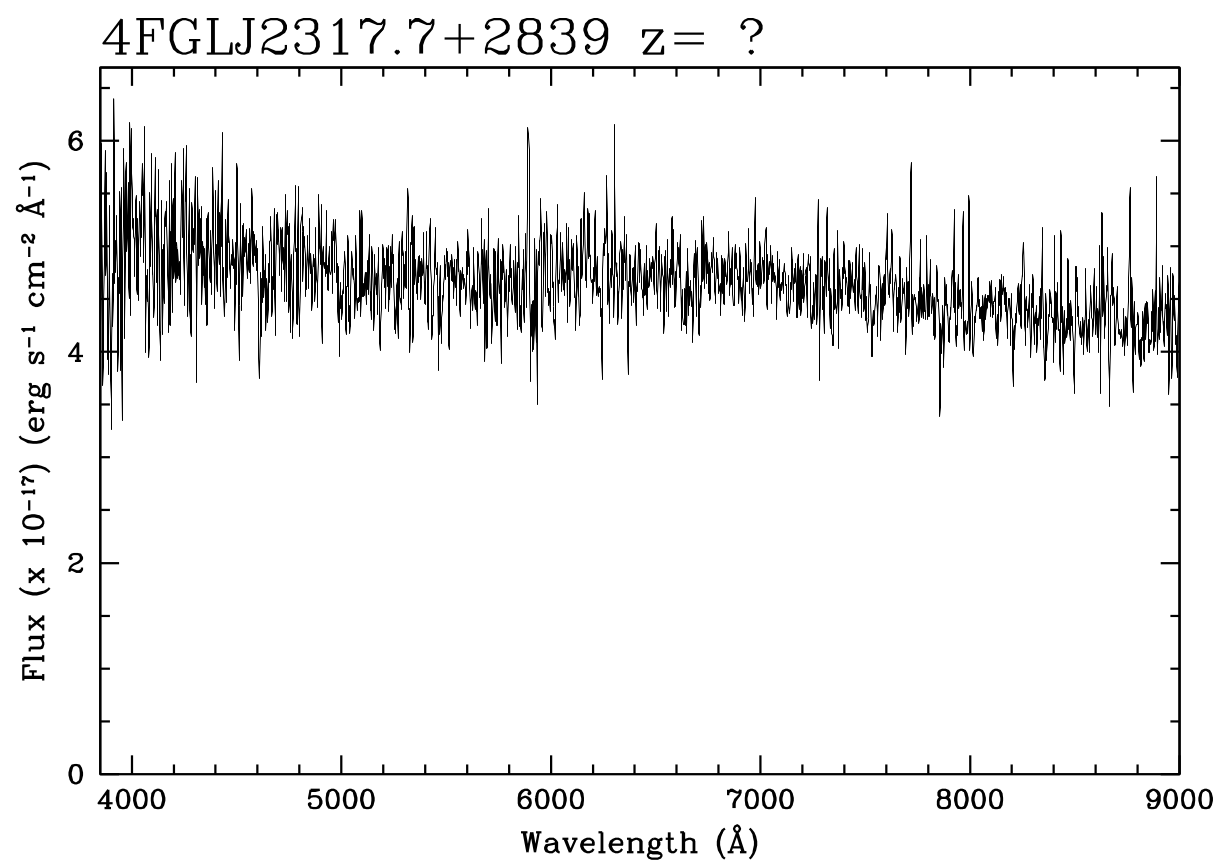}
\includegraphics[width=0.33\textwidth, angle=0]{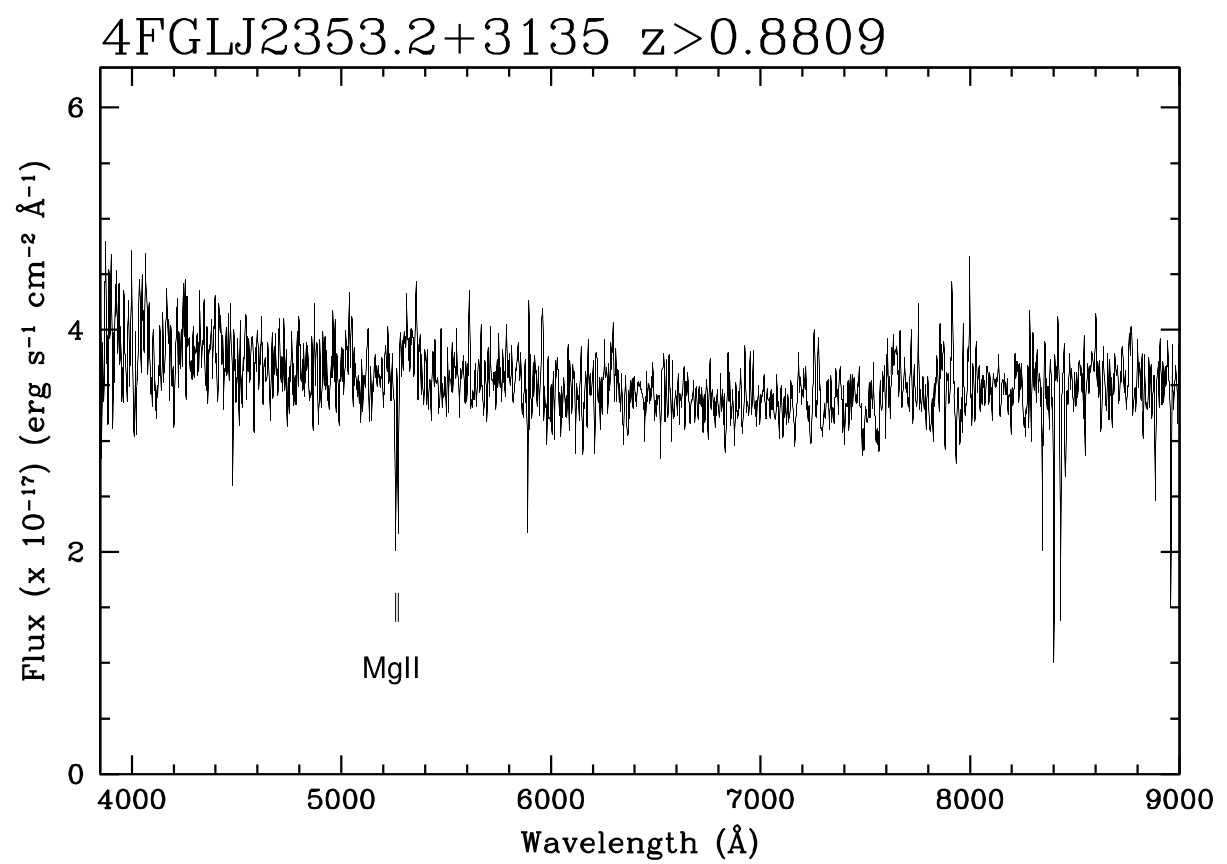}
\caption{Continued.} 
%\label{fig:spectrum}
\end{figure*}%[htbp]

\begin{figure}%[htbp]
%\vspace{2.2cm}
\hspace{-0.5cm}
\centering
\includegraphics[width=0.5\textwidth]{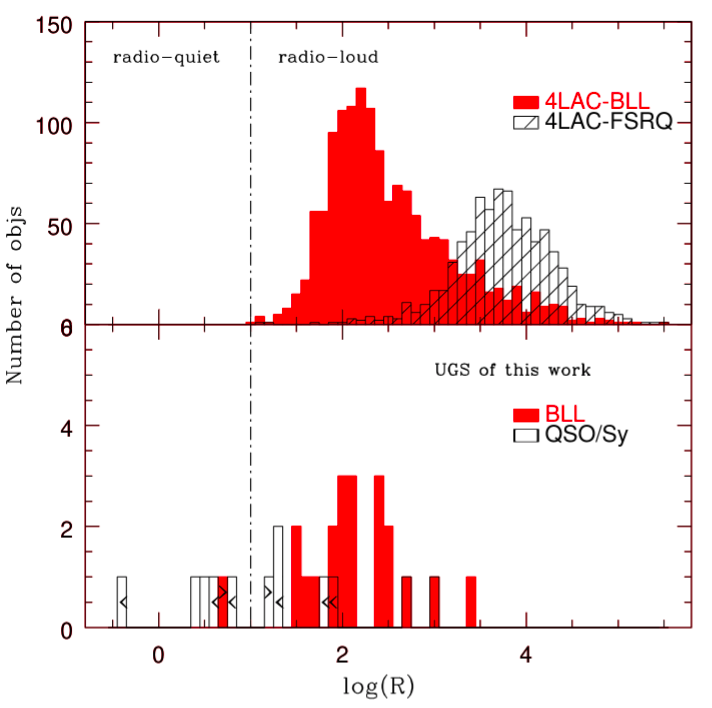}
\caption{\textit{Upper panel}: Distribution of the \textit{radio-loudness} parameter (log(R)) value for 1409 objects classified as BLL and 771 objects as FSRQ of the 4LAC catalog. \textit{Bottom panel}: Distribution of the \textit{radio-loudness} value for the counterparts of the 33 UGS of this work. The black dashed vertical line represent the \textit{radio-loudiness} parameter value (R = 10) that separates the radio-quiet from the radio-loud sources.} 
\label{fig:R_distr}
\end{figure}%[htbp]

\begin{table*}
\begin{center}
\caption{Multiwavelength fluxes and luminosities of the lower energy counterparts of the 33 UGSs} 
%(XXX objects)
%\resizebox{16cm }{!}{
%\resizebox{19cm }{!}{
\begin{tabular}{lccccccccl}
\hline 
4FGL Name & F$_{\gamma}$ &  $f^{\rm radio}_{\nu}$ & $F_{X}$ &  $f^{\rm opt}_{\nu}$ & $L_{\gamma}$ & $L_{\rm radio}$ & $L_{X}$ & M$_g$ (M$_r$) & $R$ \\% &  $\frac{F_{radio}}{F_{X}}$ & $R$ \\ 
\hline
BLL and BLG        & [$\times$ 10$^{-12}$]    &   & [$\times$ 10$^{-13}$]  & [$\times$ 10$^{-28}$] & [$\times$10$^{45}$] & [$\times$10$^{40}$] & [$\times$10$^{44}$] & \\% & $\times$ 10$^{4}$ & \\
 %\hline
%%%%BLL and BLG & &  &  & & & & & &\\
 \hline
4FGLJ0112.0+3442  & 1.4 & 41  & 2.5 & 6.3 & 0.9 & 74 & 1.5 & -22.3 (-22.7) & 450\\% & 97  & 594\\ % 6.0 \times 10^{2}$
4FGLJ0202.7+3133  & 1.0 & 15  & 3.8  & 12  & - & -     & - & - & 125 \\% & 13  & 115\\ %1.5 \times 10^{2}$\\
4FGLJ0251.1-1830  & 1.8 & 8.8 & 5.5  & 3.3 & - & -     & - & - & 265 \\% & 7.6 & 267 \\ %1.5 \times 10^{2}$\\
4FGLJ0259.0+0552  & 6.8 & 5.8 & 4.1 & 13  & - & -     & - & - & 45 \\% & 7.8 &  45\\ % 5.8 \times 10^{1}$\\
4FGLJ0838.5+4013  & 0.9 & 24  & 4.6 & 19  & 0.1 & 8.3 & 0.5 & -21.7 (-22.9) & 290\\% & 27  &  240\\ % 9.6 \times 10^{1}$\\
4FGLJ1016.1-4247  & 3.1 & 7.2 & 5.8  & 6.9 & - & -     & - & - & 100 \\% & 4.5 &  100\\ %9.9 \times 10^{1} $\\  
4FGLJ1039.2+3258  & 2.7 & 6.1 & 2.9 & 4.8 & - & -     & - & -  & 130\\% & 9.1 & 127\\ %1.3 \times 10^{2} $ \\
4FGLJ1049.8+2741  & 1.4 & 7.2 & 2.9 & 19  & 0.1 & 1.3 & 0.2 & -21.0 (-21.9) & 80\\% & 9.4 & 65\\ %3.8 \times 10^{1}$\\
4FGLJ1131.6+4657  & 1.1 & 92  & 4.5 & 36  & 0.1 & 1.2 & 0.2 & -21.3 (-22.3) & 920\\% & 95  & 657\\ %2.3 \times 10^{2}$\\
4FGLJ1146.0-0638  & 3.1 & 5.8 & 5.4  & 5.8 & 5.7 & 32 & 10 & -23.4 (-23.2) & 100 \\% & 3.2 & 175\\ %1.2 \times 10^{2}$\\
4FGLJ1410.7+7405  & 3.6 & 2.4 & 1.1 & 7.6 & - & - & - & - & 30\\% &  29  & 29\\ %2.8 \times 10^{1}$ \\
4FGLJ1544.9+3218  & 1.4 & 12  & 3.6  & 12 & - & - & - & - & 100  \\% & 11  & 92\\ %2.1 \times 10^{2}$\\
4FGLJ1554.2+2008  & 2.7 & 42  & 6.1 & 21 & 0.4 & 20 & 10 & -22.1 (-23.0) & 240\\%  & 3.2 & 262\\ %2.0 \times 10^{2}$\\
4FGLJ1555.3+2903  & 1.1 & 22  & 3.2 & 19  & 0.1 & 6.1 & 0.3 & -21.5 (-22.5) & 320\\%  & 25  & 220\\ %2.2 \times 10^{2}$\\
4FGLJ1631.8+4144  & 1.3 & 0.8 & 8.1 & 2.3 & - & - & - & - & 35\\% & 0.8 & 32\\ %2.6 \times 10^{1}$\\
4FGLJ1648.7+4834  & 1.2 & 2.5 & 6.0  & 6.3 & - & -     & - & - & 40 \\%  & 2.4 & 40\\ %7.5 \times 10^{1}$\\
4FGLJ2207.1+2222  & 1.7 & 6.5 & 2.6 & 2.5 & - & -     & - & - & 260 \\% & 9.7 & 260\\ %2.4 \times 10^{2}$\\
4FGLJ2240.3-5241  & 6.2 & 29  & 1.9 & 21  & - & -     & - & - & 135 \\% & 5.7 & 90\\ %1.2 \times 10^{2}$\\
4FGLJ2317.7+2839  & 2.8 & 4.5 & 0.8 & 5.2 & - & -     & - & - & 90 \\% & 31  & 87\\ %1.1 \times 10^{2}$\\
4FGLJ2323.1+2040  & 2.4 & 2.4 & 4.9 & 631.0 & 0.01 & 0.03 & 0.02 & -21.9 (-23.0)  & $>$5\\% & 2.4 & \red{0.4}\\
4FGLJ2353.2+3135  & 3.6 & 61  & 1.5 & 2.3 & - & -     & - & -  & 2650\\% & 230 & 4357 \\ %2.6 \times 10^{3}$\\
\hline
Objects with prominent lines & & &  & & & & & &  \\
\hline
4FGLJ0023.6-4209 & 1.1 & 3.3    & 12  & 209 & 0.01 & 0.02 & 0.08 & -21.3 (-21.9) & $>$15  \\% & 2.7    & 1.4\\ %3.6 \times 10^{-2}\\
4FGLJ0117.9+1430 & 2.9 & $<$4.2 & 4.9  & 12   & 0.1 & $<$0.03 & 0.2 & -20.2 (-20.6) & $<$65\\% & $<$3.5 & $<$38\\ %1.1 \times 10^{1} $\\ % c'è immagine LoTSS
4FGLJ0641.4+3349 & 1.3 & 1.2    & 28  & 53   & 0.1 & 0.3 & 2.3 & -22.4 (-23.1) & 2.5 \\%   & 0.3   & 2.5\\ %$2.6 \times 10^{1}$\\
4FGLJ0938.8+5155 & 1.2 & 0.6    & 0.7 & 2.8  & 0.8 & 0.06 & 0.5  & -21.5 (-21.7) & 25\\%   & 3.3    & 20\\ %1.7 \times 10^{1} $\\
4FGLJ1125.1+4811 & 0.8 & $<$0.6 & 0.4 & 2.8  & 14.5 & $<$1.5 & 7.0 & -25.1 (-25.2) & $<$20 \\%& $<$6.0 & $<$18.5\\ %$3.7 \times 10^{0}$\\ % c'è immagine LoTSS
4FGLJ1256.8+5329 & 2.7 & $<$1.6 & 1.7 & 2.1  & 14.7 & $<$1.3 & 9.1 & -23.5 (-23.8) & $<$75 \\% & $<$4.1 & $<$75\\ %$1.5 \times 10^{1}$\\ % c'è immagine LoTSS
4FGLJ1308.7+0347 & 2.0 & $<$0.2 & 7.5  & 48   & 3.3 & $<$1.0 & 13 & -25.6 (-25.5) & $<$0.5\\%  & $<$1.7 & $<$0.4\\ %$0.8 \times 10^{0}$\\
4FGLJ1346.5+5330 & 3.0 & 250    & 44  & 58   & 0.2 & 39 & 2.3 & -22.0 (-22.4) & 900\\%     & 38     & 820\\ %$8.2 \times 10^{2}$\\
4FGLJ1430.6+1543 & 0.9 & $<$1.0 & 14  & 40   & 0.1 & $<$0.2 & 1.1 & -22.1 (-22.6) & $<$4.5  \\% & $<$0.4 & $<$2.3\\ %$0.8 \times 10^{0}$\\
4FGLJ1535.9+3743 & 4.9 & 26     & 1.1 & 4.8  & 8.5 & 6.6 & 2.0 & -23.2 (-23.5) & 540\\%   & 130    & 542\\ %$5.7 \times 10^{2}$\\
4FGLJ1539.1+1008 & 2.2 & $<$1.0 & 3.1 & 17   & 0.4 & $<$0.5 & 0.5 & -22.0 (-22.3) & $<$6.0 \\% & $<$0.6 & $<$6.3\\ %$2.2 \times 10^{0}$\\
4FGLJ2030.0-0310 & 0.7 & 0.4 & 11  & 69   & 0.002 & $<$0.03 & 0.03 & -19.2 (-19.8) & 3.5 \\%  & $<$0.2 & $<$1.0\\ %$1.9 \times 10^{0}$\\ % c'è immagine RACS
\hline
\end{tabular}
%}
\label{tab:33_UGS_Flux}
\end{center}
\raggedright
\footnotesize{\textbf{Note.} Column 1: 4FGL Name of the target; Column 2: $\gamma$-ray flux (erg cm$^{-2}$ s$^{-1}$) in the 100 MeV to 100 GeV range; Column 3: Radio density flux (mJy); Column 4: X-ray flux (erg cm$^{-2}$ s$^{-1}$) in the 0.3-10 keV range; Column 5: Optical density flux (erg cm$^{-2}$ s$^{-1}$ Hz$^{-1}$) ; Column 6-7-8: $\gamma$-ray, radio and X-ray luminosity (erg s$^{-1}$); Column 9: g-band and r-band absolute magnitude of the target (derived from PANSTARRs images); Column 10: \textit{radio-loudness}  define as the ratio between radio flux density and optical flux density of the nuclear component. \\
%\textbf{Note.} mean of Mr for BLL and BLG with redshift $M_r = -22.36$}  
%\tablenotetext{}{
%\raggedright
} 
\end{table*}

\begin{table*}
\begin{center}
\caption{Statistics on Redshift, Flux and Luminosity; Comparison of UFO3 with UFO 1+2 and 4FGL-DR3 catalog } 
\begin{tabular}{lccc}
\hline 
 Sample & $< z >$   & F$_{\gamma}$   (erg cm$^{-2}$ s$^{-1}$)          & $L_{\gamma}$  (erg s$^{-1}$)       \\
%        &           &  &      \\   
        &           & $\times$ 10$^{-12}$      & $\times$ 10$^{45}$  \\
\hline
8 BLL/BLG  of this work  & 0.2  & 1.4 $\pm$ 0.4 & 0.1 $\pm$ 0.1  \\
\hline
12 QSO/Seyfert-like of this work & 0.2  & 1.3 $\pm$ 0.7 & 0.3 $\pm$ 0.3 \\
\hline
875 BLL of 4LAC  & 0.3   & 3.1 $\pm$ 1.7 & 1.2 $\pm$ 1.1  \\
\hline
792 FSRQ of 4LAC & 1.1   & 5.5 $\pm$ 3.4 & 45 $\pm$ 40 \\
%\hline
%UFO3$_{1count}$ (23 objs with z) & 0.2 $\pm$ 0.2 & 1.4 $\pm$ 0.6 & 0.4 $\pm$ 0.4  \\
\hline
%UFO1+2 (27 objs ) & 0.4 $\pm$ 0.2 & 3.0 $\pm$ 1.1 & 1.8 $\pm$ 1.3 \\
%\hline
24 BLL-UGS of Paiano et al 2017,2019 & 0.4  & 2.9 $\pm$ 1.0 & 2.0 $\pm$ 1.3  \\
\hline
3 QSO/Sy-like UGS of Paiano et al 2017,2019 & 0.3  & 4.7 $\pm$ 1.4 & 0.6 $\pm$ 0.1  \\
\hline
\end{tabular}
\label{tab:targets}
\end{center}
\raggedright
\footnotesize{\textbf{Note.} Column 1: Sample under investigation; Column 2: Median redshift; Column 3: Median energy flux from 4FGL catalog in the 100 MeV to 100 GeV range; Column 4: median luminosity. \\ 
The reported uncertainty indicates the rms of the distribution.
}  
%\tablenotetext{}{
%\raggedright
% } 
\end{table*}

%%%%%%%%%%%%%%%%%%%%%%%%%%%%%%%%%%%%%%%%%
%%%%%%%%%%%%%%%%%%%%%%%%%%%%%%%%%%%%%%%%%%%%%%%%%
%%%%%%%%%%%%%%%%%%%%%%%%%%%%%%%%%%%%%%%%%%%%%%%%%%%%%%%%%%%%%%%

\section{Notes on Individual Sources}
\label{sec:note_ind_sources}
\begin{itemize}
     
\item[] \noindent 4FGLJ0023.6-4209 - DESJ002303.74-420508.4:
In the Swift/XRT image we find the source XRTJ002303.5-420509.6 as the likely X-ray counterpart for this UGS.
It coincides with the optical source DESJ002303.74-420508.4 (g=15.6), and the radio source J002303.61-420509.57, detected from the analysis of radio RACS data covering the UGS sky region (see Fig. \ref{fig:Oskymap}). The radio source appears point-like and it has a density flux $F$ = $3.3$ mJy that corresponds to a radio-loudness R$>$15 (see Tables \ref{tab:decomp} and \ref{tab:33_UGS_Flux}). 
%Analysing RACS data, a radio emission is detected 
In the uncalibrated 6dF Galaxy Redshift survey optical spectrum, we are able to detect prominent and narrow emission lines attributed to H$_{\beta}$ 4861, [O III] 4959,5007 , H$_{\alpha}$ 6563, [N II] 6583 and [S II] 6717,6731 at $z$ = 0.053.
No broad lines are found. 
The emission line properties of this object are typical of Seyfert 2 galaxies
and from a inspection of the optical image (see Fig. \ref{fig:Oskymap}) the host galaxy appears as spiral galaxy.
Given the remarkably proximity ($z<0.053$), the $\gamma$-ray luminosity of 4FGLJ0023.6-4209 is $\sim$10$^{43}$ erg/s, one of the weakest of our sample. \\

\item[] \noindent 4FGLJ0112.0+3442 - SDSSJ011124.86+344154.6:
From the \textit{Swift/XRT} analysis, we reveal the X-ray source XRTJ011124.8+344154.1 within the \textit{Fermi} error box. 
The sources is positionally coincident with the radio source VLASS1QLCIRJ011124.83+344154.5 and the optical source SDSSJ011124.86+344154.6. 
The optical spectrum, available in the SDSS archive, shows [O II] and [O III] emission lines and the Ca II doublet absorption lines consistent with redshift $z$=$0.3397$. 
Given the power-law shape of the spectrum, the source can be classified as a BLL. 
It is worth to note that this UGS is one of three sources listed in the 4FGL that lies within the event radius of the neutrino event IceCube-230511A\footnote{https://gcn.nasa.gov/circulars/33773?page=5}.\\

\item[] \noindent 4FGLJ0117.9+1430 - SDSSJ011804.83+143158.6:
Within the \textit{Fermi} error box of this UGS, we detect the X-ray source XRTJ011804.7+143159.5 coincident with the optical source SDSSJ011804.83+143158.6 (g = 18.7). 
No radio counterpart is found in the NVSS and VLASS catalogue.  
From the RACS data we can estimate the radio flux upper limit of 4.2 mJy/beam (within 5$\sigma$).
The SDSS spectrum clearly shows prominent emission lines (H$_{\delta}$ 4102, H$_{\gamma}$ 4340, H$_{\beta}$ 4861 and [O III] 4959,5007) at z = 0.129.
We note that the Balmer lines show broad components and in particular, based on the H$_{\beta}$ 4861 width (FWHM$\sim$500 km/s) and on the line ratio ([OIII] 5007/H$_{\beta}$)$\sim$1 \citep[see e.g.][]{komossa2008}, we can classify the source as a NLSy1, according to that reported by \citet{rakshit2017}.\\

\item[] \noindent 4FGLJ0202.7+3133 - SDSSJ020242.06+313210.9:
We propose the source XRTJ020242.13+313211.4, found in the \textit{Swift}/XRT skymap, as the X-ray counterpart for this UGS, spatially coincident with the optical object SDSSJ020242.06+313210.9 (g = 18.7) and the radio source VLASS1QLCIRJ020242.03+313211.0.
The SDSS spectrum appears featureless and exhibits the characteristic power-law continuum of the BLL. \\

\item[] \noindent 4FGLJ0641.4+3349 - PANJ064111.22+334459.7:
From the \textit{Swift}/XRT imaging analysis, in the 4FGL error box of this $\gamma$-ray emitter, we find the X-ray source XRTJ064111.24+334502.0, coincident with the radio source VLASS1QLCIRJ064111.20+334459.6 (1.2 mJy) and the optical source PANJ064111.22+334459.7 (g=17.1).
The spectrum of the optical source, available in \citet{monroe2016}, shows prominent narrow emission lines due [OIII] 4959,5007 and broad  H$_{\gamma}$ 4340, H$_{\beta}$ 4861 and H$_{\alpha}$ 6564 emission lines at redshift $z$=$0.1657$. 
On the basis of the absolute magnitude, the radio-loudness $R\sim$2.5 and the emission lines properties, this object can be classified as a low redshift radio quiet QSO.\\

\item[] \noindent 4FGLJ0838.5+4013 - SDSSJ083903.08+401545.6:
Inside the \textit{Fermi} error box of this UGS we find the X-ray source XRTJ083902.9+401546.9 coincident with the radio source VLASS1QLCIRJ083903.07+401545.6 \citep{joffre2022} and the optical source SDSSJ083903.08+401545.6.  
\citet{kaur2023} and \citet{joffre2022} through a machine learning analysis proposed the X-ray object as a BLL candidate.
The SDSS spectrum is dominated by a galactic component and we can identify several absorption lines (Ca II, G-band, Mg I, and Na II) at $z$ = $0.1945$ due to old stellar population of the host galaxy. The source can be classify as a galaxy-dominated BLL in agreement with what reported in the BZCAT catalog.\\ % where, however, the association with the $\gamma$-ray emission is absent.\\

\item[] \noindent 4FGLJ0938.8+5155 - SDSSJ093834.72+515452.3:
Through the inspection of \textit{Swift}/XRT data, we find the X-ray source XRTJ093834.5+515454.7 inside the 3$\sigma$ \textit{Fermi} error ellipse. 
Analysing radio data of the LoTSS survey we found a radio detection J093834.68+515451.8 (0.6mJy) coincident with the X-ray source.
A SDSS spectrum available for the optical counterpart SDSSJ093834.72+515452.3 (g=20.3) shows prominent and narrow emission lines attributed to [O II], H$_{\gamma}$, H$_{\beta}$, [O III], H$_{\alpha}$ and [N II] consistent with $z$=$0.4168$.
A broad component is present for the H$_{\beta}$ and H$_{\alpha}$.
Based on this information, the target can be classified as a radio loud QSO.\\

\item[] \noindent 4FGLJ1039.2+3258 - SDSSJ103852.17+325651.6:
We propose the X-ray source XRTJ103852.1+325651.9 found inside the Fermi error box of this $\gamma$-ray emitter as the likely lower energy counterpart.
It is coincident with the radio source VLASS1QLCIRJ103852.17+325651.9 and the optical source SDSSJ103852.17+325651.6. 
The modest quality SDSS spectrum, reported also in \citet{demenezes2019} and associated to the IR source WISEJ103852.20+325651.7, appears featureless typical of BLL, although a possible weak signature of CaII break at $\sim$5200 can be recognized yelding a tentative redshift of 0.32.\\

\item[] \noindent 4FGLJ1049.8+2741 - SDSSJ104938.79+274213.0:
Inside the 4FGL sky region of this UGS we find the X-ray source XRTJ104938.7+274212.0, that spatially coincides with the the radio source VLASS1QLCIRJ104938.81+274213.1 and the optical source SDSSJ104938.79+274213.0 (g=18.2).
The optical spectrum provided by the SDSS survey is dominated by the component of the elliptical host galaxy (N/H=0.25 see Table \ref{tab:decomp}) and a number of relevant absorption features are present: the CaII doublet, G-band, MgI and NaI at $z$=$0.144$.\\

\item[] \noindent 4FGLJ1125.1+4811 - SDSSJ112526.27+480922.0:
We find the X-ray source XRTJ112526.0+480922.8 inside the $\gamma$-ray  $3\sigma$ error ellipse of this UGS. 
It is spatially coincident with the optical source SDSSJ112526.27+480922.0 (g=20.3).
From the analysis of LOTSS radio data, no radio counterpart is found, in agreement with \citet{gurkan2019}.
%where they claimed that this is a radio quiet source. 
In the optical spectrum, obtained by the SDSS survey, we can detect broad emission lines attributed to CIV, C III and Mg II consistent with $z$=$1.649$. 
On the basis of this spectrum and of the MWL information, we can classify this source as a radio-quiet QSO.
Within $4\sigma$ Fermi error box of this $\gamma$-ray emitter, another X-ray source is present, XRTJ112432.5+480741.0, coincident with the radio source VLASS1QLCIRJ112432.64+480739.9 and the optical source SDSSJ112432.65+480740.7 (g=22.7). An optical spectrum is available in the SDSS archive for this source. The power-law trend of the spectrum, absent of emission or absorption lines, indicates that the object is a BLL.\\

\item[] \noindent 4FGLJ1131.6+4657 - SDSSJ113142.27+470008.6:
The \textit{Swift}/XRT analysis reveals the source XRTJ113142.3+470009.2 as X-ray counterpart of this UGS.
The radio source VLASS1QLCIRJ113142.36+470009.4 and the optical source SDSSJ113142.27+470008.6 are coincident with the X-ray emission. 
The SDSS survey provides the optical spectrum with absorption lines (Ca II, G-band, MgI, Ca+Fe and NaI) due to the old stellar population of the elliptical host galaxy. The spectral lines are consistent with a redshift $z=0.1255$ and the source can be classify as a galaxy-dominated BLL. \\

\item[] \noindent 4FGLJ1256.8+5329  - SDSSJ125630.43+533204.3:
The analysis of \textit{Swift}/XRT data reveals only the X-ray source XRTJ125630.5+533202.2 within the \textit{Fermi} error box of this $\gamma$-ray emitter. 
It is coincident with the optical source SDSSJ125630.43+533204.3. The optical spectrum exhibits broad emission lines due to CIII] and MgII at $z$=$0.996$. Also the narrow (EW=11.1) emission line due to [O II] at 7439 $\AA$ is present. We can classify the source as a QSO.
From radio catalogs and the analysis of LoTSS data, no radio emission is coincident with the X-ray/optical counterpart. \\

\item[] \noindent 4FGLJ1308.7+0347 - SDSSJ130832.10+034403.9:
For this UGS, we find the X-ray source XRTJ130832.2+034405.3 slightly out of the \textit{Fermi} error ellipse reported in the 4FGL-DR3, but inside the 3$\sigma$ error box (see Sec. \ref{sec:sample}). From the analysis of the SDSS optical spectrum we can detect prominent broad emission lines attributed to Mg II, H$_{\gamma}$, H$_{\beta}$ and the doublet of narrow lines due to [O III]. The redshift of the source is $z$=$0.6193$.
From our analysis of RACS radio data and from that reported by \citet{rusinek2021} using LoTSS data, no radio object is detected coincident with the optical counterpart that can be classified as a radio quiet QSO.\\

\item[] \noindent 4FGLJ1346.5+5330 - SDSSJ134545.36+533252.3:
The X-ray counterpart for this UGS, found by our Swift/XRT analysis, is XRTJ134545.1+533252.4.
It coincides with the radio source VLASS1QLCIRJ134545.34+533252.1 and in the optical with SDSSJ134545.36+533252.3. 
The SDSS optical spectrum is a Seyfert 1-like \citep[as also proposed by ][]{wang2009} and exhibits several emission lines: narrow emission lines due to [OII], [OIII] and [OI] and the emission lines attributed to  H$_{\delta}$, H$_{\gamma}$, H$_{\beta}$ and H$_{\alpha}$ that show a very prominent broad emission component.
These lines set the source at $z$=$0.1359$.
From the radio analysis of LOFAR Two Metre Sky Survey data, \citet{pajdosz2022} show that the source exhibits a structure composed of a compact radio core, two-sided S-shaped jets and the radio luminosity characteristic of FR I radio galaxies. \\ %core-dominated 

\item[] \noindent 4FGLJ1430.6+1543 - SDSSJ143058.03+154555.6:
In the Swift/XRT skymap and within the Fermi error box of this $\gamma$-ray emitter we find the X-ray source XRTJ143057.9+154555.0 coincident with the optical source SDSSJ143058.03+154555.6.
The optical spectrum, available in the SDSS archive, shows prominent and broad emission lines such as H$_{\gamma}$, H$_{\beta}$, H$_{\alpha}$ together with the narrow lines due to the [O III] doublet at $z$=$0.1633$ and consistent with a Seyfert-1 spectrum.
%, and absorption lines such as the Ca II doublet, Mg I and Ca + Fe, .
%NVSS data
In the radio band, no counterpart is detected \citep[][]{coziol2017}. \\

\item[] \noindent 4FGLJ1535.9+3743 - SDSSJ153550.54+374055.6:
Analysing the \textit{Swift}/XRT data we find the X-ray source XRTJ153550.56+374056.8 inside the UGS \textit{Fermi} error box.
It is coincident with the radio source VLASS1QLCIRJ153550.56+374055.5 (26 mJy) and the optical source SDSSJ153550.54+374055.6 for which the spectrum is available. The spectrum shows a prominent emission line at 4551$\AA$ due to MgII, a weak and broad emission line attributed to H$_{\beta}$ and the faint and narrow line of [OIII]. The redshift is $z$=$0.6255$.
We note that also an intervening MgII absorption line is detected at $\sim$4167 ($z$=$0.4885$).
On the basis of the radio-loudness parameter (R$\sim$550) and the absolute optical magnitude (M$_g$ = -23.2) the source appears to be a radio loud quasar. \\

\item[] \noindent 4FGLJ1539.1+1008 - SDSSJ153848.47+101843.2:
From the \textit{Swift}/XRT image,  we find that the source XRTJ153848.5+101841.7 is a possible UGS X-ray counterpart. It is coincident with the optical source SDSSJ153848.47+101843.2. 
No radio emission is present \citep[][]{coziol2017}.
The SDSS spectrum displays narrow lines due to [OII], [OIII] and [SII], H$_{\beta}$ and H$_{\alpha}$ emission lines with broad and narrow components. This can be classified as a Seyfert-1-like spectrum and sets the source at $z$=$0.2345$. \\

\item[] \noindent 4FGLJ1544.9+3218 - SDSSJ154433.19+322149.1:
%This UGS is also reported in the 3HSP catalog \autocite{3HSP_catalog}. 
From the analysis of \textit{Swift}/XRT data, we propose the X-ray source XRTJ154433.1+322148.5, coincident with the radio source VLASS1QLCIRJ154433.20+322149.1  and the optical source SDSSJ154433.19+322149.1, as lower energy counterpart of the $\gamma$-ray emitter. 
In the SDSS archive the optical spectrum is available showing the power-law shape and appearing featureless. We therefore classify the source as a BLL with unknown redshift. \\

\item[] \noindent 4FGLJ1554.2+2008 - SDSSJ155424.12+201125.4:
%This UGS emitter is also reported in the 3FHL and 3HSP \autocite{3HSP_catalog} catalogs. 
The optical source SDSSJ155424.12+201125.4 and the radio source VLASS1QLCIRJ155424.15+201125.5 are spatially coincident with the X-ray object XRTJ155424.1+201125.3 found inside the positional error box of this UGS and proposed as the likely counterpart.
The optical spectrum of this source is provided by the SDSS survey and it is dominated by a galactic component with the presence of moderate non-thermal emission. 
Clear absorption features of the stellar population are detected, in particular Ca II 3934, 3968, G-band 4305, Mg I 5157, and NaI 5893 at  $z$=$0.2225$. 
We classified the source as a BLG. It is also notably that the source is one of the possible counterparts of the neutrino event IceCube-110521A \citep{Giommi_2020,Padovani_2022}.\\

\item[] \noindent 4FGLJ1555.3+2903 - SDSSJ155512.91+290329.9:
By the Swift/XRT data analysis, we found the X-ray source XRTJ155513.01+290328.0 inside the 4FGL-DR3 error box of this UGS, that coincides with the radio emitter VLASS1QLCIRJ155512.89+290330.0 and the optical source SDSSJ155512.91+290329.9. 
The SDSS spectrum is available and absorption lines due to  Ca II doublet, G-band, Mg I and NaII are clearly detected at $z$=$0.1747$, allowing us to classify the source as a galaxy-dominated BLL. \\

\item[] \noindent 4FGLJ1631.8+4144 - SDSSJ163146.72+414632.8:
XRTJ163146.8+414631.8 is the only X-ray source detected inside the 4FGL-DR3 position error box and it is coincident with the radio source VLASS1QLCIRJ163146.74+414632.7 and the optical object SDSSJ163146.72+414632.8.
In the SDSS spectrum the continuum is very flat and no emission lines are detected, allowing us to classify the source as a BLL. A possible CaII absorption doublet is present at $\sim$6800 $\AA$ yielding a tentative redshift of $z$=$0.721$. \\

\item[] \noindent 4FGLJ1648.7+4834 - SDSSJ164900.34+483411.8:
Through the Swift/XRT image analysis, we find the X-ray source XRTJ163146.8+414631.8 inside the positional error box of this UGS.
We propose the spatially coincident objects VLASS1QLCIRJ163146.74+414632.7  and
SDSSJ164900.34+483411.8 (g = 19.4) as radio and optical counterparts. 
Our optical spectrum exhibits a featureless power-law continuum typical of BLL. \\

\item[] \noindent 4FGLJ2030.0-0310 - PANJ203014.27-030722.56:
The \textit{Swift}/XRT image reveals the X-ray object XRTJ203014.3-030722.8 inside the 4FGL error box that is spatially coincident with the optical source PANJ203014.27-030722.56. 
RACS radio image reveals an hint of radio emission at the position of the X-ray source. 
As such, we observed 4FGLJ2030.0-0310 sky region with the Australia Telescope Compact Array (ATCA)\footnote{The observations were carried out over two epochs, (2023-07-04 17:25:20 -- 20:53:10 UT and 2023-07-05 13:07:30 -- 20:49:30 UT). For both observations we used PKS 1934$-$638 for primary flux and bandpass calibration, and PKS 2044$-$027 for secondary gain calibration. Data were recorded at a central frequency of 2.1\,GHz with 2\,GHz of bandwidth composed of 2048 1-MHz channels. Raw data were then edited for radio frequency interference (RFI), calibrated, and imaged following standard procedures (details in \url{https://casaguides.nrao.edu/index.php/ATCA_Tutorials}) in the Common Astronomy Software Applications for radio astronomy (CASA, version 5.1.2; \citealt{casa2022}) }. % to better constrain the radio source. 
%(see Section 2.1.2 for details)
We detected the source with a SNR $>$ 13 $\sigma$ and we measured a radio flux density of 0.40 $\pm$ 0.03\,mJy at 2.1\,GHz, corresponding to a radio-loudness R=3.5 (see Table \ref{tab:decomp} and \ref{tab:33_UGS_Flux}).
The uncalibrated 6dF Galaxy Redshift survey optical spectrum shows prominent and narrow emission lines attributed to H$_{\beta}$ 4861, [O III] 4959,5007 , H$_{\alpha}$ 6563, [N II] 6583 and [S II] 6717,6731 at $z$=$ 0.036$.
No broad lines are found, therefore the spectrum can be classified as Type 2.  
Considering the very low redshift, the absolute optical magnitude (M$_g$ = -19.2) and the host galaxy properies from the analysis of the optical image (see Tab.\ref{tab:decomp}), the source should be classified as a Seyfert-2 galaxy hosted by a dwarf galaxy. 
It is important to mention that, given its remarkably close proximity ($z<0.036$), this object has a $\gamma$-ray luminosity of approximately 2$\times$10$^{42}$ erg/s, which is one of the lowest in our sample.
\\

\item[] \noindent 4FGLJ2207.1+2222 - SDSSJ220704.10+222231.4:
The analysis of Swift-XRT imaging data reveals one X-ray object XRTJ220704.1+222231.8 in the 4FGL-DR3 error box that is spatially coincident with the optical source SDSSJ220704.10+222231.4 (g=20.4) and the radio source VLASS1QLCIRJ220704.09+222231.5. 
The SDSS optical spectrum is characterized by a power-law emission and appears featureless typical of BLL.
It is worth noting that this UGS is one of two sources listed in the 4FGL that lie within the event radius of the neutrino event IceCube-221210A\footnote{https://gcn.nasa.gov/circulars/33040?page=12}. \\
%It is worth to note that this UGS is one of the possible counterparts of the neutrino event IceCube-221210A\footnote{https://gcn.nasa.gov/circulars/33040?page=12}.

\item[] \noindent 4FGLJ2317.7+2839 - SDSSJ231740.00+283955.7:
Inside the \textit{Fermi} error box of this UGS, we detected the X-ray source XRTJ231740.1+283955.4, coincident with the radio source VLASS1QLCIRJ231740.21+283955.8 and the optical source SDSSJ231740.00+283955.7. 
The optical spectrum, available in the SDSS archive, appears featureless with a power-law continuum. This source is therefore a BLL with redshift unknown. \\

\item[] \noindent 4FGLJ2353.2+3135 - SDSSJ235319.54+313616.7:
Through the \textit{Swift}/XRT data analysis, we find that the X-ray source XRTJ235319.1+313613.4 is inside the 4FGL-DR3 position error box. 
The radio object VLASS1QLCIR J235319.50+313616.8 and the optical source SDSSJ235319.54+313616.7 are coincident with the X-ray emitter. 
The optical spectrum, available in the SDSS archive, is flat and characterized by a power-law continuum typical of BLL. 
It is clearly possible to detect an intervening absorption system due to MgII allowing us to set a redshift lower limit of $z>0.8809$. \\

\end{itemize}

%%%%%%%%%%%%%%%%%%%%%%%%%%%%%%%%%%%%%%%%

\section*{Acknowledgments}

This work has been partially supported by the ASI-INAF program I/004/11/5 and made use of data supplied by the UK Swift Science Data Centre at the University of Leicester. This research uses services or data provided by the Astro Data Lab at NSF's National Optical-Infrared Astronomy Research Laboratory. NOIRLab is operated by the Association of Universities for Research in Astronomy (AURA), Inc. under a cooperative agreement with the National Science Foundation.
We acknowledge the use of public data from the Swift data archive and we thank the Swift Observatory for scheduling and plan our proposed fill-in targets.
Moreover we acknowledge the use of data and software facilities from the SSDC, managed by the Italian Space Agency, and the United Nations ‘Open Uni verse’ initiative. 
 Funding for SDSS-III has been provided by the Alfred P. Sloan Foundation, the Participating Institutions, the National Science Foundation, and the U.S. Department of Energy Office of Science. The SDSS-III web site is http://www.sdss3.org/.
SDSS-III is managed by the Astrophysical Research Consortium for the Participating Institutions of the SDSS-III Collaboration including the University of Arizona, the Brazilian Participation Group, Brookhaven National Laboratory, Carnegie Mellon University, University of Florida, the French Participation Group, the German Participation Group, Harvard University, the Instituto de Astrofisica de Canarias, the Michigan State/Notre Dame/JINA Participation Group, Johns Hopkins University, Lawrence Berkeley National Laboratory, Max Planck Institute for Astrophysics, Max Planck Institute for Extraterrestrial Physics, New Mexico State University, New York University, Ohio State University, Pennsylvania State University, University of Portsmouth, Princeton University, the Spanish Participation Group, University of Tokyo, University of Utah, Vanderbilt University, University of Virginia, University of Washington, and Yale University.
We acknowledge the Pan-STARRS Survey.
 The Pan-STARRS1 Surveys (PS1) and the PS1 public science archive have been made possible through contributions by the Institute for Astronomy, the University of Hawaii, the Pan-STARRS Project Office, the Max-Planck Society and its participating institutes, the Max Planck Institute for Astronomy, Heidelberg and the Max Planck Institute for Extraterrestrial Physics, Garching, The Johns Hopkins University, Durham University, the Uni versity of Edinburgh, the Queen’s University Belfast, the Harvard-Smithsonian Center for Astrophysics, the Las Cumbres Observatory Global Telescope Network Incorporated, the National Central University of Taiwan, the Space Telescope Science Institute, the National Aeronautics and Space Administration under Grant No. NNX08AR22G issued through the Planetary Science Division of the NASA Science Mission Directorate, the National Science Foundation Grant No. AST-1238877, the University of Maryland, Eotvos Lorand University (ELTE), the Los Alamos National Laboratory, and the Gordon and Betty Moore Foundation.

\section*{Data availability}
 All data are incorporated into the article and its online supplementary material. \\
%The flux-calibrated and de-reddened spectra are available in our online data base ZBLLAC\footnote{\url{http://web.oapd.inaf.it/zbllac/}}.
 
%%%%%%%%%%%%%%%%%%%%%%%%%%%%%%%%%%%%%%%%%%%%%%%%%%%%%%%%%%%%%%%%%%%%%%%%%%%%%%%%%%%%%%%%%%%%%%%%%%%%%%%%%%%%%%%%%%%%%%%%%%%%%%%%%%%%%%%%%%%%%%%%%%%%%%%%%%%%%%%%%%%%%%%%%%%%%%%%%%%%%%%%%%%%%%%%%%%%%%%%%%%%%%%%%%%%%%%%%%%%%%%%%%%%%%%%%%%%%%%%%%%%%%%%%%%%%%%%%%%%%%%%%%%%%%%%%%%%%%%%%%%%%%%%%%%%%

%\newpage
\appendix
%\newpage
\section{X-RAY SKYMAPS}

In the appendix the X-ray skymaps for the 33 UGS analysed in this paper are shown. The yellow and cyan ellipses are respectively the $2\sigma$ and $3\sigma$ \textit{Fermi} $\gamma$-ray error regions. X-ray detection, found through \textit{Swift}/XRT analysis, are reported in green.
\newpage
\newpage
%%%%%%%%%%%%%   SKYMAP X
%\setcounter{figure}{0}
%\begin{figure*}%[htbp]
%\center
%   \includegraphics[width=8.7truecm]{Skymap_X/4FGLJ0023.6-4209.png}
%   \includegraphics[width=8.7truecm]{Skymap_X/4FGLJ0112.0+3442.png}
%  \caption{X-ray skymaps for 12 UGS. The yellow and cyan ellipses are respectively the $2\sigma$ and $3\sigma$ \textit{Fermi} $\gamma$-ray error regions. X-ray detection, found through \textit{Swift}/XRT analysis, are reported in green.} 
%\label{fig:Xskymap}
%\end{figure*}%[htbp]

\setcounter{figure}{0}
\begin{figure*}%[htbp]
\center
   \includegraphics[width=5.5truecm]{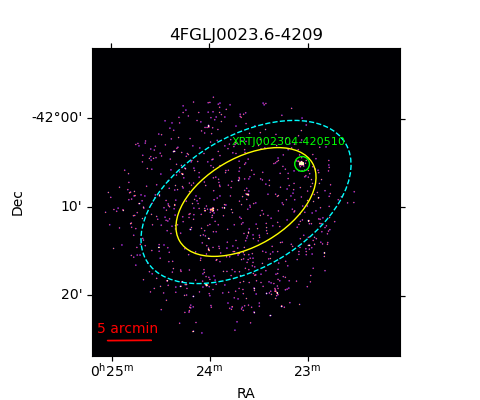}
   \includegraphics[width=5.5truecm]{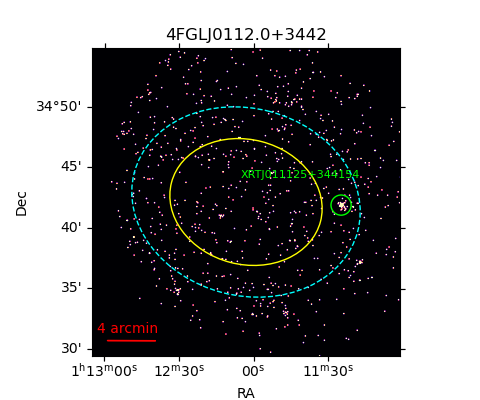}
   \includegraphics[width=5.5truecm]{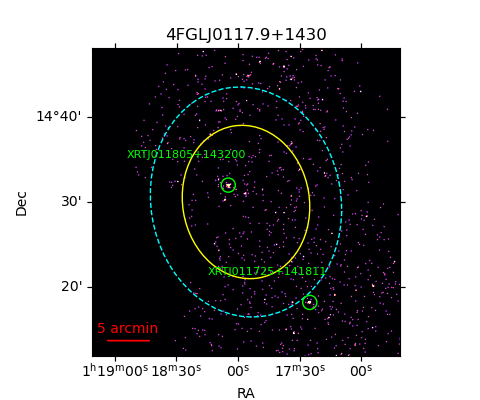}
   \includegraphics[width=5.5truecm]{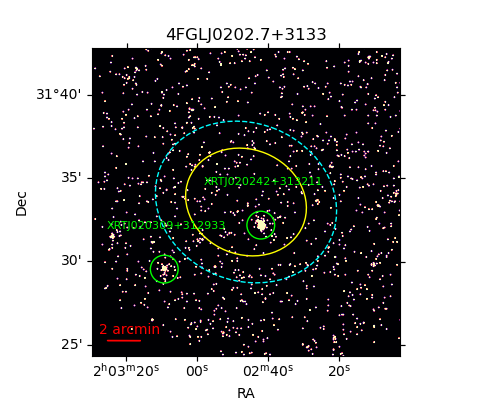}
   \includegraphics[width=5.5truecm]{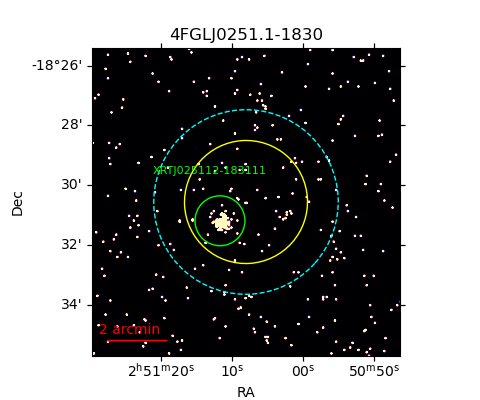}
   \includegraphics[width=5.5truecm]{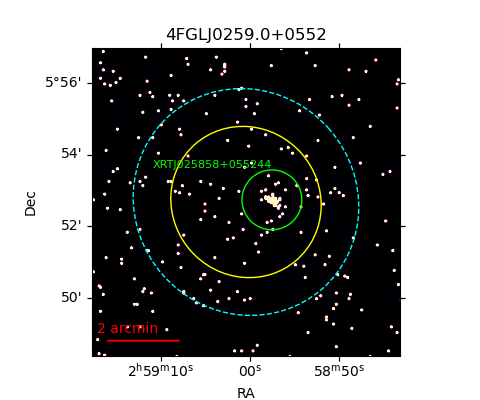}
   \includegraphics[width=5.5truecm]{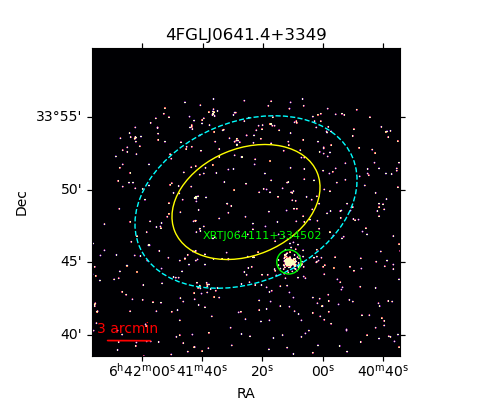}
   \includegraphics[width=5.5truecm]{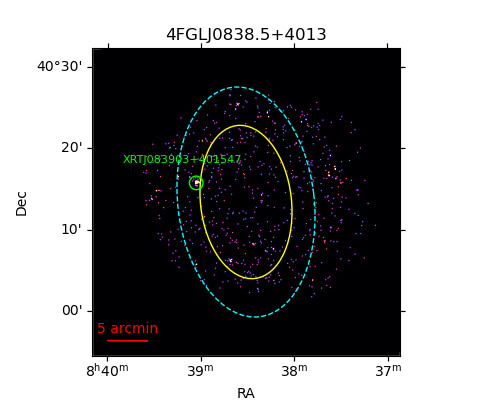}
   \includegraphics[width=5.5truecm]{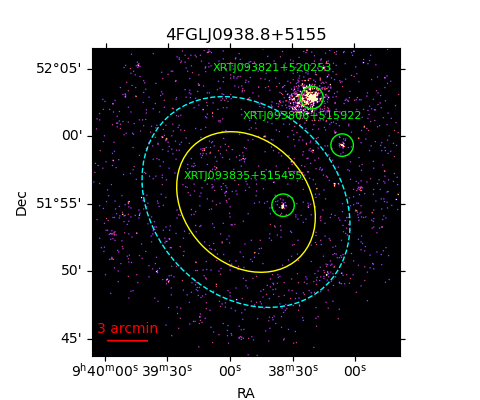}
   \includegraphics[width=5.5truecm]{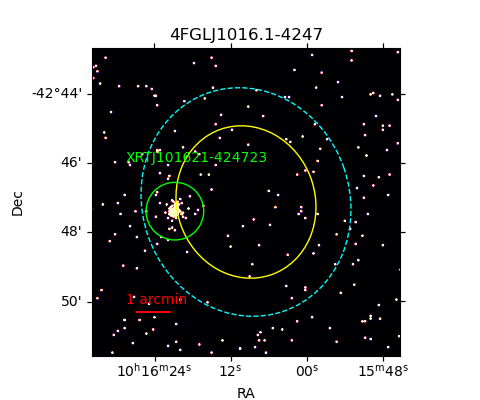}
   \includegraphics[width=5.5truecm]{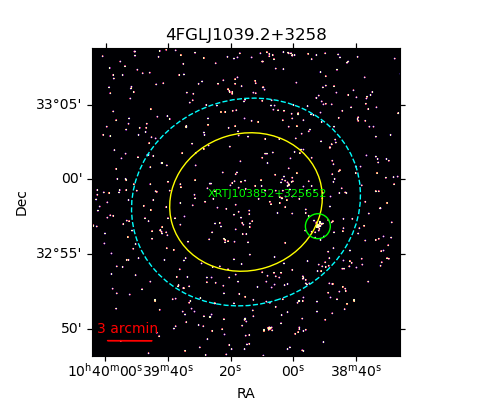}
   \includegraphics[width=5.5truecm]{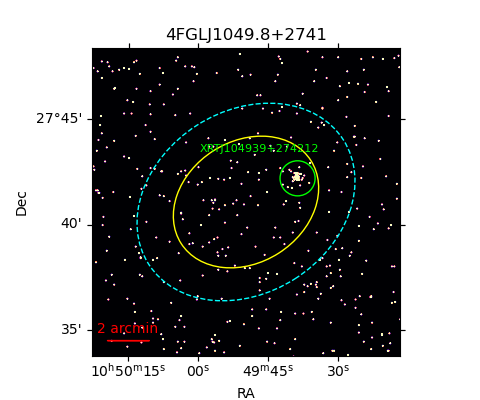}
   \includegraphics[width=5.5truecm]{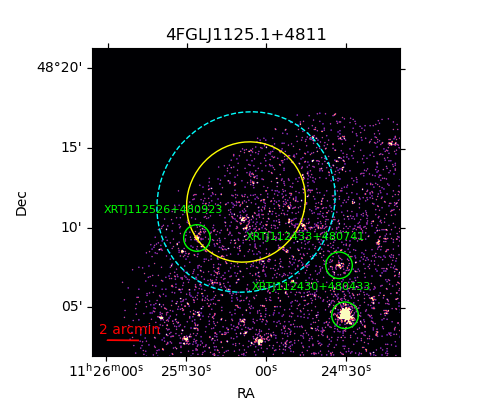}
   \includegraphics[width=5.5truecm]{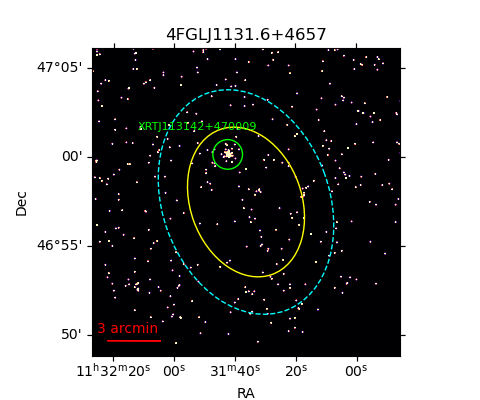}
   \includegraphics[width=5.5truecm]{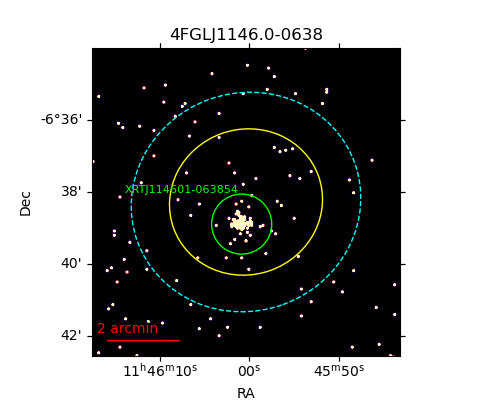}
\caption{X-ray skymaps for 33 UGS. The yellow and cyan ellipses are respectively the $2\sigma$ and $3\sigma$ \textit{Fermi} $\gamma$-ray error regions. X-ray detection, found through \textit{Swift}/XRT analysis, are reported in green.} 
\label{fig:Xskymap}
\end{figure*}%[htbp]

\setcounter{figure}{0}
\begin{figure*}%[htbp]
\center
   \includegraphics[width=5.5truecm]{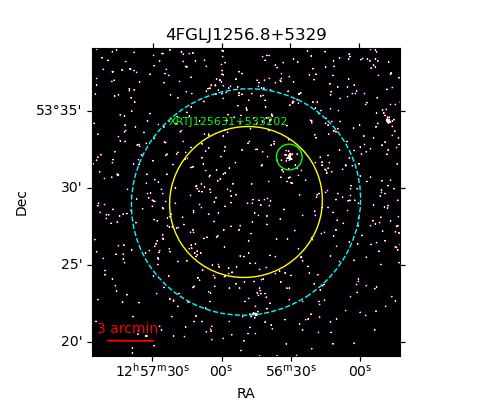}
   \includegraphics[width=5.5truecm]{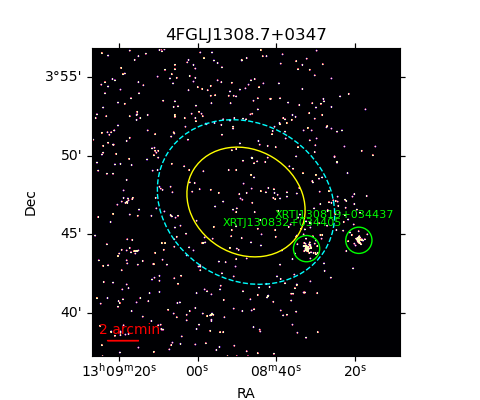}
   \includegraphics[width=5.5truecm]{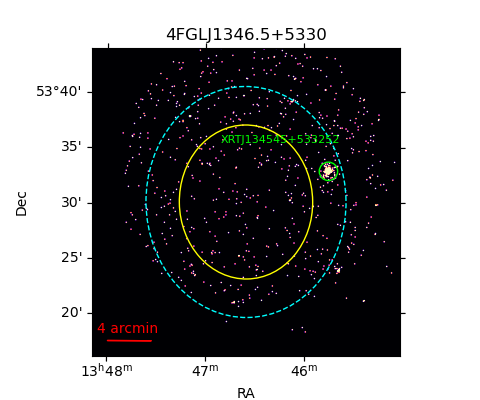}
   \includegraphics[width=5.5truecm]{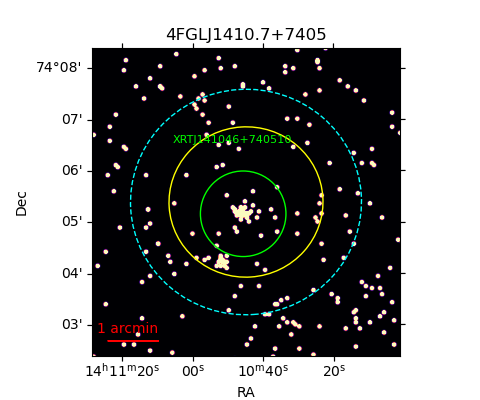}
   \includegraphics[width=5.5truecm]{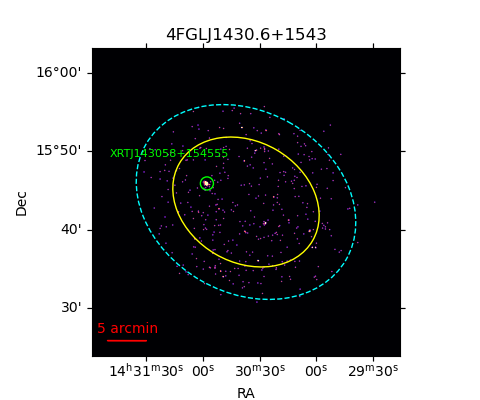}
   \includegraphics[width=5.5truecm]{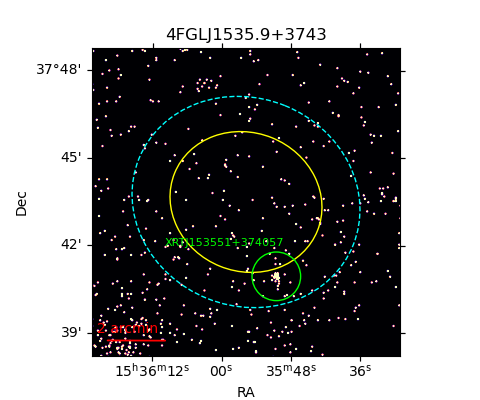}
   \includegraphics[width=5.5truecm]{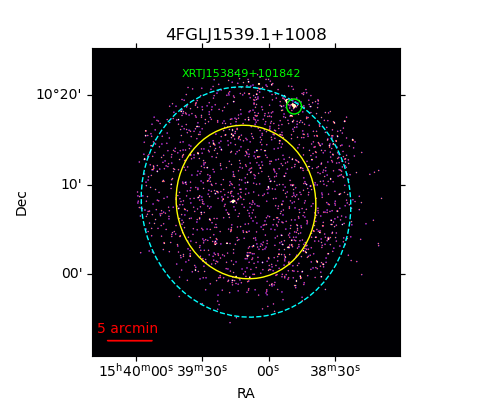}
   \includegraphics[width=5.5truecm]{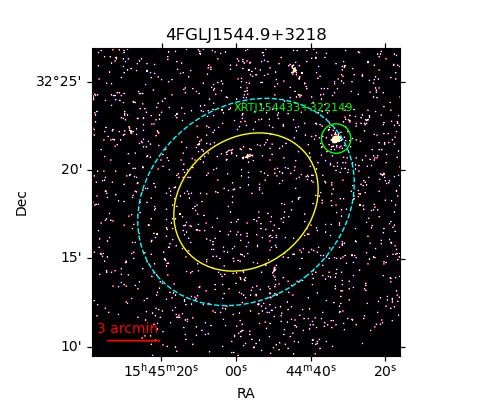}
   \includegraphics[width=5.5truecm]{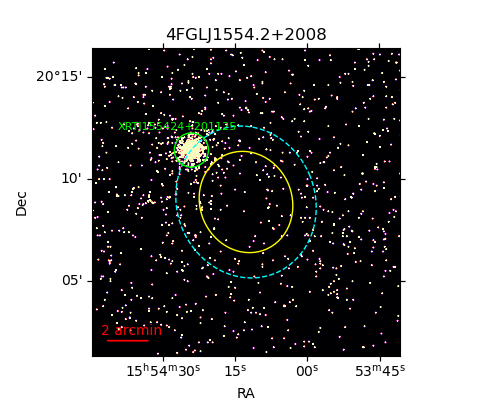}
   \includegraphics[width=5.5truecm]{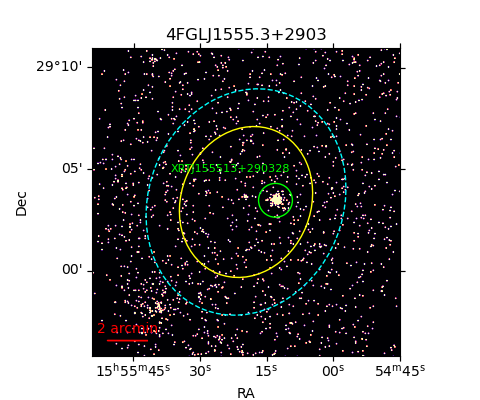}
   \includegraphics[width=5.5truecm]{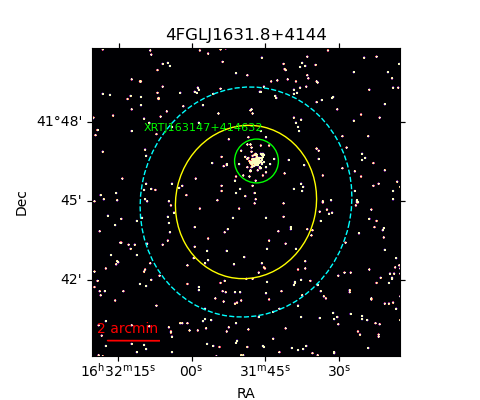}
   \includegraphics[width=5.5truecm]{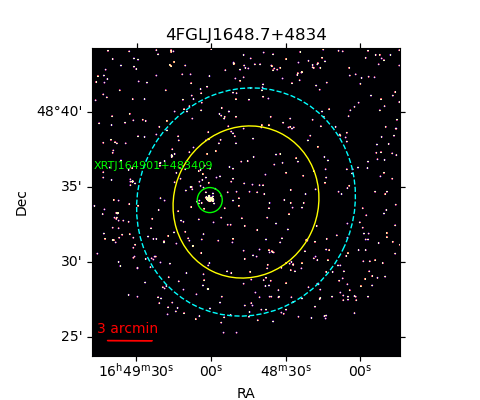}
   \includegraphics[width=5.5truecm]{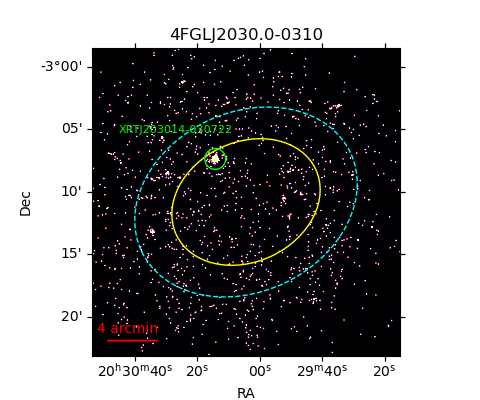}
   \includegraphics[width=5.5truecm]{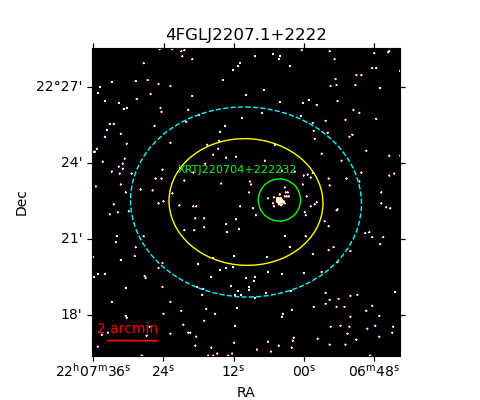}
   \includegraphics[width=5.5truecm]{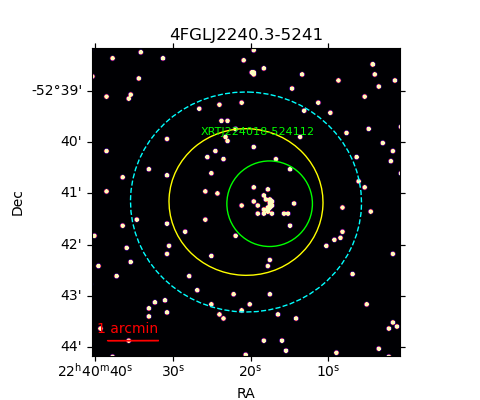}

\caption{Continued} 
%\label{fig:Xskymap}
\end{figure*}%[htbp]

\setcounter{figure}{0}
\begin{figure*}%[htbp]
\center
   \includegraphics[width=5.5truecm]{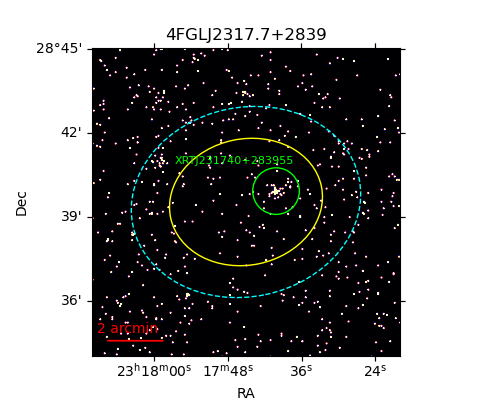}
   \includegraphics[width=5.5truecm]{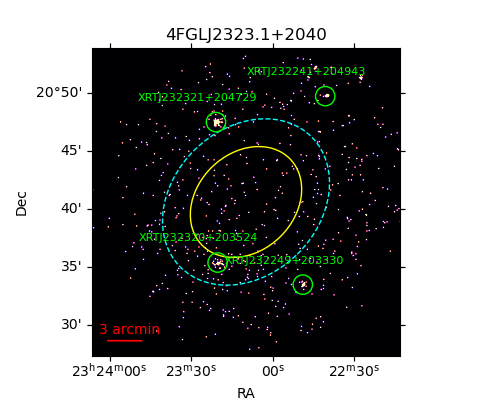}
   \includegraphics[width=5.5truecm]{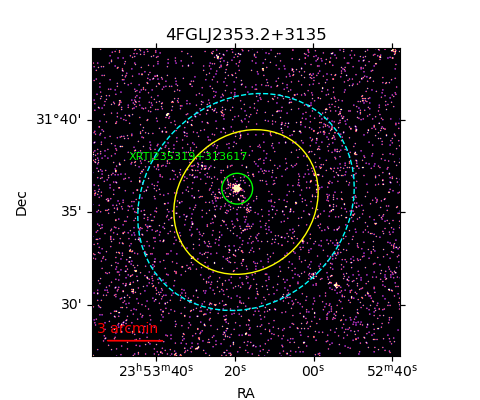}
\caption{Continued} 
%\label{fig:Xskymap}
\end{figure*}%[htbp]

%%%%%%%%%%%%%%%%%%%%%%%%%%%%%%%%%%%%%%%%%%%%%%%%%%%%%%%%%%%%%%%%%%%%

\newpage
\section{OPTICAL SKY MAPS }

In the appendix the optical skymaps for the 33 UGS analysed in this paper are shown. The green circle represent the error box of the X-ray counterpart and the red ellipses the error box of radio counterparts found or within VLASS, RACS or LoTSS catalogs.
%%%%%%%%%%%%%   SKYMAP OPT
\setcounter{figure}{1}
\begin{figure*}%[htbp]
\center
    \includegraphics[width=5.5truecm]{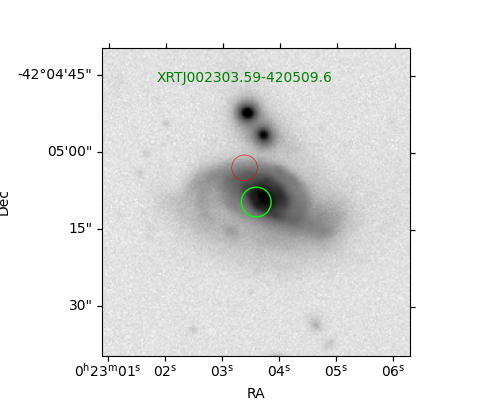}
    \includegraphics[width=5.5truecm]{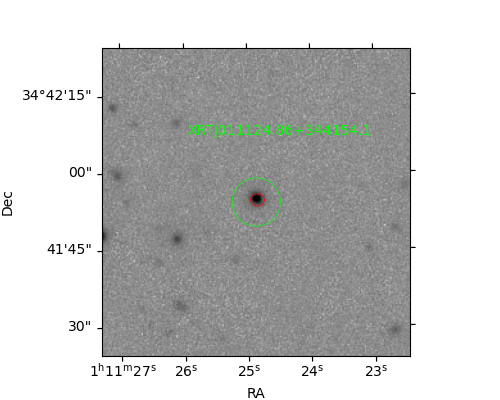}
    \includegraphics[width=5.5truecm]{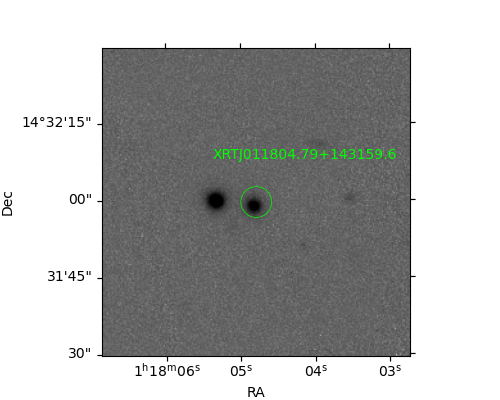}
    \includegraphics[width=5.5truecm]{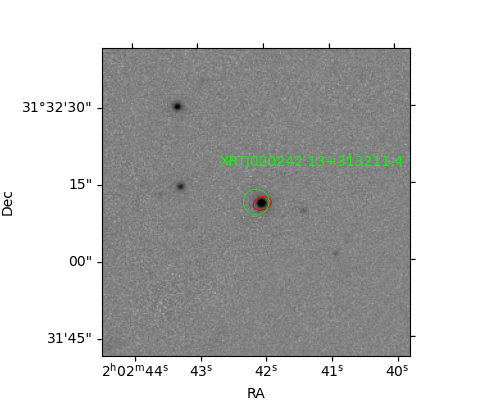}
    \includegraphics[width=5.5truecm]{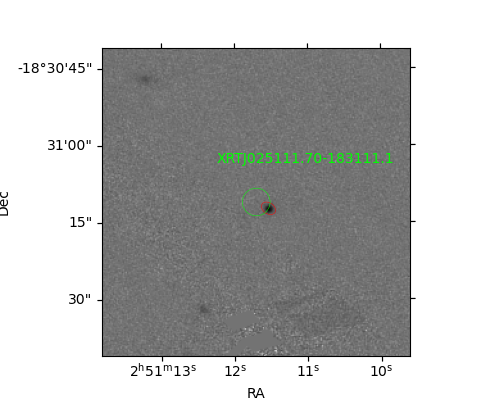}
    \includegraphics[width=5.5truecm]{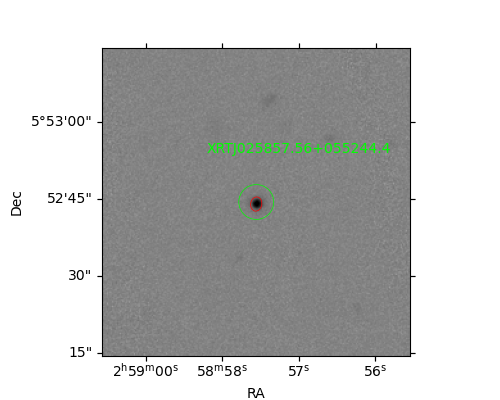}
    \includegraphics[width=5.5truecm]{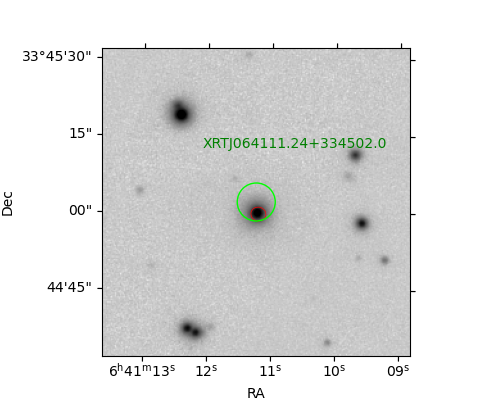}
    \includegraphics[width=5.5truecm]{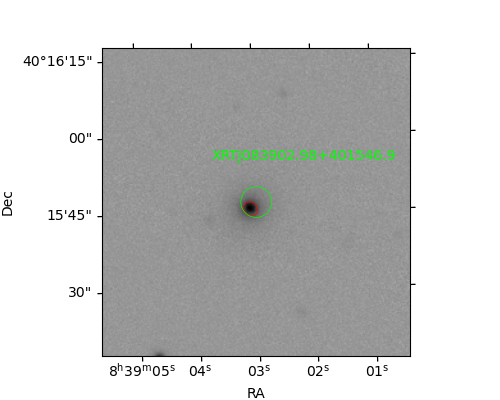}
    \includegraphics[width=5.5truecm]{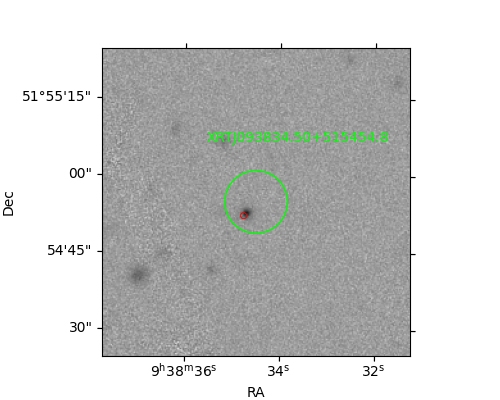}
    \includegraphics[width=5.5truecm]{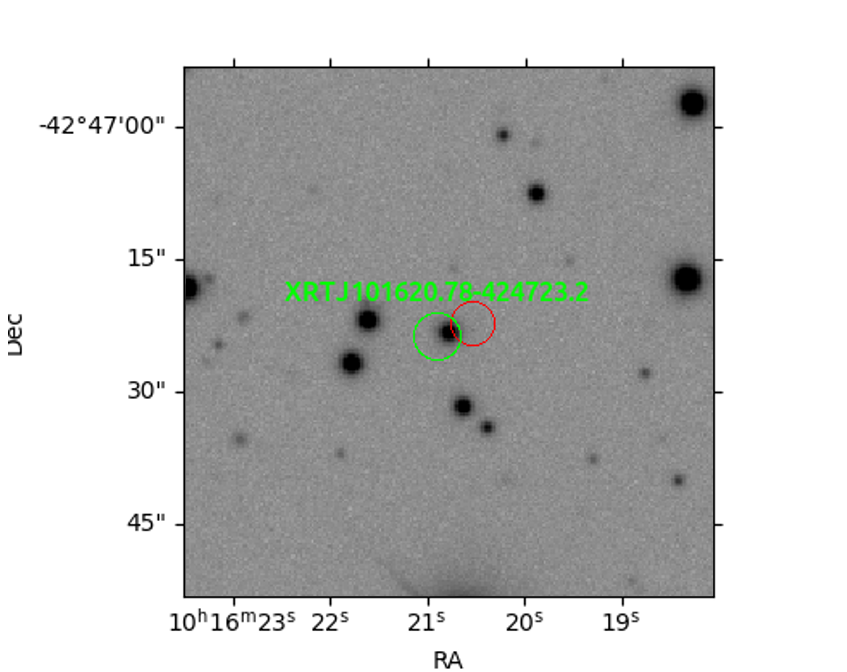}
    \includegraphics[width=5.5truecm]{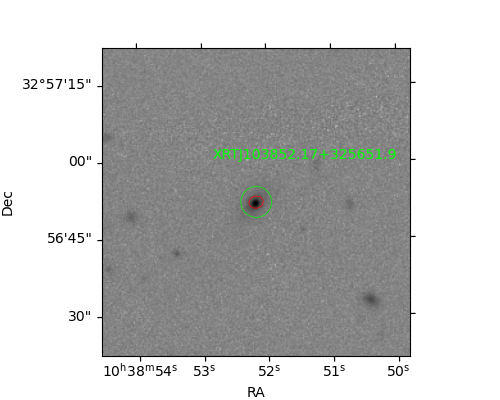}
     \includegraphics[width=5.5truecm]{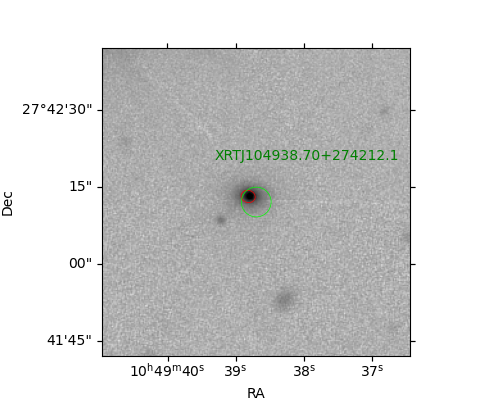}
    \includegraphics[width=5.5truecm]{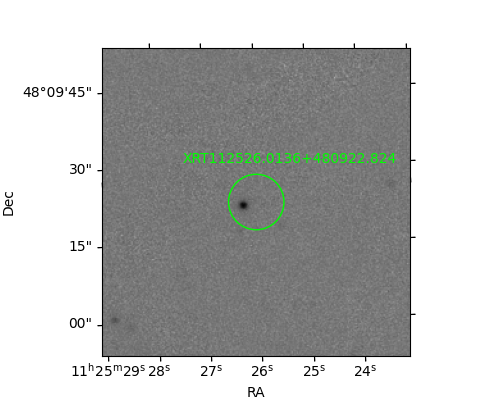}
    \includegraphics[width=5.5truecm]{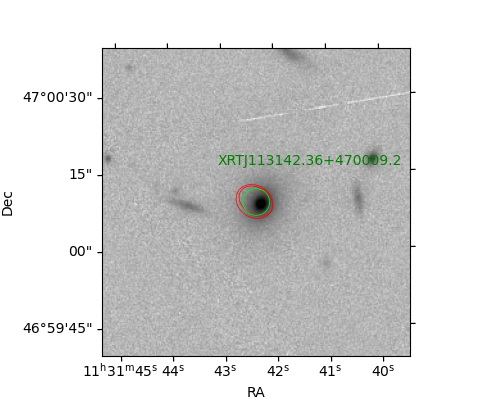}
    \includegraphics[width=5.5truecm]{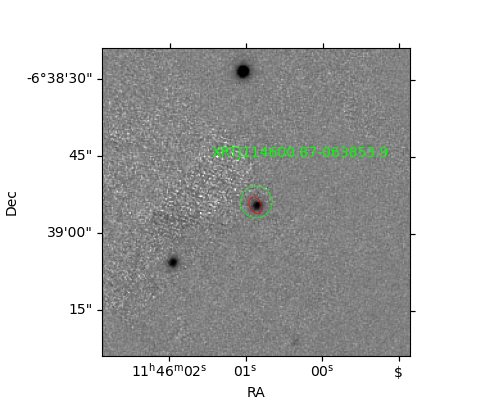}
\caption{Optical r-band skymaps for 33 UGS counterparts. The green circle represent the error box of the X-ray counterpart and the red ellipses the error box of radio counterparts found or within VLASS, RACS or LoTSS catalogs.} 
\label{fig:Oskymap}
\end{figure*}%[htbp]

\setcounter{figure}{1}
\begin{figure*}%[htbp]
\center
    \includegraphics[width=5.5truecm]{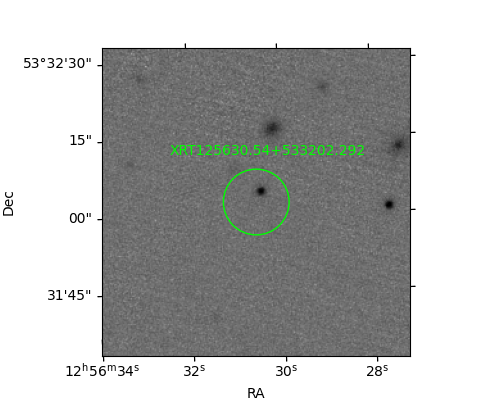}
    \includegraphics[width=5.5truecm]{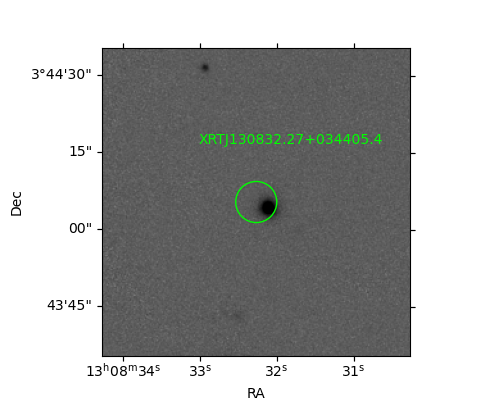}
    \includegraphics[width=5.5truecm]{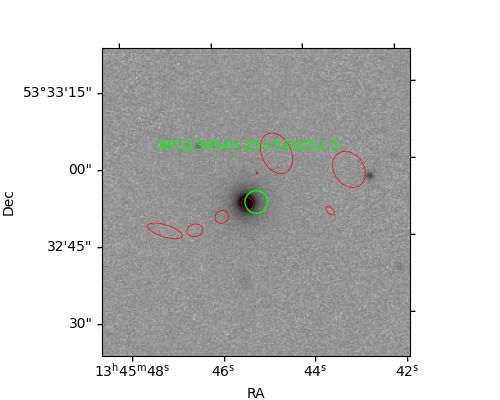}
    \includegraphics[width=5.5truecm]{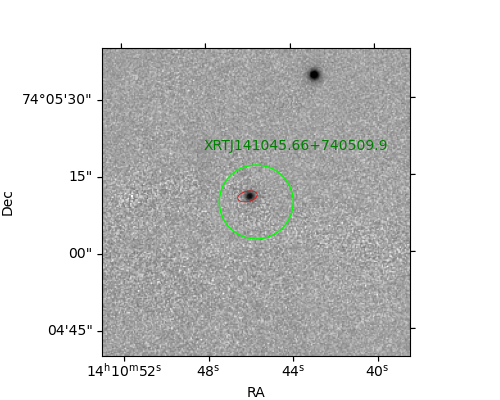}
    \includegraphics[width=5.5truecm]{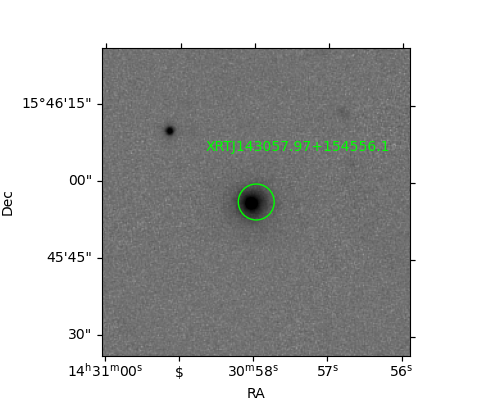}
    \includegraphics[width=5.5truecm]{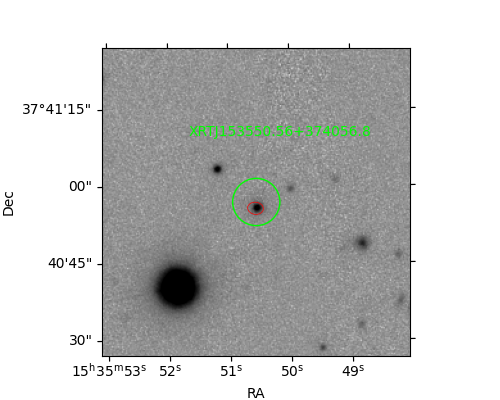}
    \includegraphics[width=5.5truecm]{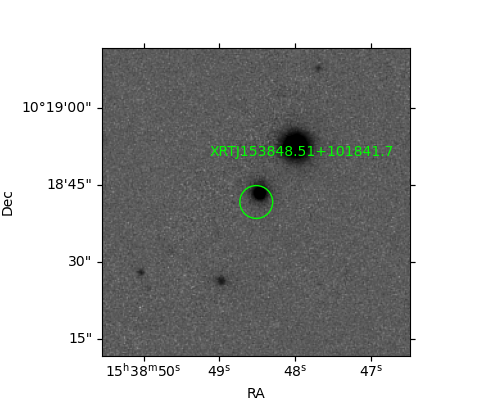}
    \includegraphics[width=5.5truecm]{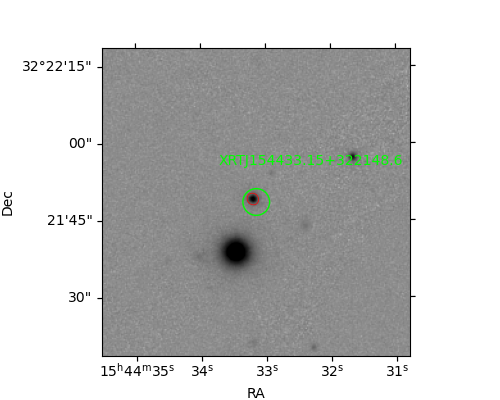}
       \includegraphics[width=5.5truecm]{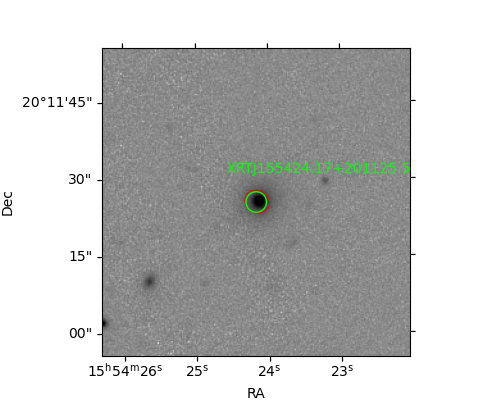}
    \includegraphics[width=5.5truecm]{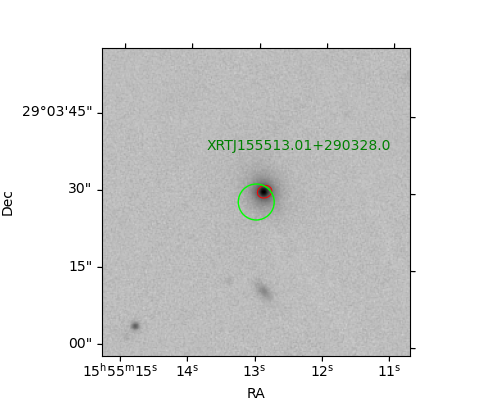}
    \includegraphics[width=5.5truecm]{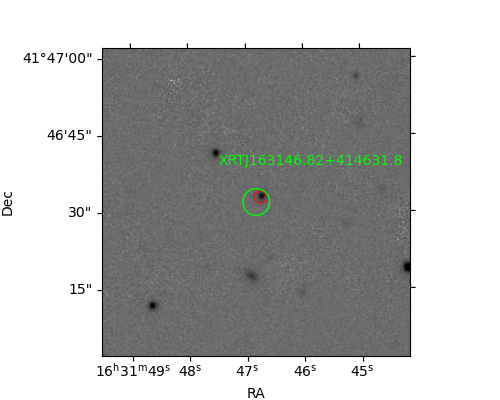}
    \includegraphics[width=5.5truecm]{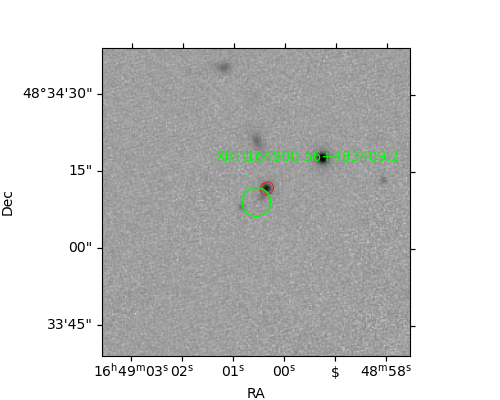}
    \includegraphics[width=5.5truecm]{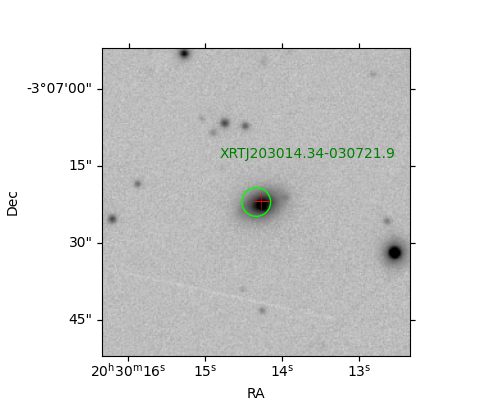}
    \includegraphics[width=5.5truecm]{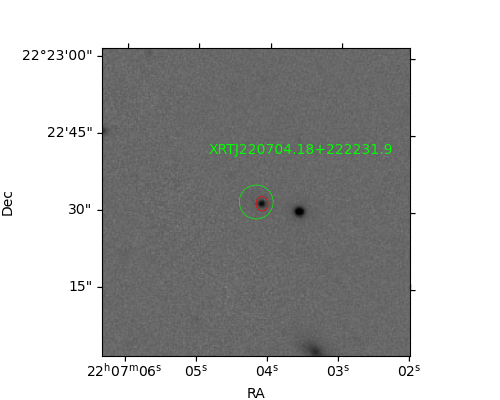}
    \includegraphics[width=5.5truecm]{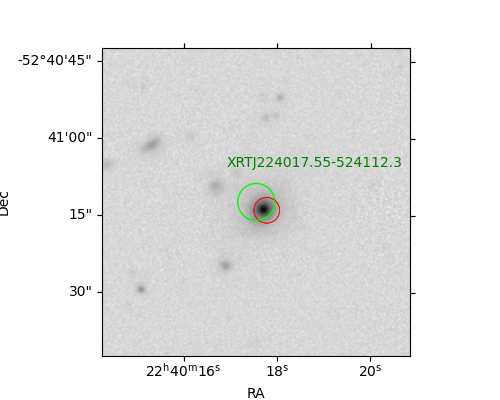}
\caption{Continued.} 
%\label{fig:Oskymap}
\end{figure*}%[htbp]

\setcounter{figure}{1}
\begin{figure*}%[htbp]
\center
    \includegraphics[width=5.5truecm]{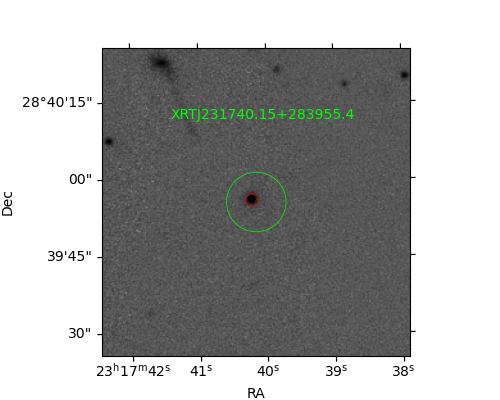}
    \includegraphics[width=5.5truecm]{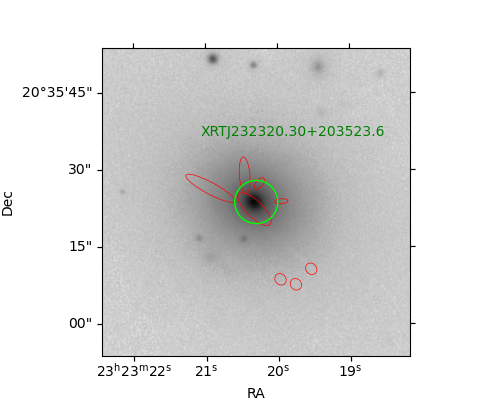}
    \includegraphics[width=5.5truecm]{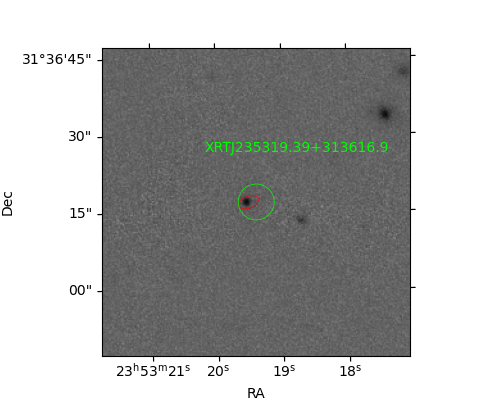}
\caption{Continued} 
%\label{fig:Oskymap}
\end{figure*}%[htbp]

\newpage

\label{lastpage}

\begin{thebibliography}{}

\bibitem[\protect\citeauthoryear{Abbott et al.}{2021}]{Abbott_2021} Abbott T.~M.~C., Adam{\'o}w M., Aguena M., Allam S., Amon A., Annis J., Avila S., et al., 2021, ApJS, 255, 20. doi:10.3847/1538-4365/ac00b3

\bibitem[\protect\citeauthoryear{Abdollahi et al.}{2020}]{4fgl} Abdollahi S., Acero F., Ackermann M., Ajello M., Atwood W.~B., Axelsson M., Baldini L., et al., 2020, ApJS, 247, 33. doi:10.3847/1538-4365/ab6bcb


\bibitem[\protect\citeauthoryear{Abdollahi et al.}{2022}]{4fgldr3} Abdollahi S., Acero F., Baldini L., Ballet J., Bastieri D., Bellazzini R., Berenji B., et al., 2022, ApJS, 260, 53. doi:10.3847/1538-4365/ac6751


\bibitem[\protect\citeauthoryear{Acero et al.}{2013}]{acero2013} Acero F., Donato D., Ojha R., Stevens J., Edwards P.~G., Ferrara E., Blanchard J., et al., 2013, ApJ, 779, 133. doi:10.1088/0004-637X/779/2/133

\bibitem[\protect\citeauthoryear{Acero et al.}{2015}]{3fgl} Acero F., Ackermann M., Ajello M., Albert A., Atwood W.~B., Axelsson M., Baldini L., et al., 2015, ApJS, 218, 23. doi:10.1088/0067-0049/218/2/23

\bibitem[\protect\citeauthoryear{Ackermann et al.}{2012}]{Ackermann_2012} Ackermann M., Ajello M., Allafort A., Antolini E., Baldini L., Ballet J., Barbiellini G., et al., 2012, ApJ, 753, 83. doi:10.1088/0004-637X/753/1/83

\bibitem[\protect\citeauthoryear{Aharonian}{2000}]{aharonian2000} Aharonian F.~A., 2000, NewA, 5, 377. doi:10.1016/S1384-1076(00)00039-7

\bibitem[\protect\citeauthoryear{Ahumada et al.}{2020}]{SDSS_2020} Ahumada R., Allende Prieto C., Almeida A., Anders F., Anderson S.~F., Andrews B.~H., Anguiano B., et al., 2020, ApJS, 249, 3. doi:10.3847/1538-4365/ab929e

\bibitem[\protect\citeauthoryear{Ajello et al.}{2014}]{Ajello_2014} Ajello M., Romani R.~W., Gasparrini D., Shaw M.~S., Bolmer J., Cotter G., Finke J., et al., 2014, ApJ, 780, 73. doi:10.1088/0004-637X/780/1/73


\bibitem[\protect\citeauthoryear{Ajello et al.}{2020}]{ajello2020} Ajello M., Di Mauro M., Paliya V.~S., Garrappa S., 2020, ApJ, 894, 88. doi:10.3847/1538-4357/ab86a6

\bibitem[\protect\citeauthoryear{Ajello et al.}{2020}]{Ajello_2020} Ajello M., Angioni R., Axelsson M., Ballet J., Barbiellini G., Bastieri D., Becerra Gonzalez J., et al., 2020, ApJ, 892, 105. doi:10.3847/1538-4357/ab791e


\bibitem[\protect\citeauthoryear{Ajello et al.}{2022}]{Ajello_2022} Ajello M., Baldini L., Ballet J., Bastieri D., Becerra Gonzalez J., Bellazzini R., Berretta A., et al., 2022, ApJS, 263, 24. doi:10.3847/1538-4365/ac9523

\bibitem[\protect\citeauthoryear{B{\"o}ttcher, Reimer, \& Zhang}{2013}]{bottcher2013} B{\"o}ttcher M., Reimer A., Zhang H., 2013, EPJWC, 61, 05003. doi:10.1051/epjconf/20136105003

\bibitem[\protect\citeauthoryear{Bruzewski et al.}{2021}]{Bruzewski_2021} Bruzewski S., Schinzel F.~K., Taylor G.~B., Petrov L., 2021, ApJ, 914, 42. doi:10.3847/1538-4357/abf73b


\bibitem[\protect\citeauthoryear{Cardelli, Clayton, \& Mathis}{1989}]{cardelli1989} Cardelli J.~A., Clayton G.~C., Mathis J.~S., 1989, ApJ, 345, 245. doi:10.1086/167900

\bibitem[\protect\citeauthoryear{CASA Team et al.}{2022}]{casa2022} Casa Team et al., 2022, PASP, 134, 1041. doi:10.1088/1538-3873/ac9642



\bibitem[\protect\citeauthoryear{Cerruti et al.}{2011}]{Cerruti2011} Cerruti M., Zech A., Boisson C., Inoue S., 2011, sf2a.conf, 555. doi:10.48550/arXiv.1111.0557

\bibitem[\protect\citeauthoryear{Cerruti}{2020}]{Cerruti2020} Cerruti M., 2020, JPhCS, 1468, 012094. doi:10.1088/1742-6596/1468/1/012094

\bibitem[\protect\citeauthoryear{Chambers et al.}{2016}]{Chambers_2016} Chambers K.~C., Magnier E.~A., Metcalfe N., Flewelling H.~A., Huber M.~E., Waters C.~Z., Denneau L., et al., 2016, arXiv, arXiv:1612.05560. doi:10.48550/arXiv.1612.05560

\bibitem[\protect\citeauthoryear{Costamante et al.}{2018}]{costamante2018} Costamante L., Cutini S., Tosti G., Antolini E., Tramacere A., 2018, MNRAS, 477, 4749. doi:10.1093/mnras/sty887

\bibitem[\protect\citeauthoryear{Coziol et al.}{2017}]{coziol2017} Coziol R., Andernach H., Torres-Papaqui J.~P., Ortega-Minakata R.~A., Moreno del Rio F., 2017, MNRAS, 466, 921. doi:10.1093/mnras/stw3164

\bibitem[\protect\citeauthoryear{Das, Gupta, \& Razzaque}{2022}]{Das2022} Das S., Gupta N., Razzaque S., 2022, icrc.conf, 1002. doi:10.22323/1.395.01002

\bibitem[\protect\citeauthoryear{D'Abrusco et al.}{2013}]{D'Abrusco_2013} D'Abrusco R., Massaro F., Paggi A., Masetti N., Tosti G., Giroletti M., Smith H.~A., 2013, ApJS, 206, 12. doi:10.1088/0067-0049/206/2/12

\bibitem[\protect\citeauthoryear{de Menezes et al.}{2019}]{demenezes2019} de Menezes R., Pe{\~n}a-Herazo H.~A., Marchesini E.~J., D'Abrusco R., Masetti N., Nemmen R., Massaro F., et al., 2019, A\&A, 630, A55. doi:10.1051/0004-6361/201936195

\bibitem[\protect\citeauthoryear{Desai et al.}{2019}]{desai2019} Desai A., Marchesi S., Rajagopal M., Ajello M., 2019, ApJS, 241, 5. doi:10.3847/1538-4365/ab01fc

\bibitem[\protect\citeauthoryear{Doert \& Errando}{2014}]{Doert_2014} Doert M., Errando M., 2014, ApJ, 782, 41. doi:10.1088/0004-637X/782/1/41

\bibitem[\protect\citeauthoryear{Evans et al.}{2009}]{Evans2009} Evans P.~A., Beardmore A.~P., Page K.~L., Osborne J.~P., O'Brien P.~T., Willingale R., Starling R.~L.~C., et al., 2009, MNRAS, 397, 1177. doi:10.1111/j.1365-2966.2009.14913.x

\bibitem[\protect\citeauthoryear{Evans et al.}{2020}]{Evans2020} Evans P.~A., Page K.~L., Osborne J.~P., Beardmore A.~P., Willingale R., Burrows D.~N., Kennea J.~A., et al., 2020, ApJS, 247, 54. doi:10.3847/1538-4365/ab7db9

\bibitem[\protect\citeauthoryear{Falcone, Stroh, \& Pryal}{2014}]{Falcone_2014} Falcone A., Stroh M., Pryal M., 2014, AAS

\bibitem[\protect\citeauthoryear{Falomo, Pian, \& Treves}{2014}]{falomo2014} Falomo R., Pian E., Treves A., 2014, A\&ARv, 22, 73. doi:10.1007/s00159-014-0073-z

\bibitem[\protect\citeauthoryear{Fronte et al.}{2023}]{Fronte_2023} Fronte L., Mazzon B., Metruccio F., Munaretto N., Doro M., Giommi P., Viale I., et al., 2023, JPhCS, 2429, 012045. doi:10.1088/1742-6596/2429/1/012045

\bibitem[\protect\citeauthoryear{Gao et al.}{2019}]{Gao2019} Gao S., Fedynitch A., Winter W., Pohl M., 2019, NatAs, 3, 88. doi:10.1038/s41550-018-0610-1

\bibitem[\protect\citeauthoryear{Garofalo et al.}{2019}]{Garofalo_2018} Garofalo D., Singh C.~B., Walsh D.~T., Christian D.~J., Jones A.~M., Zack A., Webster B., et al., 2019, RAA, 19, 013. doi:10.1088/1674-4527/19/1/13

\bibitem[\protect\citeauthoryear{Ghisellini et al.}{2017}]{Ghisellini_2017} Ghisellini G., Righi C., Costamante L., Tavecchio F., 2017, MNRAS, 469, 255. doi:10.1093/mnras/stx806

\bibitem[\protect\citeauthoryear{Giommi et al.}{2020}]{Giommi_2020} Giommi P., Glauch T., Padovani P., Resconi E., Turcati A., Chang Y.~L., 2020, MNRAS, 497, 865. doi:10.1093/mnras/staa2082

\bibitem[\protect\citeauthoryear{Giroletti et al.}{2016}]{Giroletti_2016} Giroletti M., Massaro F., D'Abrusco R., Lico R., Burlon D., Hurley-Walker N., Johnston-Hollitt M., et al., 2016, A\&A, 588, A141. doi:10.1051/0004-6361/201527817


\bibitem[\protect\citeauthoryear{Goad et al.}{2007}]{Goad2007} Goad M.~R., Tyler L.~G., Beardmore A.~P., Evans P.~A., Rosen S.~R., Osborne J.~P., Starling R.~L.~C., et al., 2007, A\&A, 476, 1401. doi:10.1051/0004-6361:20078436

\bibitem[\protect\citeauthoryear{Gordon et al.}{2021}]{Gordon_2021} Gordon Y.~A., Boyce M.~M., O'Dea C.~P., Rudnick L., Andernach H., Vantyghem A.~N., Baum S.~A., et al., 2021, ApJS, 255, 30. doi:10.3847/1538-4365/ac05c0


\bibitem[\protect\citeauthoryear{G{\"u}rkan et al.}{2019}]{gurkan2019} G{\"u}rkan G., Hardcastle M.~J., Best P.~N., Morabito L.~K., Prandoni I., Jarvis M.~J., Duncan K.~J., et al., 2019, A\&A, 622, A11. doi:10.1051/0004-6361/201833892

\bibitem[\protect\citeauthoryear{Hale et al.}{2021}]{Hale_2021} Hale C.~L., McConnell D., Thomson A.~J.~M., Lenc E., Heald G.~H., Hotan A.~W., Leung J.~K., et al., 2021, PASA, 38, e058. doi:10.1017/pasa.2021.47

\bibitem[\protect\citeauthoryear{HI4PI Collaboration et al.}{2016}]{HI4PI_coll_2016} HI4PI Collaboration, Ben Bekhti N., Flöer L., Keller R., Kerp J., Lenz D., Winkel B., et al., 2016, A\&A, 594, A116. doi:10.1051/0004-6361/201629178

\bibitem[\protect\citeauthoryear{J{\"a}rvel{\"a}, Berton, \& Crepaldi}{2021}]{jarvela2021} J{\"a}rvel{\"a} E., Berton M., Crepaldi L., 2021, FrASS, 8, 147. doi:10.3389/fspas.2021.735310


\bibitem[\protect\citeauthoryear{Joffre et al.}{2022}]{joffre2022} Joffre S., Silver R., Rajagopal M., Ajello M., Torres-Alb{\`a} N., Pizzetti A., Marchesi S., et al., 2022, ApJ, 940, 139. doi:10.3847/1538-4357/ac9797


\bibitem[\protect\citeauthoryear{Kaur et al.}{2019}]{Kaur_2019b} Kaur A., Falcone A.~D., Stroh M.~D., Kennea J.~A., Ferrara E.~C., 2019, ApJ, 887, 18. doi:10.3847/1538-4357/ab4ceb

\bibitem[\protect\citeauthoryear{Kaur, Kerby, \& Falcone}{2023}]{kaur2023} Kaur A., Kerby S., Falcone A.~D., 2023, ApJ, 943, 167. doi:10.3847/1538-4357/ac8b80

\bibitem[\protect\citeauthoryear{Kellermann et al.}{1989}]{kellermann1989} Kellermann K.~I., Sramek R., Schmidt M., Shaffer D.~B., Green R., 1989, AJ, 98, 1195. doi:10.1086/115207

\bibitem[\protect\citeauthoryear{Kerby et al.}{2021}]{kerby2021} Kerby S., Kaur A., Falcone A.~D., Eskenasy R., Hancock F., Stroh M.~C., Ferrara E.~C., et al., 2021, ApJ, 923, 75. doi:10.3847/1538-4357/ac2e91

\bibitem[\protect\citeauthoryear{Komossa}{2008}]{komossa2008} Komossa S., 2008, RMxAC, 32, 86. doi:10.48550/arXiv.0710.3326


\bibitem[\protect\citeauthoryear{Landi et al.}{2015}]{landi2015} Landi R., Bassani L., Stephen J.~B., Masetti N., Malizia A., Ubertini P., 2015, A\&A, 581, A57. doi:10.1051/0004-6361/201526221



\bibitem[\protect\citeauthoryear{Mannheim}{1993}]{mannheim1993} Mannheim K., 1993, A\&A, 269, 67. doi:10.48550/arXiv.astro-ph/9302006

\bibitem[\protect\citeauthoryear{Maraschi, Ghisellini, \& Celotti}{1992}]{maraschi1992} Maraschi L., Ghisellini G., Celotti A., 1992, ApJL, 397, L5. doi:10.1086/186531

\bibitem[\protect\citeauthoryear{Marcha et al.}{1996}]{marcha1996} Marcha M.~J.~M., Browne I.~W.~A., Impey C.~D., Smith P.~S., 1996, MNRAS, 281, 425. doi:10.1093/mnras/281.2.425

\bibitem[\protect\citeauthoryear{Marchesi, Kaur, \& Ajello}{2018}]{Marchesi_2018} Marchesi S., Kaur A., Ajello M., 2018, AJ, 156, 212. doi:10.3847/1538-3881/aae201

\bibitem[\protect\citeauthoryear{Marchesini et al.}{2020}]{Marchesini_2020} Marchesini E.~J., Paggi A., Massaro F., Masetti N., D'Abrusco R., Andruchow I., 2020, A\&A, 638, A128. doi:10.1051/0004-6361/201936928


\bibitem[\protect\citeauthoryear{Marchesini et al.}{2023}]{marchesini2023} Marchesini E.~J., Reynaldi V., Vieyro F., Saponara J., Andruchow I., L{\'o}pez I.~E., Benaglia P., et al., 2023, A\&A, 670, A91. doi:10.1051/0004-6361/202244899

%\bibitem[\protect\citeauthoryear{Massaro et al.}{2013a}]{Massaro_2013a} Massaro F., D'Abrusco R., Paggi A., Masetti N., Giroletti M., Tosti G., Smith H.~A., 2013, arXiv, arXiv:1303.3267. doi:10.48550/arXiv.1303.3267


%bibitem[\protect\citeauthoryear{Massaro et al.}{2013b}]{Massaro_2013b} Massaro F., D'Abrusco R., Paggi A., Masetti N., Giroletti M., Tosti G., Smith H.~A., et al., 2013, ApJS, 206, 13. doi:10.1088/0067-0049/206/2/13


%\bibitem[\protect\citeauthoryear{Massaro et al.}{2013c}]{Massaro_2013c} Massaro F., D'Abrusco R., Giroletti M., Paggi A., Masetti N., Tosti G., Nori M., et al., 2013, ApJS, 207, 4. doi:10.1088/0067-0049/207/1/4


%\bibitem[\protect\citeauthoryear{Massaro et al.}{2013d}]{Massaro_2013d} Massaro F., D'Abrusco R., Paggi A., Masetti N., Giroletti M., Tosti G., Smith H.~A., et al., 2013, ApJS, 209, 10. doi:10.1088/0067-0049/209/1/10

%\bibitem[\protect\citeauthoryear{Massaro et al.}{2014}]{Massaro_2014} Massaro F., Masetti N., D'Abrusco R., Paggi A., Funk S., 2014, AJ, 148, 66. doi:10.1088/0004-6256/148/4/66

\bibitem[\protect\citeauthoryear{Massaro et al.}{2015}]{Massaro_2015} Massaro F., Landoni M., D'Abrusco R., Milisavljevic D., Paggi A., Masetti N., Smith H.~A., et al., 2015, A\&A, 575, A124. doi:10.1051/0004-6361/201425119

\bibitem[\protect\citeauthoryear{Massaro et al.}{2016}]{Massaro_2016} Massaro F., {\'A}lvarez Crespo N., D'Abrusco R., Landoni M., Masetti N., Ricci F., Milisavljevic D., et al., 2016, Ap\&SS, 361, 337. doi:10.1007/s10509-016-2926-6

\bibitem[\protect\citeauthoryear{Massaro et al.}{2017}]{massaro2017} Massaro F., Marchesini E.~J., D'Abrusco R., Masetti N., Andruchow I., Smith H.~A., 2017, ApJ, 834, 113. doi:10.3847/1538-4357/834/2/113

\bibitem[\protect\citeauthoryear{Mirabal et al.}{2012}]{Mirabal_2012} Mirabal N., Fr{\'\i}as-Martinez V., Hassan T., Fr{\'\i}as-Martinez E., 2012, MNRAS, 424, L64. doi:10.1111/j.1745-3933.2012.01287.x

\bibitem[\protect\citeauthoryear{Monet et al.}{2003}]{Monet_2003} Monet D.~G., Levine S.~E., Canzian B., Ables H.~D., Bird A.~R., Dahn C.~C., Guetter H.~H., et al., 2003, AJ, 125, 984. doi:10.1086/345888

\bibitem[\protect\citeauthoryear{Monroe et al.}{2016}]{monroe2016} Monroe T.~R., Prochaska J.~X., Tejos N., Worseck G., Hennawi J.~F., Schmidt T., Tumlinson J., et al., 2016, AJ, 152, 25. doi:10.3847/0004-6256/152/1/25

\bibitem[\protect\citeauthoryear{Murase, Oikonomou, \& Petropoulou}{2018}]{Murase2018} Murase K., Oikonomou F., Petropoulou M., 2018, ApJ, 865, 124. doi:10.3847/1538-4357/aada00

\bibitem[\protect\citeauthoryear{Neeleman et al.}{2016}]{Neeleman2016} Neeleman M., Prochaska J.~X., Ribaudo J., Lehner N., Howk J.~C., Rafelski M., Kanekar N., 2016, yCat, J/ApJ/818/113. doi:10.26093/cds/vizier.18180113

\bibitem[\protect\citeauthoryear{Nolan et al.}{2012}]{2fgl} Nolan P.~L., Abdo A.~A., Ackermann M., Ajello M., Allafort A., Antolini E., Atwood W.~B., et al., 2012, ApJS, 199, 31. doi:10.1088/0067-0049/199/2/31

\bibitem[\protect\citeauthoryear{Nori et al.}{2014}]{Nori_2014} Nori M., Giroletti M., Massaro F., D'Abrusco R., Paggi A., Tosti G., Funk S., 2014, ApJS, 212, 3. doi:10.1088/0067-0049/212/1/3

\bibitem[\protect\citeauthoryear{Osterbrock}{1980}]{Osterbrock_1980} Osterbrock D.~E., 1980, NYASA, 336, 22. doi:10.1111/j.1749-6632.1980.tb15916.x

\bibitem[\protect\citeauthoryear{Padovani et al.}{2017}]{Padovani_2017} Padovani P., Alexander D.~M., Assef R.~J., De Marco B., Giommi P., Hickox R.~C., Richards G.~T., et al., 2017, A\&ARv, 25, 2. doi:10.1007/s00159-017-0102-9

\bibitem[\protect\citeauthoryear{Padovani et al.}{2022}]{Padovani_2022} Padovani P., Giommi P., Falomo R., Oikonomou F., Petropoulou M., Glauch T., Resconi E., et al., 2022, MNRAS, 510, 2671. doi:10.1093/mnras/stab3630

\bibitem[\protect\citeauthoryear{Paiano et al.}{2017a}]{Paiano_2017_TeV} Paiano S., Landoni M., Falomo R., Treves A., Scarpa R., Righi C., 2017, ApJ, 837, 144. doi:10.3847/1538-4357/837/2/144

\bibitem[\protect\citeauthoryear{Paiano et al.}{2017b}]{Paiano_hz} Paiano S., Landoni M., Falomo R., Treves A., Scarpa R., 2017, ApJ, 844, 120. doi:10.3847/1538-4357/aa7aac


\bibitem[\protect\citeauthoryear{Paiano, Franceschini, \& Stamerra}{2017c}]{paiano2017_sed} Paiano S., Franceschini A., Stamerra A., 2017, MNRAS, 468, 4902. doi:10.1093/mnras/stx749

\bibitem[\protect\citeauthoryear{Paiano et al.}{2017d}]{paiano2017_ufo1} Paiano S., Falomo R., Franceschini A., Treves A., Scarpa R., 2017, ApJ, 851, 135. doi:10.3847/1538-4357/aa9af4

\bibitem[\protect\citeauthoryear{Paiano et al.}{2019}]{paiano2019_ufo2} Paiano S., Falomo R., Treves A., Franceschini A., Scarpa R., 2019, ApJ, 871, 162. doi:10.3847/1538-4357/aaf6e4

\bibitem[\protect\citeauthoryear{Paiano et al.}{2020}]{Paiano_TeV} Paiano S., Falomo R., Treves A., Scarpa R., 2020, MNRAS, 497, 94. doi:10.1093/mnras/staa1840

\bibitem[\protect\citeauthoryear{Paiano et al.}{2021}]{Paiano_2021} Paiano S., Falomo R., Treves A., Padovani P., Giommi P., Scarpa R., 2021, MNRAS, 504, 3338. doi:10.1093/mnras/stab1034

\bibitem[\protect\citeauthoryear{Paiano et al.}{2023}]{Paiano_2023} Paiano S., Falomo R., Treves A., Padovani P., Giommi P., Scarpa R., Bisogni S., et al., 2023, MNRAS, 521, 2270. doi:10.1093/mnras/stad573


\bibitem[\protect\citeauthoryear{Pajdosz-{\'S}mierciak, {\'S}mierciak, \& Jamrozy}{2022}]{pajdosz2022} Pajdosz-{\'S}mierciak U., {\'S}mierciak B., Jamrozy M., 2022, MNRAS, 514, 2122. doi:10.1093/mnras/stac1372

\bibitem[\protect\citeauthoryear{Paliya et al.}{2021}]{Paliya_2021} Paliya V.~S., Dominguez A., Ajello M., Olmo-Garcia A., Hartmann D., 2021, yCat, J/ApJS/253/46

\bibitem[\protect\citeauthoryear{Perlmutter}{2000}]{Perlmutter2000} Perlmutter S., 2000, IJMPA, 15, 715. doi:10.1142/S0217751X00005383

\bibitem[\protect\citeauthoryear{Petrov et al.}{2013}]{Petrov_2013} Petrov L., Mahony E.~K., Edwards P.~G., Sadler E.~M., Schinzel F.~K., McConnell D., 2013, MNRAS, 432, 1294. doi:10.1093/mnras/stt550


\bibitem[\protect\citeauthoryear{Rakshit et al.}{2017}]{rakshit2017} Rakshit S., Stalin C.~S., Chand H., Zhang X.-G., 2017, ApJS, 229, 39. doi:10.3847/1538-4365/aa6971

\bibitem[\protect\citeauthoryear{Rajagopal et al.}{2021}]{Rajagopal_2021} Rajagopal M., Marchesi S., Kaur A., Dom{\'\i}nguez A., Silver R., Ajello M., 2021, ApJS, 254, 26. doi:10.3847/1538-4365/abf656


\bibitem[\protect\citeauthoryear{Rajagopal et al.}{2023}]{rajagopal2023} Rajagopal M., Marcotulli L., Labrie K., Marchesi S., Ajello M., 2023, AJ, 165, 42. doi:10.3847/1538-3881/aca1be

\bibitem[\protect\citeauthoryear{Rodrigues et al.}{2019}]{Rodrigues2019} Rodrigues X., Gao S., Fedynitch A., Palladino A., Winter W., 2019, ApJL, 874, L29. doi:10.3847/2041-8213/ab1267

\bibitem[\protect\citeauthoryear{Romani et al.}{2011}]{Romani_2011} Romani R.~W., Kerr M., Craig H.~A., Johnston S., Cognard I., Smith D.~A., 2011, ApJ, 738, 114. doi:10.1088/0004-637X/738/1/114

\bibitem[\protect\citeauthoryear{Rusinek-Abarca \& Sikora}{2021}]{rusinek2021} Rusinek-Abarca K., Sikora M., 2021, ApJ, 922, 202. doi:10.3847/1538-4357/ac2429

\bibitem[\protect\citeauthoryear{Salvetti et al.}{2017}]{Salvetti_2017a} Salvetti D., Mignani R.~P., De Luca A., Marelli M., Pallanca C., Breeveld A.~A., H{\"u}semann P., et al., 2017, MNRAS, 470, 466. doi:10.1093/mnras/stx1247

\bibitem[\protect\citeauthoryear{Salvetti et al.}{2017}]{Salvetti_2017b} Salvetti D., Chiaro G., La Mura G., Thompson D.~J., 2017, MNRAS, 470, 1291. doi:10.1093/mnras/stx1328

\bibitem[\protect\citeauthoryear{Sbarufatti et al.}{2005}]{sbarufatti2005} Sbarufatti B., Treves A., Falomo R., Heidt J., Kotilainen J., Scarpa R., 2005, AJ, 129, 559. doi:10.1086/427138

\bibitem[\protect\citeauthoryear{Schinzel et al.}{2015}]{Schinzel_2015} Schinzel F.~K., Petrov L., Taylor G.~B., Mahony E.~K., Edwards P.~G., Kovalev Y.~Y., 2015, ApJS, 217, 4. doi:10.1088/0067-0049/217/1/4

\bibitem[\protect\citeauthoryear{Schinzel et al.}{2017}]{Schinzel_2017} Schinzel F.~K., Petrov L., Taylor G.~B., Edwards P.~G., 2017, ApJ, 838, 139. doi:10.3847/1538-4357/aa6439

\bibitem[\protect\citeauthoryear{Seehars et al.}{2016}]{Seehars2016} Seehars S., Grandis S., Amara A., Refregier A., 2016, PhRvD, 93, 103507. doi:10.1103/PhysRevD.93.103507

\bibitem[\protect\citeauthoryear{Shaw et al.}{2009}]{Shaw_2009} Shaw M.~S., Romani R.~W., Healey S.~E., Cotter G., Michelson P.~F., Readhead A.~C.~S., 2009, ApJ, 704, 477. doi:10.1088/0004-637X/704/1/477

\bibitem[\protect\citeauthoryear{Shaw et al.}{2012}]{Shaw_2012} Shaw M.~S., Romani R.~W., Cotter G., Healey S.~E., Michelson P.~F., Readhead A.~C.~S., Richards J.~L., et al., 2012, ApJ, 748, 49. doi:10.1088/0004-637X/748/1/49

\bibitem[\protect\citeauthoryear{Shaw et al.}{2013}]{Shaw_2013} Shaw M.~S., Romani R.~W., Cotter G., Healey S.~E., Michelson P.~F., Readhead A.~C.~S., Richards J.~L., et al., 2013, ApJ, 764, 135. doi:10.1088/0004-637X/764/2/135

\bibitem[\protect\citeauthoryear{Shimwell et al.}{2022}]{LoTSS_2022} Shimwell T.~W., Hardcastle M.~J., Tasse C., Best P.~N., R{\"o}ttgering H.~J.~A., Williams W.~L., Botteon A., et al., 2022, A\&A, 659, A1. doi:10.1051/0004-6361/202142484


\bibitem[\protect\citeauthoryear{Stephen et al.}{2010}]{stephen2010} Stephen J.~B., Bassani L., Landi R., Malizia A., Sguera V., Bazzano A., Masetti N., 2010, MNRAS, 408, 422. doi:10.1111/j.1365-2966.2010.17123.x

\bibitem[\protect\citeauthoryear{Stroh \& Falcone}{2013}]{Stroh_2013} Stroh M.~C., Falcone A.~D., 2013, ApJS, 207, 28. doi:10.1088/0067-0049/207/2/28

\bibitem[\protect\citeauthoryear{Takahashi et al.}{2012}]{takahashi2012} Takahashi Y., Kataoka J., Nakamori T., Maeda K., Makiya R., Totani T., Cheung C.~C., et al., 2012, ApJ, 747, 64. doi:10.1088/0004-637X/747/1/64

\bibitem[\protect\citeauthoryear{Takeuchi et al.}{2013}]{takeuchi2013} Takeuchi Y., Kataoka J., Maeda K., Takahashi Y., Nakamori T., Tahara M., 2013, ApJS, 208, 25. doi:10.1088/0067-0049/208/2/25

\bibitem[\protect\citeauthoryear{Urry et al.}{2000}]{urry2000} Urry C.~M., Scarpa R., O'Dowd M., Falomo R., Pesce J.~E., Treves A., 2000, ApJ, 532, 816. doi:10.1086/308616

\bibitem[\protect\citeauthoryear{Uslenghi \& Falomo}{2011}]{uslenghi2011} Uslenghi M., Falomo R., 2011, SPIE, 8135, 813524. doi:10.1117/12.913305

\bibitem[\protect\citeauthoryear{Wang, Wang, \& Dong}{2009}]{wang2009} Wang D.-L., Wang J.-G., Dong X.-B., 2009, RAA, 9, 1078. doi:10.1088/1674-4527/9/10/002



\end{thebibliography}
\end{document}